\documentclass[
  showpacs,
  twocolumn,
  pra,
  superscriptaddress,
  notitlepage,
  nofootinbib,
  floatfix,
  longbibliography
]{revtex4-2}

\usepackage[left]{lineno}

\usepackage{amsthm}
\usepackage{etoolbox}
\usepackage{graphicx}
\usepackage{amsmath,amssymb,amsthm,mathrsfs,amsfonts,dsfont,amscd,keyval}
\usepackage{mathtools,mathrsfs}
\usepackage{xcolor}
\usepackage{hyperref}
\usepackage{url}
\usepackage{array}
\usepackage{tabularx}
\usepackage{fancyhdr}
\usepackage{enumerate}
\usepackage{multirow}
\usepackage{physics}
\usepackage{tikz}
\usepackage{booktabs}

\newcolumntype{L}[1]{>{\raggedright\arraybackslash}p{#1}}

\usetikzlibrary{shapes,arrows,positioning,decorations.pathreplacing}

\hypersetup{colorlinks=true, linkcolor=blue, citecolor=purple, urlcolor=black}

\definecolor{mygreen}{HTML}{2E7D32}  

\newcommand{\ie}{\textit{i.e.}}
\newcommand{\eg}{\textit{e.g.}}

\makeatletter
\def\l@subsubsection#1#2{} 
\newcount\AIQ@authorlinecount
\pretocmd{\frontmatter@author@produce@script}{\AIQ@authorlinecount=0\relax}{}{}
\apptocmd{\@author@present@script}{%
  \advance\AIQ@authorlinecount\@ne
  \ifnum\AIQ@authorlinecount=6 \\\fi
  \ifnum\AIQ@authorlinecount=11 \\\fi
  \ifnum\AIQ@authorlinecount=15 \\\fi
}{}{}
\makeatother

\begin{document}

\title{When AI meets quantum information: A comprehensive review}

\author{Min Chen}
\email{mic380@pitt.edu}
\affiliation{Department of Computer Science, University of Pittsburgh, Pittsburgh, Pennsylvania 15260, USA}
\author{Yu Gan}
\affiliation{Department of Computer Science, University of Pittsburgh, Pittsburgh, Pennsylvania 15260, USA}
\author{Xin Jin}
\affiliation{Department of Computer Science, University of Pittsburgh, Pittsburgh, Pennsylvania 15260, USA}
\author{Yuqing Li}
\affiliation{Department of Computer Science, University of Pittsburgh, Pittsburgh, Pennsylvania 15260, USA}
\author{Junqi Wang}
\affiliation{Department of Computer Science, University of Pittsburgh, Pittsburgh, Pennsylvania 15260, USA}
\author{Zeguan Wu}
\affiliation{Department of Computer Science, University of Pittsburgh, Pittsburgh, Pennsylvania 15260, USA}
\author{Yunfei Wang}
\affiliation{Joint Center for Quantum Information and Computer Science, NIST and University of Maryland, College Park, MD 20742, USA}
\affiliation{Joint Quantum Institute, NIST and University of Maryland, College Park, MD 20742, USA}
\author{Bingzhi Zhang}
\affiliation{Ming Hsieh Department of Electrical and Computer Engineering, University of Southern California, Los Angeles, CA, USA}
\author{Priyam Srivastava}
\affiliation{Department of Informatics and Networked Systems, School of Computing and Information, University of Pittsburgh, Pittsburgh, Pennsylvania 15260, USA}
\author{Tianlong Chen}
\affiliation{Department of Computer Science, The University of North Carolina at Chapel Hill, Chapel Hill, United States}
\author{Ankit Kulshrestha}
\affiliation{Fujitsu Research of America, Santa Clara, CA 95054, USA}
\author{Yuan Liu}
\affiliation{Department of Electrical and Computer Engineering, North Carolina State University, Raleigh, North Carolina 27606, United States}
\affiliation{Department of Computer Science, North Carolina State University, Raleigh, North Carolina 27606, United States}
\affiliation{Department of Physics and Astronomy, North Carolina State University, Raleigh, North Carolina 27695, United States}
\author{Juan Jos\'e Mendoza-Arenas}
\affiliation{Department of Mechanical Engineering and Materials Science, University of Pittsburgh, Pittsburgh, PA 15261, USA}
\affiliation{Department of Physics and Astronomy, University of Pittsburgh, PA 15261, USA}
\author{Kaushik P. Seshadreesan}
\affiliation{Department of Informatics and Networked Systems, School of Computing and Information, University of Pittsburgh, Pittsburgh, Pennsylvania 15260, USA}
\author{Sarvagya Upadhyay}
\affiliation{Fujitsu Research of America, Santa Clara, CA 95054, USA}
\author{Xueyue Zhang}
\affiliation{Department of Applied Physics and Applied Mathematics, Columbia University, New York, NY 10027, USA}
\author{Quntao Zhuang}
\affiliation{Ming Hsieh Department of Electrical and Computer Engineering, University of Southern California, Los Angeles, CA, USA}
\affiliation{Department of Physics and Astronomy, University of Southern California, Los Angeles, CA, USA}
\author{Junyu Liu}
\email{junyuliu@pitt.edu}
\affiliation{Department of Computer Science, University of Pittsburgh, Pittsburgh, Pennsylvania 15260, USA}

\date{\today}

\begin{abstract}
Artificial intelligence (AI) and quantum information (QI) are rapidly co-evolving. AI is becoming a practical tool for learning, designing, controlling, and verifying quantum systems, while QI offers new computational models, representational structures, and learning-theoretic questions for AI. This survey reviews the interface from both directions. In the \textit{AI for QI} direction, we organize recent progress around the central tasks of extracting information from limited measurements, training and discovering quantum algorithms, stabilizing noisy hardware, automating experimental and programming workflows, and extending learning-based methods to sensing and networking. In the \textit{QI for AI} direction, we examine how quantum computation and quantum-inspired structures affect learning through algorithmic speedups, expressivity, trainability, generalization, neural-network design, and tensor-network representations. We close by identifying cross-cutting challenges in reproducibility, scalability, hardware realism, and co-design, arguing that progress will depend on tighter integration of theory, experiment, and hybrid quantum--classical systems.
\end{abstract}

\maketitle

\renewcommand{\thesection}{\Roman{section}}
\renewcommand{\thesubsection}{\Alph{subsection}}
\renewcommand{\thesubsubsection}{\arabic{subsubsection}}

\clearpage
\begingroup
\small
\setcounter{tocdepth}{2}
\makeatletter
\let\l@title\@gobbletwo
\let\l@abstract\@gobbletwo
\let\l@paragraph\@gobbletwo
\makeatother
\tableofcontents
\endgroup
\newpage

\section{Introduction}

Artificial intelligence (AI) and quantum information (QI) science and technology are two rapidly advancing fields whose development is increasingly intertwined, though in an asymmetric way. AI already provides powerful tools for modeling complex systems and supporting inference, optimization, design, and discovery. Quantum computing seeks computational advantages from superposition, interference, and entanglement, with the long-term goal of solving problems that remain intractable for classical hardware. Related QI technologies, including quantum sensing and quantum networking, exploit quantum resources to enhance measurement, discrimination, communication, and distributed information processing. The interaction between these areas is becoming central to future progress: AI is emerging as an essential tool for building, controlling, and operating quantum devices, optimizing quantum sensors, and designing quantum networking protocols, while quantum computing and quantum-inspired methods are being explored as new computational and representational resources for machine learning and AI.

This interaction is driven by a common technical reality: both AI and QI must reason about complex systems through limited and noisy observations. A quantum device or protocol rarely exposes the object of interest directly. The relevant state, signal, or noise process has to be inferred from measurements and then acted on under tight experimental constraints. This is why AI methods now appear throughout the quantum-technology stack, supporting device characterization, variational training, hardware control, sensing, networking, and increasingly autonomous workflows. The reverse direction starts from a different but related question: whether quantum mechanics offers useful computational or representational structure for AI. Quantum computers naturally manipulate high-dimensional linear-algebraic objects, while quantum models force one to think carefully about encoding, entanglement, and what information can actually be read out from measurement. Even when this does not lead to an immediate speedup, it can suggest new ways to analyze model capacity, generalization, and efficient classical architectures.

This overlap has become especially timely. On the AI side, large language models have moved beyond prompt-conditioned text generation toward tool use, long-horizon planning, and autonomous scientific workflows, to the point that recent work treats agentic science as an emerging research paradigm~\citep{wang2024survey,jiang2025adaptation,wei2025ai}. AlphaEvolve~\citep{novikov2025alphaevolve} and Aurora~\citep{bodnar2025foundation} illustrate the same shift from another angle: AI is beginning to participate directly in algorithm design and scientific prediction workflows, with a growing role across the full scientific pipeline. On the quantum side, some of the newest signals appear directly at the AI--quantum interface. Quantum oracle sketching~\citep{zhao2026exponential} revisits a long-standing concern in quantum machine learning by showing that small quantum processors can give exponential space advantages for learning tasks on massive classical data without assuming full quantum random access memory (QRAM). Recent calibration benchmarks~\citep{cao2026qcaleval} and AI-based surface-code pre-decoders~\citep{chamberland2026fast} point to a growing role for AI in quantum calibration and error-correction workflows. Recent demonstrations in networked quantum interferometry~\citep{stas2026entanglement} and integrated photonic quantum optics~\citep{clark2026integrated} also show that QI technologies are developing across sensing, networking, hardware platforms, and quantum processors. This convergence motivates a more comprehensive review of the interface between AI and QI.

\textbf{Position of this work relative to prior surveys.} Recent reviews cover important parts of this literature. Broad overviews document AI across the quantum-computing workflow \citep{alexeev2025artificial} and the wider bidirectional notion of quantum AI \citep{klusch2024qai,acampora2025quantum}. Other recent surveys focus on specific subproblems, including quantum machine learning \citep{wang2024comprehensive,du2025quantum}, the complexity of learning quantum states \citep{anshu2024survey}, AI methods for representing and characterizing quantum systems \citep{du2025artificial}, machine learning for estimation and control of quantum systems \citep{Ma_2025}, barren plateaus in variational quantum models \citep{larocca2025barren,cunningham2025investigating}, and tensor-network or low-rank methods for machine learning \citep{yan2025tensor,borsoi2025low}.

This review connects the two directions, \emph{AI for QI} and \emph{QI for AI}, through recurring questions about learning, representation, optimization, control, and resource constraints. On the \emph{AI for QI} side, we focus on how learning-theoretic, optimization-theoretic, and control-theoretic ideas help characterize quantum systems, train quantum models, stabilize noisy hardware, and design quantum sensing and networking protocols. On the \emph{QI for AI} side, we examine algorithmic speedups together with how quantum models and quantum-inspired constructions alter representation power, generalization behavior, and architectural design. We also highlight tensor networks as an important bridge because they connect quantum many-body structure, efficient representation, and practical machine learning through explicit multilinear models.

\textbf{Organization.} The review proceeds as follows. Section~\ref{sec:preliminaries} collects the quantum-mechanical, machine-learning, and computational preliminaries used throughout the paper. Section~\ref{sec:ai4quantum} surveys \emph{artificial intelligence for quantum information}, covering statistical learning of quantum systems, AI-theoretic perspectives on quantum algorithms, AI-assisted quantum algorithm discovery, learning, correction, and control of noisy quantum systems, autonomous quantum workflow orchestration, and AI-enabled quantum sensing and networking. Section~\ref{sec:quantum4ai} turns to \emph{quantum computing and quantum-inspired methods for artificial intelligence}. There we discuss quantum algorithmic speedups for classical machine learning, conditions and limits of quantum learning benefits, quantum neural networks, quantum-inspired analyses of classical neural networks, and tensor-network methods for machine learning. Section~\ref{sec:crosscutting} identifies cross-cutting challenges, and Section~\ref{sec:conclusion} concludes.

Across all sections, our central argument is that the interface between AI and QI is an emerging theory and practice of learning, representation, and control under quantum constraints. For convenience, Table~\ref{tab:notation_summary} summarizes the globally recurring notation used throughout this review.

\section{Preliminaries}
\label{sec:preliminaries}

\subsection{Quantum States, Measurements, and Variational Models}
\label{subsec:preliminaries}

This section collects background on how quantum states and measurements are represented, how dynamics and noise are modeled, and how parameterized circuits become learnable objects under finite-shot data.

We begin with the state space. An $n$-qubit system is described on the Hilbert space $\mathcal{H}=(\mathbb{C}^{2})^{\otimes n}$. Its physical state is represented by a density operator $\rho$ satisfying $\rho\succeq 0$ and $\Tr(\rho)=1$. Pure states are the special case $\rho=\ket{\psi}\bra{\psi}$, whereas mixed states describe either classical uncertainty or entanglement with degrees of freedom that have been ignored. For a bipartite system $AB$, the state of subsystem $A$ is the reduced density operator $\rho_A=\Tr_B(\rho_{AB})$. This notion of reduction is important throughout the section, because later discussions of locality, data embeddings, and barren plateaus often depend on whether reduced states remain structured or become close to maximally mixed.

Observables are represented by Hermitian operators $O$, and their expectation values are given by
\begin{equation}
    \langle O\rangle_{\rho}=\Tr(O\rho).
\end{equation}
Measurements are more generally described by a positive-operator-valued measure (POVM) $\{M_s\}$, where each $M_s\succeq 0$ and $\sum_s M_s=\mathbb{I}$. The probability of observing outcome $s$ is given by the Born rule
\begin{equation}
    p_{\rho}(s)=\Tr(M_s\rho).
\end{equation}
The key practical point is that quantum information is not accessed directly. What one obtains experimentally is a collection of classical samples distributed according to $p_\rho(s)$. Thus, in essentially all AI-for-quantum settings, the learning algorithm infers latent quantum objects from partial and noisy measurement data, without direct access to $\rho$ itself \citep{helstrom1969quantum,holevo2011probabilistic}.

Dynamics and noise enter through maps on quantum states. In closed systems, evolution is generated by a Hamiltonian $H$ and acts unitarily as $\rho\mapsto U\rho U^\dagger$ with $U=e^{-itH}$. In open systems and on real hardware, evolution is more generally modeled by a completely positive trace-preserving map, or quantum channel, denoted $\mathcal E$. A quantum channel is a linear map that sends density operators to density operators, and it admits the Kraus form
\begin{equation}
    \mathcal E(\rho)=\sum_a K_a \rho K_a^\dagger,
    \qquad
    \sum_a K_a^\dagger K_a=\mathbb I.
\end{equation}
In this review, the word \emph{process} is used in essentially the same sense, namely an input-output map acting on quantum states. In continuous-time descriptions one often writes $\dot\rho=\mathcal L(\rho)$, where $\mathcal L$ is a Liouvillian generator. This distinction matters because the tasks surveyed below range from learning static quantum properties to learning unknown dynamics, noise models, and control responses on NISQ devices \citep{preskill2018quantum}.

In this language, \emph{quantum noise} refers to departures from ideal unitary evolution, such as decoherence, dissipation, and control imperfections, and these effects are often absorbed into an effective channel or Liouvillian description. A particularly common source of experimental error is \emph{state preparation and measurement} (SPAM) error, meaning that the prepared input state and the recorded measurement outcome may both differ systematically from the intended idealized model. This matters throughout the review because realistic learning and control problems often require simultaneous inference of target dynamics and nuisance noise processes.

Another recurring object is the parameterized quantum circuit (PQC), or ansatz. A typical PQC is a unitary family $U(\boldsymbol{\theta})$ built from elementary gates whose angles are collected in a parameter vector $\boldsymbol{\theta}=(\theta_1,\ldots,\theta_p)$. In many applications one starts from an input state $\rho_{\mathrm{in}}$, applies $U(\boldsymbol{\theta})$, and reads out an observable $O$. This produces the expectation-valued model
\begin{equation}
    f(\boldsymbol{\theta})=\Tr\!\left[O\,U(\boldsymbol{\theta})\rho_{\mathrm{in}}U^\dagger(\boldsymbol{\theta})\right].
\end{equation}
In quantum machine learning, one often inserts a data-encoding circuit before or inside the ansatz, so the model may also depend on an input $x$ through a feature state $\ket{\phi(x)}$ or an encoded density operator $\rho(x)$. In variational quantum algorithms, the same structure appears with a task-specific cost observable in place of a supervised-learning label.

Once the expectation value above is inserted into a classical loss function $\mathscr{L}(\boldsymbol{\theta})$, training proceeds by a hybrid quantum--classical loop. The classical optimizer updates $\boldsymbol{\theta}$, while the quantum device is repeatedly queried to estimate $f(\boldsymbol{\theta})$ and, when needed, its gradients. This is where the statistical character of the problem becomes unavoidable: these quantities are estimated through finitely many circuit repetitions. As a result, optimization quality depends on how many shots are available and how noisy the hardware is.

This setup explains several recurring themes in Section~\ref{sec:ai4quantum}. Some subsections focus on \emph{inference} from measurement data, some on the \emph{optimization} of variational models, and others on \emph{robustness} under finite-shot and noisy conditions. In all three cases, the quantum object of interest is only indirectly accessible, and the available data, feedback, and optimization signals are statistical and resource-limited. One-off local indices in theorem statements, asymptotic bounds, or subsection-specific examples are defined in place. In particular, we reserve $\mathcal{L}$ for Liouvillian generators and use $\mathscr{L}$ for loss or optimization objectives.

\subsection{AI and ML Preliminaries}
\label{subsec:ai-preliminaries}

This review uses a small set of standard AI and ML ideas repeatedly. In supervised learning, one observes a training set
\(\mathcal{S}_{\mathrm{tr}}=\{(x_i,y_i)\}_{i=1}^{N}\) drawn from an underlying data distribution \(\mathcal{D}\), chooses a hypothesis or model \(f_{\boldsymbol{\theta}}\) from a parameterized family, and fits the parameters by minimizing an empirical objective
\begin{equation}
    \widehat R(\boldsymbol{\theta})
    = \frac{1}{N}\sum_{i=1}^{N}
    \ell\!\left(f_{\boldsymbol{\theta}}(x_i),y_i\right),
\end{equation}
where \(\ell\) is a per-example loss. The corresponding population risk is
\begin{equation}
    R(\boldsymbol{\theta})
    = \mathbb{E}_{(x,y)\sim\mathcal{D}}
    \!\left[\ell\!\left(f_{\boldsymbol{\theta}}(x),y\right)\right].
\end{equation}
The gap
\begin{equation}
    \mathcal{G}(\boldsymbol{\theta})
    = R(\boldsymbol{\theta})-\widehat R(\boldsymbol{\theta})
\end{equation}
is the basic object in generalization analysis~\citep{vapnik1995nature,shalev2014understanding}. In practice, one often optimizes a regularized objective
\begin{equation}
    \widehat{\boldsymbol{\theta}}
    \in \arg\min_{\boldsymbol{\theta}}
    \left[
        \widehat R(\boldsymbol{\theta})
        + \lambda_{\mathrm{reg}}\operatorname{Reg}(\boldsymbol{\theta})
    \right],
\end{equation}
where \(\operatorname{Reg}\) encodes a structural preference such as sparsity, smoothness, locality, or small norm. Learning guarantees often bound the probability that the generalization gap exceeds a tolerance,
\begin{equation}
    \Pr\!\left[
        \mathcal{G}(\widehat{\boldsymbol{\theta}})>\varepsilon
    \right]
    \le \delta,
\end{equation}
with sample complexity controlled by three factors. The first is the target accuracy \(\varepsilon\), and the second is the allowed failure probability \(\delta\). The third factor is the part that changes from setting to setting: it measures how hard the model class or prediction task is. In classical PAC (Probably Approximately Correct) learning this role is played by the VC (Vapnik-Chervonenkis) dimension \(d_{\mathrm{VC}}\), which is a combinatorial measure that quantifies the capacity, or complexity, of a hypothesis class. In classical-shadow prediction it is often the shadow norm, or the locality \(k\) of the observables. In variational-model generalization it can be the number of active gates or trainable parameters. In quantum machine learning (QML) generalization bounds it can be a margin or an effective dimension computed from the trained model and the data. For unsupervised, generative, or system-identification tasks, labels may be absent and the objective may be a likelihood, reconstruction error, divergence, or prediction loss for measured trajectories. This empirical-risk language underlies our discussions of quantum tomography, Hamiltonian and channel learning, QML benchmarks, and tensor-network learning models.

Training is the process of finding useful parameters under finite computation, data, and measurement budgets. Gradient-based methods use derivatives of \(\widehat R\) or of a task-specific cost, while derivative-free methods, Bayesian optimization, evolutionary search, and reinforcement learning can be preferable when gradients are unavailable, expensive, noisy, or delayed. In this review, \emph{trainability} refers to whether useful optimization signal can be obtained and exploited with feasible resources; \emph{sample efficiency} refers to how much experimental or training data are needed; and \emph{inductive bias} refers to structural assumptions built into the model, such as locality, symmetry, sparsity, temporal memory, or graph structure. These notions recur in the analysis of barren plateaus, quantum kernels, neural decoders, autonomous calibration, sensing, and networking.

Several model families appear throughout the review. A \emph{kernel method} represents data through pairwise similarities \(k(x,x')\), producing a Gram matrix
\begin{equation}
    \mathbf{K}_{ij}=k(x_i,x_j)
\end{equation}
on the training set and reducing many supervised tasks to regularized optimization in the induced feature space~\citep{scholkopf2002learning,hofmann2008kernel}. A typical kernel predictor has the form
\begin{equation}
    f_{\boldsymbol{\alpha}}(x)
    = \sum_{i=1}^{N} \alpha_i k(x,x_i),
    \qquad
    \boldsymbol{\alpha}
    =(\mathbf{K}+\lambda_{\mathrm{reg}}I_N)^{-1}\mathbf{y}
\end{equation}
in kernel ridge regression. In quantum kernel methods, the similarity is often estimated from overlaps of quantum feature states, for example \(k(x,x')=|\langle\phi(x)|\phi(x')\rangle|^2\).

A \emph{neural network} is a parameterized composition of linear maps and nonlinear or attention-based transformations. A feed-forward example can be written as
\begin{equation}
    \begin{aligned}
    h_0 &= x,\\
    h_\ell &= \sigma_\ell(W_\ell h_{\ell-1}+b_\ell),\\
    f_{\boldsymbol{\theta}}(x) &= W_L h_{L-1}+b_L,
    \end{aligned}
\end{equation}
where the trainable parameters \(\boldsymbol{\theta}\) collect the weights and biases. Different neural architectures encode different biases: CNNs (Convolutional Neural Networks) emphasize locality, RNNs (Recurrent Neural Networks) and sequence models emphasize temporal structure, Transformers use attention to model long-range dependencies, and graph neural networks propagate information along graph edges. In a Transformer block, the basic attention operation is
\begin{equation}
    \operatorname{Attn}(Q_{\mathrm{att}},K_{\mathrm{att}},V_{\mathrm{att}})
    =
    \operatorname{softmax}\!\left(
        \frac{Q_{\mathrm{att}}K_{\mathrm{att}}^{T}}{\sqrt{d_{\mathrm{att}}}}
    \right)V_{\mathrm{att}},
\end{equation}
where \(K_{\mathrm{att}}\) denotes attention keys and should not be confused with the Gram matrix \(\mathbf{K}\). Tangent-kernel analyses study training by linearizing \(f_{\boldsymbol{\theta}}\) around its initialization. The classical neural tangent kernel (NTK)~\citep{jacot2018neural},
\begin{equation}
    \Theta^{\mathrm{NTK}}_{ij}(\boldsymbol{\theta})
    =
    \nabla_{\boldsymbol{\theta}} f_{\boldsymbol{\theta}}(x_i)^{T}
    \nabla_{\boldsymbol{\theta}} f_{\boldsymbol{\theta}}(x_j),
\end{equation}
motivates the quantum neural tangent kernel (QNTK) discussion later in the review.
For graph-structured inputs, a typical message-passing layer updates a node representation by aggregating information from its neighbors,
\begin{equation}
    h_v^{(\ell+1)}
    =
    \psi_\ell\!\left(
        h_v^{(\ell)},
        \bigoplus_{u\in\mathcal{N}(v)}
        \phi_\ell(h_v^{(\ell)},h_u^{(\ell)},e_{uv})
    \right),
\end{equation}
where \(\mathcal{N}(v)\) is the neighborhood of node \(v\), \(e_{uv}\) is an edge feature, \(\oplus\) is a permutation-invariant aggregation operation, and \(\phi_\ell,\psi_\ell\) are learnable maps.

Generative models learn a distribution \(p_{\boldsymbol{\theta}}(x)\) or a sampler whose outputs resemble the data distribution. For explicit density models, maximum-likelihood training minimizes
\begin{equation}
    \widehat R_{\mathrm{NLL}}(\boldsymbol{\theta})
    =
    -\frac{1}{N}\sum_{i=1}^{N}
    \log p_{\boldsymbol{\theta}}(x_i).
\end{equation}
The NLL is the negative average log-likelihood of the dataset, so minimizing $\widehat{R}_{\text {NLL }}$ encourages the model to assign high probability density to the observed samples. For sequence models, including most large language models, the learned distribution is often factorized autoregressively as
\begin{equation}
    p_{\boldsymbol{\theta}}(x_{1:T})
    =
    \prod_{t=1}^{T}
    p_{\boldsymbol{\theta}}(x_t\mid x_{<t}).
\end{equation}
This category includes explicit density models, implicit samplers, Born-machine-style models, and diffusion models that generate samples through learned denoising dynamics~\citep{ho2020ddpm}. A simplified diffusion-style denoising objective has the form
\begin{equation}
    \min_{\boldsymbol{\theta}}
    \mathbb{E}_{t,x_0,\boldsymbol{\xi}}
    \!\left[
        \left\|
            \boldsymbol{\xi}
            - \boldsymbol{\xi}_{\boldsymbol{\theta}}(x_t,t)
        \right\|_2^2
    \right],
\end{equation}
where \(x_t\) is a noised version of a data point and \(\boldsymbol{\xi}_{\boldsymbol{\theta}}\) is a learned denoiser. Foundation models and large language models are high-capacity generative models trained on broad corpora and adapted through prompting, fine-tuning, retrieval, or tool use. These ideas appear below in discussions of quantum generative models, tensor-network generative models, scientific foundation models, autonomous programming, and self-driving laboratory workflows.

For losses derived from negative log-likelihoods, an important curvature object is the Fisher information matrix. Given a dataset $\mathcal{D}=\left\{x_i\right\}_{i=1}^M$, the empirical Fisher matrix is defined as
\begin{equation}
    \tilde{F}_{\mu \nu}(\boldsymbol{\theta})=\frac{1}{M} \sum_{i=1}^M \partial_\mu \log p_{\boldsymbol{\theta}}\left(x_i\right) \partial_\nu \log p_{\boldsymbol{\theta}}\left(x_i\right)~,
\end{equation}
equivalently as the sample average of outer products of the score function. It captures how sensitively the model distribution changes under infinitesimal parameter variations and is used in natural-gradient methods to precondition updates according to the local information geometry of the model. In quantum variational settings, a related object is the \emph{quantum Fisher information matrix} (QFIM), which measures the distinguishability of nearby parameterized quantum states. For a normalized pure-state family $\{\ket{\psi(\boldsymbol{\theta})}\}$, writing $\ket{\partial_\mu\psi}:=\partial_\mu\ket{\psi(\boldsymbol{\theta})}$, one common convention is
\begin{equation}
\begin{aligned}
F^{\mathrm{Q}}_{\mu\nu}(\boldsymbol{\theta})
&=4\,\mathrm{Re}\!\Big[
\braket{\partial_\mu\psi|\partial_\nu\psi}
\\
&\qquad -\braket{\partial_\mu\psi|\psi}\braket{\psi|\partial_\nu\psi}
\Big].
\end{aligned}
\end{equation}
which is four times the Fubini--Study metric, or equivalently four times the real part of the quantum geometric tensor~\citep{stokes2020quantum}. Thus, the empirical Fisher $\tilde{F}$ used in supervised-learning losses and the QFIM used in quantum natural-gradient methods are distinct but analogous metric-like objects.

Several AI-for-quantum settings are sequential decision problems. Reinforcement learning (RL) models an agent interacting with an environment through states, actions, transitions, and rewards, often formalized as a Markov decision process with transition law \(P_{\mathrm{RL}}(z_{t+1}\mid z_t,a_t)\), reward \(r_t\), and discount factor \(\gamma_{\mathrm{RL}}\). A policy \(\pi(a\mid z)\) maps an observed state or belief state \(z\) to an action distribution, and training seeks a policy with high expected return
\begin{equation}
    J(\pi)
    =
    \mathbb{E}_{\pi}\!\left[
        \sum_{t\ge 0}\gamma_{\mathrm{RL}}^{t} r_t
    \right].
\end{equation}
For a fixed policy, the value function satisfies the Bellman relation
\begin{equation}
    V^{\pi}(z)
    =
    \mathbb{E}_{a\sim\pi(\cdot\mid z),\,z'\sim P_{\mathrm{RL}}(\cdot\mid z,a)}
    \!\left[
        r(z,a)+\gamma_{\mathrm{RL}}V^{\pi}(z')
    \right].
\end{equation}
When the agent observes only partial information, the problem is naturally partially observable; this is common in quantum networking, online calibration, and feedback control. Agentic AI systems extend this closed-loop view to language-model-driven workflows: a model or collection of models plans, calls tools, observes outputs, updates memory, and revises actions over multiple steps~\citep{wang2024survey,jiang2025adaptation}. This terminology is used below for autonomous quantum programming, self-driving laboratories, and formal-verification workflows.

\begin{table*}[t]
\centering
\caption{Notation Summary.}
\label{tab:notation_summary}
{\scriptsize
\renewcommand{\arraystretch}{1.0}
\setlength{\tabcolsep}{4pt}
\newcommand{\symA}[1]{\parbox[t]{0.12\textwidth}{\raggedright #1}}
\newcommand{\symB}[1]{\parbox[t]{0.27\textwidth}{\raggedright #1}}
\newcommand{\symC}[1]{\parbox[t]{0.48\textwidth}{\raggedright #1}}
\begin{tabular*}{0.94\textwidth}{@{\extracolsep{\fill}}lll}
\toprule
\symA{\textbf{Symbol}} & \symB{\textbf{Meaning}} & \symC{\textbf{Remarks}} \\
\midrule
\symA{$n$} & \symB{number of qubits} & \symC{The Hilbert-space dimension typically scales as $2^n$.} \\
\symA{$\mathcal{H}$} & \symB{Hilbert space} & \symC{For $n$ qubits, $\mathcal{H}=(\mathbb{C}^{2})^{\otimes n}$.} \\
\symA{$\rho$} & \symB{density operator / quantum state} & \symC{The central latent object in state-learning and variational settings.} \\
\symA{$\rho_A$} & \symB{reduced state of subsystem $A$} & \symC{Obtained from $\rho_{AB}$ by partial trace over subsystem $B$.} \\
\symA{$O$} & \symB{Hermitian observable} & \symC{Expectation values are written as $\langle O\rangle_\rho=\Tr(O\rho)$.} \\
\symA{$\{M_s\}$} & \symB{POVM} & \symC{The measurement operators associated with classical outcomes $s$.} \\
\symA{$s$} & \symB{classical measurement outcome} & \symC{Drawn according to the Born-rule distribution $p_\rho(s)$.} \\
\symA{$p_\rho(s)$} & \symB{measurement-outcome distribution} & \symC{Defined by $p_\rho(s)=\Tr(M_s\rho)$.} \\
\symA{$\Tr$} & \symB{trace} & \symC{Used for expectations, Born probabilities, and reduced states.} \\
\symA{$H$} & \symB{Hamiltonian} & \symC{Generator of closed-system unitary dynamics.} \\
\symA{$\mathcal{L}$} & \symB{Liouvillian / Lindbladian generator} & \symC{Reserved throughout the paper for open-system continuous-time dynamics.} \\
\symA{$\mathcal{E}$} & \symB{quantum channel} & \symC{A completely positive trace-preserving map.} \\
\symA{$U$} & \symB{unitary evolution or circuit} & \symC{Includes both dynamical evolution and parameterized circuits.} \\
\symA{$t$} & \symB{evolution time} & \symC{Used in dynamical prediction and simulation formulas.} \\
\symA{$\boldsymbol{\theta}$} & \symB{trainable parameter vector} & \symC{Used for PQCs, Hamiltonian models, and other parameterized hypotheses.} \\
\symA{$p$} & \symB{number of trainable parameters} & \symC{The length of $\boldsymbol{\theta}$ when that quantity is used explicitly.} \\
\symA{$x$} & \symB{classical input / data point} & \symC{Used in quantum feature maps, kernels, and supervised learning models.} \\
\symA{$y_i$, $\mathbf{y}$} & \symB{target label and label vector} & \symC{$\mathbf{y}=(y_1,\ldots,y_N)^\top$ in supervised settings.} \\
\symA{$\mathcal{D}$} & \symB{data distribution} & \symC{Used for population-risk and generalization statements.} \\
\symA{$\mathcal{S}_{\mathrm{tr}}$} & \symB{training set} & \symC{Usually written as $\{(x_i,y_i)\}_{i=1}^{N}$ in supervised settings.} \\
\symA{$\ell$} & \symB{per-example loss} & \symC{Used to define empirical and population risks.} \\
\symA{$\widehat R$, $R$, $\mathcal{G}$} & \symB{empirical risk, population risk, and generalization gap} & \symC{The gap is written as $\mathcal{G}=R-\widehat R$ when it is useful to name it.} \\
\symA{$d_{\mathrm{VC}}$} & \symB{VC dimension} & \symC{Used for PAC-style sample-complexity statements; other sections use quantities such as margins, shadow norms, or active-parameter counts.} \\
\symA{$\lambda_{\mathrm{reg}}$, $\operatorname{Reg}$} & \symB{regularization strength and penalty} & \symC{Used in classical and quantum learning objectives when structural bias is imposed explicitly.} \\
\symA{$\tilde y_j$} & \symB{observed measurement-derived datum} & \symC{Used for experimentally observed classical values in system-identification losses.} \\
\symA{$y_j(\theta)$} & \symB{model prediction for datum $j$} & \symC{The prediction generated by a parameterized Hamiltonian or Liouvillian model.} \\
\symA{$f(\boldsymbol{\theta})$, $f_{\boldsymbol{\theta}}(x)$} & \symB{model output or predictor} & \symC{May denote an expectation-valued model or a supervised predictor, depending on context.} \\
\symA{$p_{\boldsymbol{\theta}}(x)$} & \symB{learned data distribution} & \symC{Used for generative models, including Born-machine and diffusion-style settings.} \\
\symA{$\mathscr{L}$} & \symB{loss or optimization objective} & \symC{Reserved throughout the paper for training losses, empirical risks, and related objectives.} \\
\symA{$\pi(a\mid z)$} & \symB{policy in sequential decision-making} & \symC{Maps an observed state or belief state $z$ to an action distribution.} \\
\symA{$z_t$, $a_t$, $r_t$} & \symB{state or observation, action, and reward} & \symC{Used in RL, feedback control, and quantum-network decision problems.} \\
\symA{$P_{\mathrm{RL}}$, $J(\pi)$, $V^\pi$} & \symB{RL transition law, return, and value function} & \symC{Used for MDP and POMDP formulations of control and networking.} \\
\symA{$\gamma_{\mathrm{RL}}$} & \symB{RL discount factor} & \symC{Weights future rewards; the subscript avoids conflict with other local uses of $\gamma$.} \\
\symA{$\mathbf{F}$} & \symB{quantum Fisher information matrix (QFIM)} & \symC{Appears in quantum natural-gradient methods.} \\
\symA{$\mathcal{Q}_{\mu\nu}$} & \symB{quantum geometric tensor (QGT)} & \symC{Its real part gives the Fubini--Study metric tensor.} \\
\symA{$g_{\mu\nu}$} & \symB{Fubini--Study metric tensor} & \symC{Defines the local geometry of the variational state manifold.} \\
\symA{$k(x,x')$} & \symB{kernel function} & \symC{Usually estimated from quantum feature states or overlaps.} \\
\symA{$\mathbf{K}$, $\boldsymbol{\alpha}$} & \symB{Gram matrix and kernel coefficients} & \symC{$\mathbf{K}_{ij}=k(x_i,x_j)$; $\boldsymbol{\alpha}$ denotes the coefficients in a kernel predictor.} \\
\symA{$\Theta^{\mathrm{NTK}}_{ij}$, $\Theta_{ij}(\boldsymbol{\theta})$} & \symB{NTK and QNTK entries} & \symC{Measure parameter-space sensitivity correlations between inputs $x_i$ and $x_j$.} \\
\symA{$|\psi\rangle$} & \symB{pure quantum state vector} & \symC{Used for pure states; mixed states are represented by~$\rho$.} \\
\symA{$N$} & \symB{system size or dataset size} & \symC{Denotes matrix dimension in QLSA contexts or sample count in learning contexts.} \\
\symA{$\mathcal{O}(\cdot)$, $\Omega(\cdot)$} & \symB{asymptotic upper and lower bounds} & \symC{Standard Bachmann--Landau notation for complexity scaling.} \\
\symA{$\lambda_i$} & \symB{eigenvalue} & \symC{Eigenvalues of a matrix or Hamiltonian; central to QLSA discussions.} \\
\symA{$D$} & \symB{problem or Hilbert-space dimension} & \symC{Used for the ambient dimension when that quantity is global to the problem.} \\
\symA{$\varepsilon$} & \symB{target accuracy / approximation error} & \symC{Used in complexity and approximation guarantees.} \\
\symA{$\delta$} & \symB{failure probability / confidence parameter} & \symC{Used in sample-complexity and concentration bounds.} \\
\symA{$|b\rangle$, $|x\rangle$} & \symB{encoded input and solution states} & \symC{Standard notation in quantum linear-system algorithms.} \\
\symA{$U_A$} & \symB{block-encoding of matrix $A$} & \symC{Satisfies $(\langle 0| \otimes I)\, U_A\, (|0\rangle \otimes I)=A/\alpha$.} \\
\symA{$\kappa$} & \symB{condition number} & \symC{Governs the complexity of quantum linear-system algorithms.} \\
\symA{$S(\rho_A)$} & \symB{von Neumann entanglement entropy} & \symC{Used in the discussion of entanglement structure and scaling laws.} \\
\symA{$\mathcal{T}$, $A^{(k)}$, $r_k$, $\chi$} & \symB{tensor-network object, local cores, bond dimensions, and truncation rank} & \symC{Used in the tensor-network section; these symbols are fixed there with their standard meanings.} \\
\bottomrule
\end{tabular*}}
\end{table*}

\subsection{Computational Access Models for Quantum Methods in AI}
\label{subsec:qc-preliminaries}

Section~\ref{subsec:preliminaries} emphasized physical states, measurements, noise, and variational learning on quantum hardware. For quantum methods in AI, quantum mechanics plays a different role. In the algorithmic part of the review, QC is treated as a computational model for AI-relevant linear algebra, learning, and optimization tasks; later subsections also consider quantum-inspired models and analyses. The key questions are how classical data are encoded into quantum states, how matrices and operators are accessed, and what information can actually be extracted from the quantum output.

A common primitive is state preparation. A vector $\mathbf{b}\in\mathbb{C}^N$ may be encoded as the normalized quantum state
\begin{equation}
|b\rangle=\frac{1}{\|\mathbf{b}\|}\sum_{i=1}^{N} b_i |i\rangle .
\end{equation}
This amplitude-encoding viewpoint allows an $n=\log_2 N$ qubit register to represent an $N$-dimensional vector compactly. At the same time, it makes clear why data loading is such a central issue in QC-for-AI: a speedup based on manipulating $|b\rangle$ is meaningful only if the state can itself be prepared efficiently, or if one is given an input model that grants comparably efficient oracle access~\citep{dervovic2018quantum}.

A second recurring primitive is matrix access. Quantum algorithms typically access a matrix $A$ through sparse-access queries or a \emph{block-encoding} $U_A$ satisfying
\begin{equation}
(\langle 0| \otimes I)\, U_A\, (|0\rangle \otimes I) = A/\alpha
\end{equation}
for some normalization factor $\alpha$~\citep{chakraborty2019block,low2019hamiltonian}. Once such access is available, techniques such as Hamiltonian simulation, linear combination of unitaries (LCU), and quantum singular value transform (QSVT) can implement matrix functions $f(A)$, turning inversion, spectral filtering, and related operations into reusable quantum subroutines.

Finally, quantum outputs are usually not full classical objects. What one often obtains is a quantum state, a sample distribution, or an expectation value such as $\langle O\rangle_{\psi}=\langle \psi|O|\psi\rangle$. Recovering all $N$ entries of a vector or all entries of a matrix would typically destroy a polylogarithmic speedup. For that reason, the complexity claims in Section~\ref{sec:quantum4ai} must always be read together with their access and readout assumptions, not in isolation.

These computational preliminaries explain the organization of Section~\ref{sec:quantum4ai}. Some subsections study quantum algorithmic speedups for linear algebra and optimization, some ask when quantum data or quantum feature maps yield genuine learning advantages, and others analyze variational or tensor-network models inspired by quantum structure. What unifies them is that QC enters as a structured linear-algebraic resource whose power depends critically on state preparation, operator access, and measurement-limited readout.

\section{Artificial Intelligence for Quantum Information}
\label{sec:ai4quantum}

\subsection{Statistically Learning Quantum Systems}
\label{subsec:learnability}

A central topic in \emph{AI for QI} is learning quantum systems from measurement data. In realistic experiments, the state $\rho$ and its governing dynamics, such as a Hamiltonian $H$ for closed systems or a Liouvillian generator $\mathcal{L}$ for open systems, are not observed directly. They are inferred from finite, noisy, and often incomplete classical outcomes produced by quantum measurements. Using the measurement formalism introduced in Sec.~\ref{subsec:preliminaries}, a POVM $\{M_s\}$ generates an outcome $s$ according to
\begin{equation}
    s \sim p_{\rho}(s), \qquad p_{\rho}(s)=\Tr(M_s\rho),
\end{equation}
where $p_{\rho}(s)$ is the probability of observing outcome $s$ when the system is in state $\rho$, and $\Tr$ denotes the trace. This is the Born rule written in sampling form. For projective measurements $M_s=\Pi_s$ and a pure state $\rho=\ket{\psi}\bra{\psi}$, it becomes
\begin{equation}
    p_{\rho}(s)=\Tr(\Pi_s\rho)=\bra{\psi}\Pi_s\ket{\psi},
\end{equation}
and, for $\Pi_s=\ket{s}\bra{s}$, further reduces to $p_{\rho}(s)=|\langle s \mid \psi\rangle|^2$.

This sampling view turns quantum system identification into a statistical-inference problem: the learner must infer latent quantum structure from finite data under measurement constraints. It also connects state and process inference to \emph{quantum learning theory}, which studies what quantum objects can be learned, under which access models, and with what resource costs~\cite{anshu2024survey}. Therefore the organizing principle is statistical learning from measurement data. The analogy with classical statistical learning~\cite{vapnik1995nature,shalev2014understanding} is therefore helpful but limited, because quantum measurements impose constraints that ordinary classical datasets do not. Table~\ref{tab:learning_correspondence} summarizes this correspondence.

We use this lens to separate three related goals. First, state reconstruction aims to learn a representation of $\rho$ itself. Second, property prediction estimates selected observables without reconstructing the full density matrix. Third, dynamical learning infers Hamiltonians, Liouvillians, or channels from input-output or time-resolved data. The discussion below first treats state reconstruction and property prediction, using quantum state tomography, structured ans\"atze, and classical-shadow methods as representative examples. It then turns to dynamical learning, including Hamiltonian, Liouvillian, and channel learning.

\begin{table*}[t]
\centering
\renewcommand{\arraystretch}{1.5}
\setlength{\tabcolsep}{8pt}
\caption{Conceptual correspondence between statistical learning and quantum system inference. The analogy is approximate: quantum measurements introduce constraints absent in classical settings.}
\label{tab:learning_correspondence}
{\small
\newcommand{\colA}[1]{\parbox[t]{0.20\textwidth}{\raggedright #1}}
\newcommand{\colB}[1]{\parbox[t]{0.38\textwidth}{\raggedright #1}}
\newcommand{\colC}[1]{\parbox[t]{0.28\textwidth}{\raggedright #1}}
\begin{tabular*}{0.94\textwidth}{@{\extracolsep{\fill}}lll}
\hline\hline
\colA{\textbf{Statistical learning}} & \colB{\textbf{Quantum system inference}} & \colC{\textbf{Example}} \\
\hline
\colA{Unknown target}
  & \colB{Quantum state, dynamical generator, or channel}
  & \colC{$\rho$,\; $H$,\; $\mathcal{L}$,\; $\mathcal{E}$} \\[6pt]

\colA{Data distribution}
  & \colB{Distribution from measurement}
  & \colC{$p_{\rho}(s)=\Tr(M_s\rho)$} \\[6pt]

\colA{Sample}
  & \colB{Single-shot measurement outcome}
  & \colC{$s \sim p_{\rho}(s)$} \\[6pt]

\colA{Sufficient statistic / summary}
  & \colB{Classical representation derived from measurement outcomes}
  & \colC{Pauli expectation values, classical shadows} \\[6pt]

\colA{Hypothesis class}
  & \colB{Structured family of candidate quantum objects}
  & \colC{Rank-$r$ states, $k$-local Hamiltonians, tensor-network states, sparse Lindbladians, Pauli channels} \\[6pt]

\colA{Prediction target}
  & \colB{Expectation value or dynamical prediction from the learned object}
  & \colC{$\Tr(O\rho)$,\;
    $\Tr\!\big[O\,e^{-iHt}\rho\, e^{iHt}\big]$,\;
    $\Tr\!\big[O\,e^{t\mathcal{L}}(\rho)\big]$} \\[6pt]

\colA{Loss / risk}
  & \colB{Discrepancy between predicted and observed measurement outcomes}
  & \colC{Squared error, negative log-likelihood, trace distance} \\[6pt]

\colA{Sample complexity}
  & \colB{Number of state copies or measurement shots needed for a target accuracy}
  & \colC{$\Theta(4^n/\epsilon^2)$ for full tomography; $O(\max_i\|O_{i,0}\|_{\mathrm{shadow}}^2\log(M/\delta)/\epsilon^2)$ for fixed-observable prediction, with $O_{i,0}=O_i-\Tr(O_i)\mathbb{I}/2^n$} \\[6pt]

\colA{Inductive bias}
  & \colB{Physical or structural constraint used to regularize inference}
  & \colC{Locality, sparsity, positivity, low rank, bounded interaction range} \\
\hline\hline
\end{tabular*}
}
\end{table*}

\textbf{Quantum state tomography, classical shadows, and prediction of physical properties.} The most direct learning task is full quantum state tomography: reconstruct an unknown state from measurement data. This task quickly becomes infeasible for many-body systems. An $n$-qubit state is described by a $2^n \times 2^n$ density matrix, so a generic state has on the order of $4^n$ real parameters. Correspondingly, a standard upper-bound scaling for the copy complexity of full reconstruction is
\begin{equation}
    N_{\mathrm{full}} = O\!\left(\frac{4^n}{\epsilon_{\mathrm{rec}}^2}\right),
\end{equation}
where $\epsilon_{\mathrm{rec}}$ is the target reconstruction error in trace distance, \ie, one seeks an estimate $\hat{\rho}$ such that $\tfrac{1}{2}\|\rho-\hat{\rho}\|_1 \le \epsilon_{\mathrm{rec}}$~\cite{haah2017sample}.

This scaling motivates tomography methods that exploit structure. A standard example is compressed sensing. For rank-$r$ states, the number of randomly chosen measurement settings or Pauli expectation values can scale as~\cite{gross2010quantum,flammia2012quantum}
\begin{equation}
m_{\mathrm{CS}} = O\!\big(r D\, \mathrm{polylog}(D)\big),
\end{equation}
where $m_{\mathrm{CS}}$ is the compressed-sensing measurement complexity, $r$ is the rank of $\rho$, $D=2^n$ is the Hilbert-space dimension, and $\mathrm{polylog}(D)$ denotes a polynomial in $\log D$. Here $m_{\mathrm{CS}}$ counts settings or expectation values rather than total experimental shots; finite-sample implementations also include accuracy and confidence factors. The important point is that, under a low-rank assumption, the setting count can be reduced relative to generic $O(D^2)$-parameter tomography when $r \ll D$. Other structured reconstructions, including tensor-network and locally constrained tomography, use entanglement or locality assumptions instead of low rank in the full Hilbert space~\cite{cramer2010efficient,lanyon2017efficient}. These methods are examples of inductive bias.

Many experiments do not require a full density matrix. They ask instead for selected properties of $\rho$, such as an energy $E=\Tr(H\rho)$, a magnetization or local order parameter $m=\Tr(M\rho)$, or a two-point correlation function $C_{ij}=\Tr(O_{ij}\rho)$. Shadow tomography formalizes this property-prediction task~\cite{aaronson2020shadow}: the goal is to estimate many specified observables without reconstructing the entire state. The classical-shadow framework of~\citet{huang2020predicting} gives a practical randomized-measurement procedure for this task. Each randomized measurement produces a single classical snapshot, and the collection of snapshots can be reused to estimate many observables with provable guarantees.

Concretely, a classical-shadow snapshot $\hat{\rho}$ is designed to be an unbiased estimate of the unknown state: $\mathbb{E}[\hat{\rho}]=\rho$. Therefore, for any observable $O$, the quantity $\hat{o}:=\Tr(O\hat{\rho})$ is a noisy one-shot estimate of the desired value $\Tr(O\rho)$. This leads to the central question: \emph{how many such snapshots are needed before the averaged estimates are accurate for all target observables?} In the resulting sample bound, only the traceless part of each observable matters. The identity component is already known from $\Tr(\rho)=1$: if $O_{i,0}:=O_i-\Tr(O_i)\mathbb{I}/2^n$, then $\Tr(O_i\rho)=\Tr(O_{i,0}\rho)+\Tr(O_i)/2^n$. Thus only $O_{i,0}$ contributes to the statistical difficulty. For a target set $\{O_i\}_{i=1}^M$, the classical-shadow theorem gives the sample bound~\cite{huang2020predicting}
\begin{equation}
N_{\mathrm{shadow}} =
O\!\left(
\frac{\log(M/\delta)}{\epsilon_{\mathrm{obs}}^2}
\max_i \|O_{i,0}\|_{\mathrm{shadow}}^2
\right),
\end{equation}
where $N_{\mathrm{shadow}}$ is the number of copies or randomized measurements needed to estimate all $\Tr(O_i\rho)$ to additive error $\epsilon_{\mathrm{obs}}$ with success probability at least $1-\delta$.

The task-dependent quantity in this bound is the shadow norm. It measures how much a one-shot estimate can fluctuate under the chosen randomized-measurement ensemble. A convenient expression is
\begin{equation}
\|O\|_{\mathrm{shadow}}^2 := \max_{\sigma}\, \mathbb{E}_{\sigma}\!\left[\hat{o}^{\,2}\right],
\qquad
\hat{o}:=\Tr(O\hat{\rho}),
\end{equation}
where $\mathbb{E}_{\sigma}$ means that the snapshot is generated from input state $\sigma$, including the randomness in the unitary, the measurement outcome, and the construction of $\hat{\rho}$~\cite{huang2020predicting}. The maximization over $\sigma$ makes the quantity a worst-case guarantee, independent of the unknown state. In words, a small shadow norm means that a single snapshot gives a relatively stable estimate. A large shadow norm means that the one-shot estimates have larger worst-case fluctuations, so more independent snapshots must be averaged to reach the same target error $\epsilon_{\mathrm{obs}}$ and failure probability $\delta$.

A useful way to interpret this abstract norm is to look at the measurement ensemble most often used in near-term experiments: random single-qubit Pauli measurements. In this protocol, each qubit is independently measured in a randomly chosen $X$, $Y$, or $Z$ basis. A Pauli string is a tensor product of single-qubit Pauli operators and identities, such as $Z_1X_3Y_7$. Its weight $k$ is the number of non-identity factors, equivalently the number of qubits on which the operator acts nontrivially. This set of qubits is the support of the Pauli string. For this measurement ensemble, a weight-$k$ Pauli string $O$ with eigenvalues $\pm1$ has $\|O\|_{\mathrm{shadow}}^2 = 3^{k}$~\cite{huang2020predicting}. Combining this with the sample bound above shows why the measurement cost depends on $k$ rather than directly on the total number of qubits $n$: one pays for the number of qubits involved in the observable, not for all qubits in the device. Thus a two-qubit correlation can remain cheap to estimate even in a large system, whereas a Pauli string acting nontrivially on $O(n)$ qubits can still require exponentially many snapshots. For observables that are sums of many Pauli terms, the coefficients and number of terms enter through the shadow norm. Thus classical shadows give dimension-efficient prediction for structured observable families, not a dimension-free solution to arbitrary state reconstruction.

The same shadow representation can also be used as input to classical machine learning models~\cite{huang2022provably,huang2021power}. Instead of reconstructing one particular state, one converts measurement data from many related quantum states into compact classical features and trains a supervised model to predict physical properties of unseen states drawn from the same family. Under the distributional and property-class assumptions in these works, the sample complexity is controlled by the structure of the target property and need not scale with the full Hilbert-space dimension~\cite{huang2020predicting,huang2022provably}. This should be read as a structured property-learning guarantee, not as a dimension-free guarantee for arbitrary properties.

Neural-network quantum state tomography takes a complementary route. Rather than fixing a linear shadow estimator, it parametrizes the state itself with a flexible model. In neural-network quantum state tomography~\cite{torlai2018neural}, for example, a restricted Boltzmann machine or related neural ansatz is trained directly on measurement outcomes to approximate the target state. Such models can capture states that are not well matched to simple low-rank or sparse ans\"atze, but they typically do not provide the same general closed-form statistical guarantees as compressed-sensing or shadow-based frameworks. Neural-shadow methods~\cite{wei2024neuralshadow} aim to combine these advantages by using shadow-derived losses together with neural state models, retaining some measurement-efficiency benefits of shadows under controlled assumptions while increasing representational flexibility.

\textbf{Learning dynamical laws from measurement data.}
Static inference is only part of the problem. Many experiments aim to learn the law that generates the observed dynamics. In the simplest closed-system setting, one seeks an unknown Hamiltonian
\begin{equation}
  H(\boldsymbol{\theta})
    = \sum_{\mu=1}^{p} \theta_\mu\, P_\mu,
\end{equation}
where $\{P_\mu\}$ is a known operator basis and the coefficients $\theta_\mu$ are inferred from time-resolved measurement data. This expansion also shows why structure is essential. If the basis is unrestricted, then $p$ can be $\Theta(4^n)$, so full identification generically has exponential parameter and sample requirements. Physically relevant Hamiltonians typically have locality, sparsity, bounded interaction range, or related structure that reduces the effective complexity. Local-Hamiltonian learning exploits such assumptions~\cite{bairey2019learning}. For certain geometrically local Hamiltonians, protocols~\cite{stilckfranca2024efficient,hangleiter2024robustly} achieve sample complexity polynomial in the system size~$n$ and inverse target accuracy, under specified locality, control, and measurement-access assumptions, using experimentally accessible product-state preparations and local measurements.

Practical Hamiltonian-learning methods differ mainly in how data are chosen and how candidate models are fit. Bayesian and adaptive protocols~\cite{wiebe2014hamiltonian} choose input states, evolution times, and measurements to distinguish competing Hamiltonian hypotheses, then update a posterior distribution after each round of data. Simulator-based fitting protocols compare measured data with predictions from parameterized models, using energies, variances, equilibrium states, or dynamical responses; these are best viewed as model calibration or validation unless the identifiability assumptions are specified~\cite{gu2024practical,kokail2019self}.

A common supervised formulation casts Hamiltonian learning as empirical risk minimization over dynamical data. Given input states $\rho_j$, evolution times $t_j$, measured observables $O_j$, and measured expectation values $\tilde y_j$, a candidate Hamiltonian $H(\theta)$ predicts
\begin{equation*}
y_j(\theta)=
\Tr\!\left[
O_j\, e^{-iH(\theta)t_j}\rho_j e^{iH(\theta)t_j}
\right],
\end{equation*}
for datum $j$. One then minimizes
\begin{equation}
\mathscr{L}_{\mathrm{Ham}}(\theta)
=
\frac{1}{N}\sum_{j=1}^{N}
\left(
\tilde y_j-y_j(\theta)
\right)^2 .
\end{equation}
Here the structured family $\{H(\theta)\}$ is the hypothesis class, and $y_j(\theta)$ is the model prediction for the $j$th experiment.

Open-system learning follows the same template, but the physical constraints are stronger. For Markovian dynamics, the candidate Hamiltonian is replaced by a parameterized Liouvillian $\mathcal{L}_{\theta}$. To ensure that $e^{t\mathcal{L}_{\theta}}$ is a physical channel, the hypothesis class is usually restricted to a GKLS/Lindblad form with positive rates, or to another parametrization that guarantees complete positivity and trace preservation. The prediction becomes
\begin{equation}
y_j(\theta)=
\Tr\!\left[
O_j\, e^{t_j\mathcal{L}_{\theta}}(\rho_j)
\right],
\end{equation}
with empirical risk
\begin{equation}
\mathscr{L}_{\mathrm{open}}(\theta)
=
\frac{1}{N}\sum_{j=1}^{N}
\left(
\tilde y_j-y_j(\theta)
\right)^2 .
\end{equation}
For time-ordered trajectories or repeated-shot outcome counts, the squared-error loss is usually replaced by a negative log-likelihood derived from the corresponding probabilistic model. This formulation also makes robustness part of the learning problem. Dissipation, decoherence, and state preparation and measurement (SPAM) errors can be statistically confounded with the effective dynamics unless they are modeled jointly or calibrated independently. Self-consistent characterization methods such as gate-set tomography address related SPAM and gauge-freedom issues~\cite{merkel2013selfconsistent,greenbaum2015introduction}. Recent large-scale examples include learning sparse Pauli-Lindblad models~\cite{vandenberg2024paulilindblad}, weakly dissipative Liouvillian learning~\cite{olsacher2025liouvillian}, and Lindblad learning from time-series data~\cite{vandenberg2025largescale}.

An even broader view treats the unknown object as a quantum channel $\mathcal{E}$, \ie, a completely positive trace-preserving input-output map. Hamiltonian and Liouvillian dynamics are then structured special cases. As recalled in sSec.~\ref{subsec:preliminaries}, closed-system evolution generated by $H$ defines the channel $\rho\mapsto e^{-itH}\rho e^{itH}$, while a continuous-time open-system model generated by $\mathcal{L}$ induces the channel family $\mathcal{E}_t=e^{t\mathcal{L}}$. General channel learning also covers effective noise, control imperfections, and black-box processes that may not admit a simple generator description. The prediction rule is
\begin{equation}
y_j(\mathcal{E})=\Tr\!\left[O_j\, \mathcal{E}(\rho_j)\right],
\end{equation}
so the learner selects a channel hypothesis from finite input-output data. This channel-learning viewpoint connects quantum process tomography~\cite{chuang1997prescription,altepeter2003ancilla,mohseni2006direct}, shadow-based channel learning~\cite{levy2024shadowqpt}, and process learning without input control~\cite{fanizza2024process}. It is complementary to gate-set tomography and randomized benchmarking, which handle SPAM/gauge freedoms or report operational error rates rather than reconstructing a single CPTP map~\cite{greenbaum2015introduction,magesan2011scalable}. Across these settings, the central object is an unknown input-output map that can extend beyond a generator of time evolution.

Across state, property, and dynamics learning, three issues recur. The first is the fundamental limit question: \emph{what can be learned, under which measurement model, and with how many samples?} For full state tomography, near-optimal worst-case bounds~\cite{haah2017sample} clarify the copy complexity needed to reconstruct an entire unknown state to a target accuracy. Shadow tomography~\cite{aaronson2020shadow}, classical shadows~\cite{huang2020predicting}, and learning from noisy quantum experiments~\cite{huang2022foundations} give complementary limits for more restricted prediction tasks. More broadly, learnability can depend sharply on the allowed data-collection operations. Exponential separations can appear between protocols that allow joint measurements or quantum memory and protocols that measure each copy separately and process only classical outcomes~\cite{huang2022foundations,chen2022exponential}.

The second issue is data acquisition. In many classical learning problems, the learner receives a fixed dataset. Quantum experiments often allow partial control over how data are gathered. Adaptive and active learning protocols~\cite{wiebe2014hamiltonian,lange2023adaptive,quek2021adaptive} exploit this control by choosing measurement bases, probe states, or evolution times online to maximize information gain or reduce posterior uncertainty. Such protocols are especially natural in tomography and Hamiltonian learning, where measurement design can strongly affect statistical efficiency.

The third issue is robustness. Noise and SPAM errors are not merely experimental nuisances; they are part of the learnability question. Under specified noise and measurement-depth assumptions, work on learning from noisy quantum experiments and on robust ultra-shallow or shallow-shadow protocols~\cite{huang2022foundations,farias2024robust,hu2025shallow} shows how shadow-based methods can preserve useful bias, variance, and sample-complexity control. These results also clarify how noise changes estimator bias, variance, and feasible measurement depth. The common message is that learnability depends jointly on the target task, the structure of the quantum object, the measurement model, and the available computational resources.

\subsection{AI-Theoretic Perspectives on Quantum Algorithms}
\label{subsec:optimization}

A central challenge in QC is optimization over high-dimensional design spaces, especially variational circuits such as VQEs~\cite{peruzzo2014variational} and QAOA~\cite{farhi2014quantum}. These problems are intrinsically hard for two reasons: First, their loss landscapes are often nonconvex and can exhibit barren plateaus \citep{cerezo2021variational,mcclean2018barren}. Second, even when useful directions exist, finite-shot estimation and hardware noise make them difficult to resolve \citep{thanasilp2024exponential}. This subsection uses AI theory to explain the learning dynamics of these variational quantum algorithms (VQAs). We therefore treat a VQA as a \emph{learnable system}. The four themes below develop this viewpoint by examining whether usable optimization signal exists, how quantum geometry should shape updates, what effective function class the model induces, and how finite shots and hardware noise change the picture \citep{larocca2025barren,ragone2024lie,fontana2024characterizing,stokes2020quantum,holevo2011probabilistic,helstrom1969quantum,meyer2021fisher,fitzek2024optimizing,thanasilp2024exponential,barligea2025scalability}.

\subsubsection{Trainability in Quantum Learning}
\label{subsubsec:trainability}

In classical machine learning, trainability usually refers to whether an optimizer can find parameters with low training loss using feasible           computational resources The same idea applies to VQAs and QML, but quantum training has additional constraints. The ansatz and cost construction shape the loss landscape, expectation values and gradients must be estimated from finitely many measurement shots, and the data encoding can change gradient scaling or even create dataset-dependent optimization pathologies~\citep{thanasilp2023subtleties}. In the gradient-based setting considered here, a model is trainable when useful optimization signals remain resolvable and the total resources needed to find good parameters scale at most polynomially with problem size and target precision.

\textbf{Barren Plateau.} The central threat to trainability in quantum settings is the \emph{barren plateau} (BP) phenomenon~\citep{mcclean2018barren}. The BP phenomenon can be made precise by looking at gradients at random initializations. If, for every trainable parameter, these gradients have zero mean and an exponentially small variance as the number of qubits grows, then the cost function is said to exhibit a BP~\citep{mcclean2018barren,cerezo2021cost}. Formally, for a parameterized cost function $C(\boldsymbol{\theta})$, this means that the following two conditions hold for all variational parameters $\theta_\nu \in \boldsymbol{\theta}$, with parameters $\boldsymbol{\theta}\sim p(\boldsymbol{\theta})$ drawn from a specified initialization distribution:
\begin{align}
&\mathbb{E}_{\boldsymbol{\theta}\sim p(\boldsymbol{\theta})}[\partial_\nu C(\boldsymbol{\theta})] = 0, \label{eq:bp-mean}\\
&\mathrm{Var}_{\boldsymbol{\theta}\sim p(\boldsymbol{\theta})}[\partial_\nu C(\boldsymbol{\theta})] \in \mathcal{O}(1/\alpha^n), \quad \alpha > 1, \label{eq:bp-var}
\end{align}
where $\partial_\nu C(\boldsymbol{\theta}) = \partial C(\boldsymbol{\theta})/\partial \theta_\nu$, $n$ is the number of qubits, and $\alpha>1$ is a constant independent of $n$. In many standard random-initialization settings, including common Pauli-rotation parameterizations with symmetric sampling, the mean gradient is zero~\citep{mcclean2018barren,cerezo2021cost}. Together with the exponentially vanishing variance, this implies that for typical initializations the gradients are exponentially concentrated near zero as the system size grows~\citep{mcclean2018barren,cerezo2021cost,thanasilp2023subtleties}. This exponential suppression renders gradient-based optimization ineffective, as an exponential number of measurement shots is needed to resolve the vanishingly small gradient signal from statistical noise~\citep{arrasmith2021effect,thanasilp2023subtleties}.

\citet{thanasilp2023subtleties} show that BP results also apply to supervised QML models. Under suitable assumptions, if the underlying linear expectation values exhibit a BP, then common QML losses such as mean squared error and negative log-likelihood inherit the same exponential gradient suppression. Familiar BP mechanisms, including global cost functions and deep unstructured circuits, therefore remain problematic in supervised learning settings. The same work identifies a second failure mode termed the \emph{dataset-induced BP}: highly entangling embeddings can make local reduced states nearly maximally mixed, causing gradients to concentrate even when the measurement is local and the QNN is shallow.
The same work also identifies a related issue: the curvature information used by natural-gradient methods can suffer from the same problem. Under the negative-log-likelihood assumptions of \citet{thanasilp2023subtleties}, the empirical Fisher entries obey
\begin{equation}
\mathbb{E}_{\boldsymbol{\theta}\sim p(\boldsymbol{\theta})}
\!\left[
|\tilde{F}_{\mu\nu}(\boldsymbol{\theta})|
\right]
\in \mathcal{O}(1/\alpha^n),
\end{equation}
where $\tilde{F}_{\mu\nu}(\boldsymbol{\theta})$ is the $(\mu,\nu)$ entry of the empirical Fisher information matrix, $p(\boldsymbol{\theta})$ is the initialization distribution, $n$ is the number of qubits, and $\alpha>1$ is independent of $n$. Exponentially small Fisher entries are exponentially hard to estimate reliably, which limits the usefulness of natural-gradient preconditioning in this regime.

Later work extends the above conclusions beyond gradient-based QML. \citet{thanasilp2024exponential} prove analogous exponential concentration in quantum kernel methods, and \citet{wang2025predictive} report predictive collapse in deep data re-uploading models on high-dimensional data. Symmetry or covariance can avoid or mitigate such concentration in structured settings~\citep{henderson2025quantum,agliardi2025mitigating}. Overall, BPs are a family of mechanisms controlled by the ansatz, observable or cost construction, initialization, locality, and noise processes. A recent survey \cite{cunningham2025investigating} synthesizes these factors together with the main mitigation strategies.

\textit{Leveraging Dynamical Lie Algebras and Adjoint Representations to Characterize BPs.}
Recent works highlight a shift from empirical diagnosis toward criteria for predicting when BPs arise and how their severity scales with system size \citep{larocca2025barren,cunningham2025investigating}. Two representative directions are as follows. First, the \emph{dynamical Lie algebra} (DLA) viewpoint~\citep{ragone2024lie} characterizes an ansatz by the smallest Lie algebra generated by the circuit's elementary gate generators under commutators. Using this object, \citet{ragone2024lie} derive variance expressions for sufficiently deep circuits and show that BP onset can be inferred from the algebraic growth of the ansatz. Second, the adjoint-representation viewpoint~\citep{fontana2024characterizing} analyzes the same algebraic structure in the Heisenberg picture, where the circuit transforms observables by conjugation, $O \mapsto U^\dagger O U$. For the class they call Lie-algebra-supported ans\"atze, in which the relevant observable is supported on the dynamical Lie algebra $\mathfrak{g}$, \citet{fontana2024characterizing} express the gradient variance in terms of the decomposition $\mathfrak{g}=\bigoplus_\alpha \mathfrak{g}_\alpha\oplus\mathfrak{c}$ into simple ideals and a center. In the Haar or 2-design setting, their result has the form
\begin{equation}
\mathrm{Var}[\partial_\nu \langle O\rangle_{\rho_{\mathrm{in}}}]
=
\sum_{\alpha}
\frac{
\|G_{\nu,\alpha}\|_{\mathrm{K}}^2
\|O_{\alpha}\|_{\mathrm{F}}^2
\|\rho_{\mathrm{in},\alpha}\|_{\mathrm{F}}^2
}{d_{\mathfrak{g}_\alpha}^2},
\end{equation}
where $G_{\nu,\alpha}$, $O_\alpha$, and $\rho_{\mathrm{in},\alpha}$ denote the components of the parameter generator, observable, and input state associated with the ideal $\mathfrak{g}_\alpha$, and $d_{\mathfrak{g}_\alpha}=\dim(\mathfrak{g}_\alpha)$. This formula makes the intuition concrete: gradients become small when the relevant support is spread over exponentially large algebraic components, or when the generator, state, and observable have little overlapping support on the same components. When locality, symmetry, or problem structure keeps these components small and aligned, useful gradient signal can persist.

\textit{Constructive BP-free regimes and mitigation strategies.}
Several recent works identify constructive BP-free regimes by adding architectural or initialization structure to expressive ans\"atze. \citet{park2024hamiltonian} show that a Hamiltonian variational ansatz (HVA) can avoid BPs while staying close to the underlying Hamiltonian structure. \citet{zhang2024absence} demonstrate BP-free behavior in finite local-depth circuits (FLDCs) that can still support long-range entanglement, reinforcing the idea that locality-structured architectures can remain trainable at scale. \citet{yao2025avoiding} propose a related entanglement-based strategy that steers the effective ensemble away from overly mixing regimes.

Beyond such BP-free regimes, practical mitigation methods target initialization, sparsity, and optimizer design. BEINIT~\citep{kulshrestha2022beinit} initializes gate parameters from a data-dependent Beta distribution and adds perturbations during gradient descent, empirically reducing the chance that a QNN becomes trapped in a BP. QAdaPrune~\citep{kulshrestha2024qadaprune} adaptively prunes redundant or weakly contributing variational parameters. The resulting sparse parameter sets can match the unpruned circuit and sometimes improve trainability when the original circuit stalls.

Gradient-free training provides another optimizer-level mitigation route. Although BP-induced concentration can also limit gradient-free methods~\citep{arrasmith2021effect}, the learned meta-optimizer of \citet{kulshrestha2023learning} proposes QNN parameters directly without estimating gradients on the quantum device and reports improved minima with fewer circuit evaluations in the studied settings. More generally, \citet{chen2025taming} construct a modified parameterized quantum circuit by inserting a trainable gadget layer into an existing circuit. Each gadget couples a system qubit to an ancilla initialized in $\ket{0}$ through a single-qubit operation and three trainable two-qubit rotations, $R_{XX}$, $R_{YY}$, and $R_{ZZ}$, after which the ancilla is discarded; the resulting ansatz is therefore a parameterized quantum channel rather than a purely unitary circuit. The construction preserves expressivity because setting the gadget parameters to zero recovers the original PQC, while suitable placement of the gadget layer yields inverse-polynomial lower bounds on the loss variance and on the gradient variance of parameters after the gadget layer for local observables. They also introduce an activation step that adds extra gadgets around parameters before the gadget layer when those parameters remain difficult to train. 

Together, these works show that BP avoidance can come from architectural structure, initialization, sparsification, controlled perturbations, or optimizer design that limits unnecessary mixing while preserving task-relevant correlations.

\textbf{Traps and Trainability Tradeoffs.}
BPs are only one aspect of optimization hardness. Another is the presence of \emph{traps}, \ie, suboptimal local minima or other stationary regions where local optimization can stall even if gradients do not vanish exponentially. \citet{nemkov2025barren} show that BP landscapes can be ``swamped with traps,'' so avoiding vanishing gradients alone is not sufficient for trainability. At a higher level, \citet{cerezo2025does} identify a tension between BP-free design and quantum advantage: under certain notions of \emph{provable} BP absence, circuits may become easier to simulate classically. This suggests that trainability and possible quantum speedups should be analyzed together. \citet{garcia2025quantum} connect the discussion to modern AI theory by showing that broad classes of deep, random QNNs converge in function space to Gaussian processes, reinforcing a concentration-based view of trainability at large scale.

Taken together, the recent literature places BPs inside a broader picture of trainability. Besides, they also include traps and poor local minima, and the practical difficulty of resolving weak signals under finite measurement budgets \citep{larocca2025barren,cunningham2025investigating,nemkov2025barren,thanasilp2024exponential,singkanipa2025beyond}.
From this perspective, an important goal for future research is to predict trainability \emph{before} large-scale optimization begins.

\subsubsection{Geometry-Aware Optimization}
\label{subsubsec:geometry}

This subsection focuses on one prominent class of optimization: geometry-aware optimization. We ask and survey how the geometry of the quantum state manifold can inform parameter updates. Natural-gradient preconditioning is the main example discussed below, while the broader geometric viewpoint also involves QGT/QFIM-based diagnostics, metric estimation, and geometry-driven variational dynamics.
A parameterized quantum circuit (PQC) generates a smooth family of quantum states $\ket{\psi(\boldsymbol{\theta})}$, so optimization can be viewed in parameter space and on the underlying manifold of quantum states \citep{provost1980riemannian,kolodrubetz2017geometry,stokes2020quantum}. For a normalized pure-state family $\{\ket{\psi(\boldsymbol{\theta})}\}_{\boldsymbol{\theta}}$, with $\boldsymbol{\theta}=(\theta_1,\ldots,\theta_p)$, the \emph{quantum geometric tensor} (QGT) is defined as
\begin{equation}
\mathcal{Q}_{\mu\nu}(\boldsymbol{\theta})
=
\bra{\partial_\mu \psi(\boldsymbol{\theta})}
\Bigl(I-\ket{\psi(\boldsymbol{\theta})}\bra{\psi(\boldsymbol{\theta})}\Bigr)
\ket{\partial_\nu \psi(\boldsymbol{\theta})},
\end{equation}
where $\ket{\partial_\mu \psi(\boldsymbol{\theta})}:=\partial \ket{\psi(\boldsymbol{\theta})}/\partial \theta_\mu$ \citep{provost1980riemannian,kolodrubetz2017geometry}. Its real part,
\begin{equation}
g_{\mu\nu}(\boldsymbol{\theta}) := \mathrm{Re}\,\mathcal{Q}_{\mu\nu}(\boldsymbol{\theta}),
\end{equation}
is the \emph{Fubini--Study metric tensor}, which measures the infinitesimal distance between neighboring quantum states on the state manifold. Equivalently, it defines the line element $ds_{\mathrm{FS}}^2=\sum_{\mu,\nu} g_{\mu\nu} d\theta_\mu d\theta_\nu$. For pure states, this metric is directly related to the \emph{quantum Fisher information matrix} (QFIM) \citep{stokes2020quantum} up to a constant factor.
This geometric viewpoint leads to the \emph{quantum natural gradient} (QNG)~\citep{stokes2020quantum}: QNG rescales the gradient by the inverse QFIM so that the update follows steepest descent with respect to the state-space metric, schematically
$\Delta\boldsymbol{\theta}\propto -\mathbf{F}^{-1}\nabla_{\boldsymbol{\theta}}\mathscr{L}$,
where $\mathbf{F}$ denotes the QFIM and $\mathscr{L}$ denotes the training loss, or more generally the optimization objective.
The main intuition is that apparent plateaus in parameter space can partly reflect poor conditioning of the map $\boldsymbol{\theta}\mapsto\ket{\psi(\boldsymbol{\theta})}$. Geometry-aware updates aim to correct this mismatch.
The main drawback is cost: in both classical and quantum settings, natural-gradient methods require estimating, storing, and inverting a Fisher-type metric, which can become computationally expensive, measurement-expensive, or numerically unstable in high dimensions \citep{martens2020new,stokes2020quantum,gomez2025efficient}.

Recent work develops this geometric viewpoint along several complementary directions.
On the optimizer-design side, \emph{qBang}~\citep{fitzek2024optimizing} is a representative recent example that combines QFIM or QGT-based preconditioning with momentum to accelerate navigation of flat energy landscapes in VQA objectives.
On the estimation side, because forming a full QFIM can be measurement-expensive, structure-exploiting protocols are being developed. For example, \citet{gomez2025efficient} study \emph{commuting-block circuits}, in which the relevant generators can be grouped into mutually commuting blocks, and propose a QFIM-estimation protocol that exploits this structure to reduce the number of required state preparations.
Complementarily, robust hardware-facing methods for estimating Fisher-type geometric quantities have also advanced: \citet{vitale2024robust} demonstrate robust estimation of the \emph{quantum Fisher information} (QFI, \ie, the single-parameter or directional version of the QFIM) on a quantum processor, emphasizing robustness to experimental imperfections.
More broadly, the same geometric ideas also appear in variational time-evolution methods. The connection to QITE is that exact imaginary-time evolution is generally nonunitary and leaves the variational ansatz class, so variational QITE projects this flow back onto the tangent space of the parameterized state manifold. This projection is measured with the Fubini--Study/QFIM geometry, leading to metric-dependent parameter updates closely related to QNG. \citet{gacon2024variational} formulate variational quantum time evolution without computing the QGT explicitly, directly targeting a key bottleneck in geometry-driven updates. Along related lines, \citet{chen2025analytic} develop an analytic theory of \emph{quantum imaginary time evolution} (QITE) by making this connection explicit: they relate QITE, in the continuous-time variational limit, to quantum natural-gradient dynamics and use that geometric formulation to analyze its update dynamics. A separate but practically important question is whether geometry-aware optimization remains useful on noisy hardware. In that direction, \citet{dell2025quantum} study QNG on noisy platforms using the quantum approximate optimization algorithm (QAOA) as a case study, providing evidence and diagnostics for the robustness of metric-preconditioned optimization under realistic noise.

The discussion above suggests several concrete future directions for geometry-aware optimization. First, optimizer design should move beyond applying QNG as a standalone preconditioner and study how geometric information can be combined with momentum, adaptive stepsizes, shot allocation, and other noise-aware optimization heuristics~\citep{fitzek2024optimizing,dell2025quantum}. Second, metric estimation remains a central practical challenge: future methods need cheaper ways to estimate, approximate, or avoid explicitly forming QGT/QFIM objects by exploiting circuit structure, locality, commuting blocks, or low-rank approximations~\citep{gomez2025efficient,vitale2024robust,gacon2024variational}. Third, the link between geometry-aware optimization and variational time evolution deserves further clarification, especially for QITE and related methods where the update rule can be viewed as projecting nonunitary dynamics onto a parameterized state manifold~\citep{gacon2024variational,chen2025analytic}. Finally, trainability claims for geometry-aware updates should be tested under finite shots, hardware noise, and numerical ill-conditioning, since the usefulness of geometric preconditioning ultimately depends on whether the relevant metric information is resolvable on realistic devices~\citep{vitale2024robust,dell2025quantum}.

\subsubsection{Kernel-Based Quantum Learning}
\label{subsubsec:kernels}

Kernel methods provide another way to analyze quantum learning~\cite{scholkopf2002learning,hofmann2008kernel}. For fixed kernels, standard models such as kernel ridge regression and support vector machines reduce training to convex optimization. In \emph{quantum kernel methods} (QKMs), one defines a \emph{quantum feature map} $x \mapsto \ket{\phi(x)}$ via a state-preparation circuit. For normalized pure-state embeddings, a common choice is the \emph{fidelity kernel}
\begin{equation}
  k(x,x') \;=\; \big|\langle \phi(x)\,|\,\phi(x')\rangle\big|^2 .
\end{equation}
This setup makes a uniquely quantum constraint explicit. Once a kernel $k(\cdot,\cdot)$ is fixed, the downstream classical learning problem is standard and typically reduces to regularized convex optimization~\citep{scholkopf2002learning,hofmann2008kernel}. In quantum kernel methods, the training kernel matrix or Gram matrix must be estimated from finitely many quantum measurements. As a result, statistical performance depends jointly on the inductive bias of the kernel and the shot complexity required to resolve pairwise similarities to useful precision~\citep{thanasilp2024exponential}.

\textbf{Exponential concentration.}
A key recent result is that quantum kernels can exhibit \emph{exponential concentration} with the number of qubits~\citep{thanasilp2024exponential}. Let $k(x,x')$ denote the kernel value for a pair of inputs $x,x'\sim\mathcal{D}$ drawn from a data distribution $\mathcal{D}$, and let $n$ be the number of qubits in the corresponding encoded feature states. The kernel is said to be probabilistically exponentially concentrated~\citep{thanasilp2024exponential} toward a data-independent value $\mu$ if, for every fixed resolution threshold $\delta>0$,
\begin{align}
\Pr_{x,x'\sim\mathcal{D}}\!\left[\,|k(x,x')-\mu|\ge \delta\,\right]
&\le \frac{\beta}{\delta^2}, \\
\beta &\in \mathcal{O}(1/b^n), \qquad b>1,
\end{align}
where $\mu$ is the concentration value, $\beta$ controls the exponentially small tail, and the probability is taken over input pairs. As noted, an equivalent variance-based criterion is
\begin{equation}
\mathrm{Var}_{x,x'\sim\mathcal{D}}[k(x,x')] \in \mathcal{O}(1/b^n),
\end{equation}
where the variance is again over input pairs~\citep{thanasilp2024exponential}. For fidelity-type kernels, where $0\le k(x,x')\le 1$, this means that kernel values over different input pairs become increasingly indistinguishable as $n$ grows; in many relevant cases they concentrate around a constant or even an exponentially small value.
If the intrinsic spread of $k(x,x')$ across input pairs is exponentially small, then a polynomial number of measurement shots cannot resolve kernel values to a precision finer than that spread. Measurement noise then overwhelms the remaining data dependence, and the off-diagonal entries of the estimated Gram matrix become effectively indistinguishable unless one uses exponentially many measurements~\citep{thanasilp2024exponential}.
This loss of variation is especially damaging because kernel methods are completely determined by the Gram matrix $\mathbf{K}$ on the training set. Once $\mathbf{K}$ is fixed, training reduces to solving a convex problem (\eg, kernel ridge regression or support vector machines), and predictions for a new input depend only on its similarities to the training points \citep{scholkopf2002learning,hofmann2008kernel}.
If the off-diagonal entries of $\mathbf{K}$ become effectively data-independent under feasible shot budgets, the learned predictor becomes insensitive to input variation, yielding a ``trivial'' classifier or regressor even though the optimization itself is easy \citep{thanasilp2024exponential}.
This highlights a uniquely quantum bottleneck: the \emph{estimability} of the kernel matrix under finite-shot measurement constraints can limit performance even when the downstream optimization problem is convex \citep{thanasilp2024exponential,agliardi2025mitigating,henderson2025quantum}.

\textbf{Structured quantum kernels: symmetry, generalization, and implementability.}
Recent work treats concentration as a kernel-design problem. \citet{kubler2021inductive} argue that useful quantum kernels need problem-specific inductive bias together with large implicit feature spaces. Symmetry is one important source of such bias. If the data distribution or target function is invariant, or equivariant, under a group action, then a symmetry-aligned kernel can treat symmetry-related inputs consistently and suppress task-irrelevant variation. A \emph{covariant kernel}~\citep{agliardi2025mitigating,henderson2025quantum} implements this idea by making the feature map transform compatibly with the group action. Informally, if a group element $g$ maps $x$ to $g\cdot x$, then the encoded features transform in a matched way and the kernel respects the symmetry, often through a relation such as $k(g\!\cdot\! x,g\!\cdot\! x')=k(x,x')$. \citet{agliardi2025mitigating} and \citet{henderson2025quantum} show that such symmetry-structured kernels can mitigate or avoid exponential concentration while retaining trainability on structured data.

Once a kernel remains informative and estimable, the next question is generalization. A fixed quantum kernel leads to an ordinary kernel-regression or kernel-classification problem on its Gram matrix, so phenomena from classical kernel learning can reappear in quantum feature spaces. For a training set $\{(x_i,y_i)\}_{i=1}^{N}$, a kernel predictor \(f(x)=\sum_{i=1}^{N}\alpha_i k(x,x_i)\) interpolates the data when there exists \(\boldsymbol{\alpha}=(\alpha_1,\ldots,\alpha_N)^\top\) such that \(f(x_i)=y_i\) for every training point. Equivalently, \(\mathbf{K}\boldsymbol{\alpha}=\mathbf{y}\), where \(\mathbf{y}=(y_1,\ldots,y_N)^\top\).

\begin{figure}[t]
\centering
\includegraphics[width=0.82\linewidth]{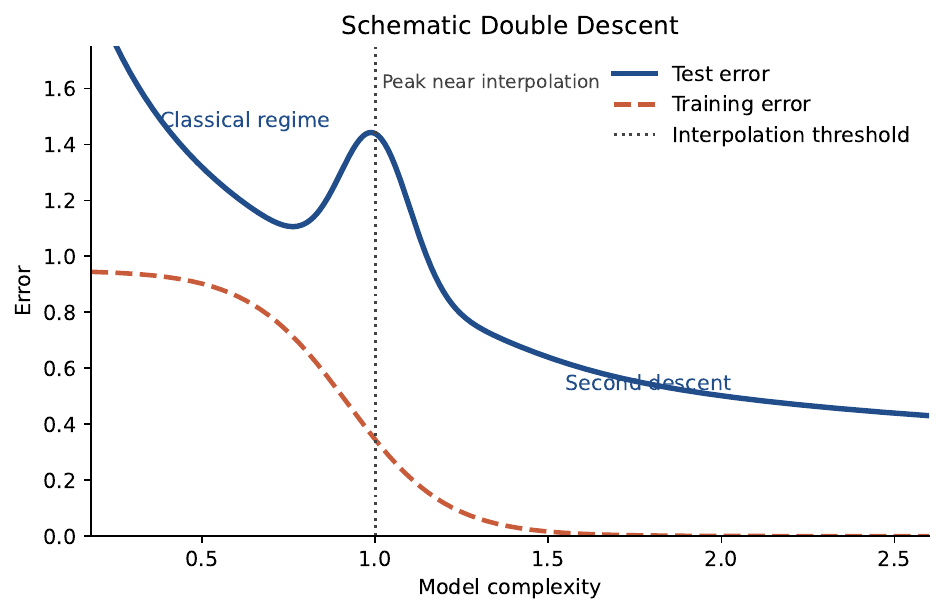}
\caption{Schematic illustration of double descent in a kernel-learning setting. As model complexity approaches the interpolation threshold, the test error can peak before decreasing again in a deeper interpolating regime.}
\label{fig:double-descent-schematic}
\end{figure}

The regime at or beyond this interpolation threshold is often called overparameterized~\citep{kempkes2026double,tomasi2025benign}. In this regime, \emph{double descent}~\citep{kempkes2026double,tomasi2025benign} means that test error can vary nonmonotonically with model complexity: it decreases at first, rises near the interpolation threshold, and then decreases again deeper in the interpolating regime. Figure~\ref{fig:double-descent-schematic} gives a schematic example. \citet{kempkes2026double} demonstrate this behavior in quantum kernel methods, while \citet{tomasi2025benign} study \emph{benign overfitting} with quantum kernels and provide numerical evidence that interpolating solutions can generalize well in suitable regimes.

A related theoretical question is \emph{representability}~\citep{gil2024expressivity}: can a desired kernel function be realized by a quantum feature map? The usual setup encodes each input $x$ into a quantum state $\ket{\phi(x)}$ and defines the kernel from state overlaps, for example \(k(x,x') = |\langle\phi(x)\mid\phi(x')\rangle|^2\). In this setting, \citet{gil2024expressivity} show that any kernel has an embedding-based quantum realization in principle, and they identify broad kernel classes with efficient embedding constructions.

Recent work extends this discussion in several directions. \citet{shin2025new} introduce \emph{entangled tensor kernels}, showing that embedding quantum kernels belong to a broader structural class that helps clarify their inductive bias and possible routes to dequantization. \citet{kadri2025towards} study \emph{operator-valued} quantum kernels for structured outputs and richer learning tasks. \citet{rodriguez2025neural} propose \emph{neural quantum kernels}, where a QNN is trained first and the learned representation is then turned into a problem-informed kernel.

These results make it useful to distinguish \emph{representability in principle} from \emph{efficient implementability}. The former asks whether some quantum embedding reproduces the desired kernel. The latter asks whether the embedding can be realized with feasible circuit depth, width, and measurement cost. A kernel can therefore be representable in theory while remaining impractical to implement or estimate. Quantum kernels should be evaluated along all three axes: representability, estimability under feasible shot budgets, and statistical behavior in the induced overparameterized regime.

\begin{figure}[t]
\centering
\includegraphics[width=0.95\linewidth]{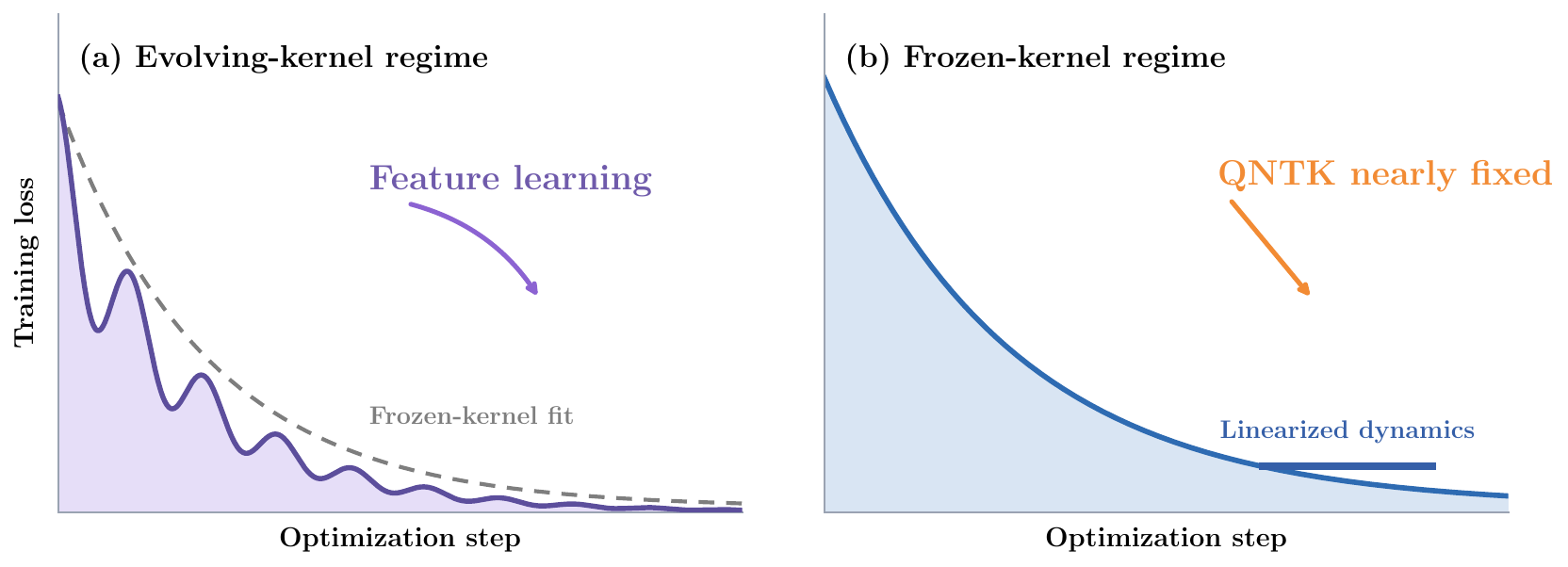}
\caption{Schematic loss trajectories in two QNTK regimes. Right: in the frozen-kernel regime, the QNTK remains approximately constant during early training, so the loss follows the smooth decay predicted by the linearized tangent-kernel dynamics. Left: when the kernel evolves with the parameters, the training trajectory departs from the frozen-kernel picture; this weakly nonlinear regime enables representation learning and can no longer be captured by a fixed-kernel approximation alone.}
\label{fig:qntk_schematic}
\end{figure}

\textbf{Connection to tangent kernel theory: the Quantum Neural Tangent Kernel (QNTK).} The kernel viewpoint also connects naturally to \emph{tangent-kernel} analyses of VQAs, where one studies training dynamics through the local sensitivity of the model output with respect to parameters.
In this literature, the \emph{Quantum Neural Tangent Kernel} (QNTK)~\cite{liu2022representation} is a central diagnostic that generalizes the neural tangent kernel philosophy~\cite{jacot2018neural,arora2019exact} from classical deep learning to PQCs. Following \citet{liu2022representation,liu2023analytic}, for a training set $\{x_i\}_{i=1}^{N}$ of size $N$, a parameter vector $\boldsymbol{\theta}=(\theta_1,\ldots,\theta_p)$ with $p$ variational parameters, and a model output $f_{\boldsymbol{\theta}}(x)$, the QNTK is defined as the data-dependent kernel matrix $\mathbf{\Theta}(\boldsymbol{\theta})\in\mathbb{R}^{N\times N}$ with entries
\begin{equation}
\Theta_{ij}(\boldsymbol{\theta}) = \sum_{\ell=1}^{p}
\frac{\partial f_{\boldsymbol{\theta}}(x_i)}{\partial \theta_\ell}
\frac{\partial f_{\boldsymbol{\theta}}(x_j)}{\partial \theta_\ell},
\end{equation}
so each entry measures the similarity of the parameter-space sensitivities induced by inputs $x_i$ and $x_j$. It is sometimes useful to consider a scalar proxy for the overall gradient signal strength during training. For the same parameter vector $\boldsymbol{\theta}$ and an error functional $e(\boldsymbol{\theta})=\langle O\rangle - O_0$, where $O$ is the measured observable and $O_0$ is the target value, one such proxy at training step $t$ is
\begin{equation}
K(t) = \sum_{\ell=1}^{p} \left(\frac{\partial e(\boldsymbol{\theta})}{\partial \theta_\ell}\right)^2\bigg|_{\boldsymbol{\theta}=\boldsymbol{\theta}(t)},
\end{equation}
which is the squared $\ell_2$-norm of the gradient of the training error. Although this scalar quantity is not the full QNTK, it quantifies how strongly the error responds to joint perturbations across all parameters and therefore serves as a proxy for the rate at which gradient descent can reduce the error in early training \citep{liu2022representation,liu2023analytic}.
Figure~\ref{fig:qntk_schematic} summarizes the two dynamical pictures that recur in the QNTK literature: an early-time frozen-kernel regime, where training is governed by an almost constant tangent kernel, and an evolving-kernel regime, where the kernel changes with the parameters and representation learning becomes possible.

QNTK theory began by adapting classical neural tangent kernel theory to VQAs. \citet{liu2022representation} introduced QNTK as a tool for analyzing training dynamics. \citet{liu2023analytic} then developed an analytic theory for \emph{wide quantum neural networks} in an \emph{overparametrized regime}. In their setting, the number of parameters is large enough that the convergence rate satisfies $\gamma=\mathcal{O}(1)$, which permits exact solutions for the training dynamics and a detailed description of the optimization trajectory. Related work uses the same tangent-space view to study parameter redundancy. \citet{wang2022symmetric} show that circuit symmetries can create redundant directions in parameter space and that symmetry-aware pruning can improve parameter efficiency without substantially degrading performance. \cite{tang2022graphqntk} extends QNTK to graph-structured data through GraphQNTK.

Recent QNTK work studies concentration, noise, and training dynamics beyond the earliest regime. \citet{Liu_2024} use QNTK to relate \emph{quantum laziness}, meaning exponential suppression in the number of qubits, to \emph{barren plateaus}, meaning flat loss landscapes with vanishing gradients. They clarify the distinction between these phenomena and demonstrate noise resilience in overparametrized regimes. \citet{Yu_2024} show that \emph{circuit expressivity}, the ability of a circuit family to represent diverse quantum states, can itself lead to QNTK concentration and thereby connect expressibility measures to training dynamics.

These results also make a practical point. Even when the tangent kernel predicts a useful optimization signal, that signal is built from gradients and Jacobian entries that must be estimated from finitely many measurements. If those quantities cannot be estimated accurately, the optimizer cannot reliably use the tangent-kernel information. QNTK-based learning therefore faces the same estimability constraint as overlap-based quantum kernels~\citep{Yu_2024,thanasilp2024exponential}.

Other work extends QNTK beyond the initial frozen-kernel picture. \citet{zhang2024dynamical} identify a dynamical transition in deep QNNs when the target value crosses the minimum achievable value. Their analysis reveals a duality between QNTK and total-error dynamics described by generalized Lotka--Volterra equations. This transition separates \emph{frozen-kernel} and \emph{frozen-error} phases and gives a more detailed picture of late-time VQA training. \citet{zhang2025quantum} find similar transitions for quantum-data-driven learning, while \citet{zhang2024curse} derive limits on learning random quantum data within the QNTK framework. Beyond standard VQA training, \citet{chen2025analytic} develop a QNTK-based analytic theory for quantum imaginary time evolution. On the practical side, \citet{scala2025towards} use QNTK for QNN diagnostics, and \citet{huang2026quantum} extend QNTK ideas to quantum-enhanced neural contextual bandits. These works make QNTK a useful tool for connecting circuit design, optimization dynamics, and trainability.

QNTK theory has important limitations. It applies primarily to the early phase of training, when the circuit remains close to its initialization. The dynamical phase transition found by \citet{zhang2024dynamical} shows that late-time behavior can be more complex, which matters for applications that require the full optimization trajectory. Most QNTK analyses also assume noiseless circuits, while real quantum hardware is noisy. \citet{Liu_2024} show noise resilience in overparametrized regimes, but the general relationship between noise, QNTK, and trainability remains open.

Across these developments, stronger representational power can also amplify concentration and measurement burden. Claims about trainable or generalizable quantum kernels should therefore specify whether the kernel is \emph{estimable}: how many measurements are needed to distinguish inputs reliably. They should also identify the inductive biases, such as symmetry or covariance, that keep kernels both expressive and estimable \citep{thanasilp2024exponential,agliardi2025mitigating,henderson2025quantum,gil2024expressivity}.

\subsubsection{Scalability Under Noise}
\label{subsubsec:noise_scalability}

A final AI-theoretic question is the most operational one for near-term QC: \emph{does variational training remain scalable once finite-shot fluctuations and hardware noise are unavoidable?}
In practice, VQAs optimize expectation values estimated from a finite number of circuit repetitions (``shots'') and, on real devices, under noisy gates and decoherence.
From a learning-theoretic viewpoint, scalability then depends on three concrete questions: \emph{can the cost and its gradients still be estimated accurately enough with a reasonable number of shots? Can optimization remain stable when those estimates are noisy? How strongly does hardware noise reshape the loss landscape itself?}

\textbf{How noise limits scalability.}
Finite-shot estimation makes both the loss and its gradients random variables. When gradients are small, estimator variance can dominate the effective training dynamics. Recent analyses show that these fluctuations can determine the practical scaling of the quantum--classical training loop, linking convergence to shot budgets and step-size choices \citep{scriva2024challenges}. Systematic studies of variational optimization under finite shots and hardware noise also identify regimes in which noise swamps the gains from deeper circuits or more parameters unless the measurement budget grows accordingly \citep{barligea2025scalability}.

Hardware noise also changes the objective being optimized by altering the prepared state and the induced cost function. More realistic noise models can steer the dynamics in systematic ways. \citet{singkanipa2025beyond} show that noise can induce barren-plateau-like suppression and drive training toward noise-dependent fixed points, while \citet{nemkov2025barren} emphasize that plateau regimes can be ``swamped with traps,'' meaning that many local structures may stall optimization even when gradients are not exactly zero. Practical failure can therefore come from both weak signals and trap-rich landscapes.

Recent algorithms respond by making the hybrid pipeline more measurement-efficient and noise-aware. \citet{ding2024random} propose random coordinate descent for parameterized quantum circuits, reducing gradient-estimation overhead by updating only a subset of parameters. \citet{menickelly2022latency} argue that optimizers should be evaluated under latency and measurement-execution constraints, turning iteration complexity into realistic wall-clock complexity for hybrid loops. \citet{liang2024artificial} introduce adaptive shot allocation, treating the shot budget as a learnable resource-allocation problem, and \citet{hao2025end} develop a few-shot QAOA protocol for obtaining high-quality parameters with limited measurements.

Another direction is geometry-aware preconditioning. \citet{dell2025quantum} study \emph{quantum natural gradient} (QNG) for QAOA and report improved robustness relative to vanilla gradient descent across hardware-relevant noise models. These results suggest that trainability claims in the noiseless setting should be paired with an explicit account of shot complexity and noise-induced landscape deformation. Noise is not always detrimental: \citet{liu2025stochastic} show that controlled stochastic noise can help escape barren plateaus and flat regions, acting as a regularizer in certain regimes. Noise-aware algorithm design therefore includes both mitigation and beneficial stochasticity.

Taken together, recent progress makes ``scalability under noise'' a question about signal resolution. As the system grows, the cost and gradient signals must remain \emph{resolvable} with feasible measurement resources. Optimizer design and measurement allocation must also be coordinated so that the training loop preserves signal-to-noise ratio and avoids noise-induced plateaus or trap-dominated regimes \citep{scriva2024challenges,barligea2025scalability,singkanipa2025beyond,nemkov2025barren,ding2024random,liang2024artificial,dell2025quantum}.

\subsection{AI for Quantum Algorithm Discovery}

AI for quantum algorithm discovery can happen at various hierarchical levels, including circuit, architecture, abstract mathematical algorithms, or co-design among them. Early work by Williams and Gray demonstrated that automated search could be used to design quantum circuits, framing circuit construction as an optimization problem over possible gate sequences rather than as a purely manual task \cite{williams1998automated}. Although the circuits studied in such early work were necessarily small, the key conceptual contribution remains important: quantum circuit design can be treated as a search problem over a combinatorial design space. This perspective provides a foundation for later approaches based on genetic programming, reinforcement learning, differentiable architecture search, and symbolic rewriting.

Evolutionary methods continue to be relevant in modern quantum circuit generation. Stein and Färber revisit genetic programming for quantum circuit discovery and explicitly incorporate measures related to quantum advantage into the generation process \cite{stein2025incorporating}. This is an important shift from simply searching for circuits that reproduce a desired input-output behavior or minimize gate count. By including quantum advantage as part of the objective, the search process is encouraged to identify circuits whose structure is not merely classically reproducible or trivially simulable. This suggests that AI-based circuit generation should not only optimize syntactic circuit properties, but also include higher-level criteria tied to computational usefulness, expressivity, or potential advantage.

\textbf{Reinforcement learning for circuit discovery.} Reinforcement learning has recently emerged as one of the most active approaches for automated quantum circuit discovery. In this setting, an agent sequentially constructs or modifies a circuit by selecting actions such as adding gates, applying transformations, or choosing higher-level circuit components. Zen et al. apply this idea to the discovery of fault-tolerant logical state-preparation circuits, showing that reinforcement learning can be used to search for circuits that satisfy nontrivial constraints imposed by quantum error correction \cite{gqpr-dgz7}. This is especially relevant for fault-tolerant quantum computing, where useful circuits must obey logical encoding constraints, avoid propagating errors catastrophically, and often optimize depth or resource overhead. In contrast to generic circuit synthesis, this work demonstrates AI-driven discovery in a setting where the target circuits are directly connected to error-corrected quantum computation.

A major challenge for reinforcement learning-based circuit discovery is scalability. Searching gate by gate becomes increasingly difficult as the number of qubits, circuit depth, and constraint complexity grow. Olle \textit{et al.} address this issue by introducing “gadgets” as higher-level action primitives for reinforcement learning \cite{olle2025scaling}. Rather than forcing the agent to rediscover useful subcircuits repeatedly from elementary gates, the search can operate over reusable circuit motifs. This hierarchical approach reduces the effective search depth and makes it possible to discover more complex circuits. A related direction is the automated discovery of such gadgets themselves, where repeated or useful substructures are identified and promoted to new building blocks for subsequent learning \cite{yevtushenko2025automated}. Together, these works point toward a form of compositional circuit discovery, in which AI systems gradually build libraries of reusable quantum subroutines.

\textbf{Co-designing circuit and architecture.} Beyond direct circuit construction, reinforcement learning has also been applied to circuit synthesis, transpilation, and optimization. Kremer \textit{et al.} study practical and efficient quantum circuit synthesis and transpiling with reinforcement learning, targeting tasks where the goal is to produce circuits that satisfy architectural or compilation constraints while remaining resource efficient \cite{kremer2024practical}. This line of work is important because algorithm discovery cannot be separated from implementability: a theoretically attractive circuit may become impractical after compilation to a real hardware topology or fault-tolerant gate set. Similarly, Riu \textit{et al.} combine reinforcement learning with ZX-calculus, allowing agents to optimize quantum circuits through graph-rewriting rules rather than only through gate-level transformations \cite{Riu2025reinforcement}. This symbolic perspective is promising because ZX-calculus provides a mathematically structured representation in which non-obvious circuit equivalences can be exploited during search.

Another direction is differentiable quantum architecture search. Zhang \textit{et al.} formulate quantum circuit architecture design as a differentiable optimization problem, enabling gradient-based search over parameterized circuit structures \cite{Zhang_2022}. This approach is particularly relevant for VQAs, where the choice of ansatz strongly affects trainability, expressivity, and hardware efficiency. Compared with reinforcement learning or genetic programming, differentiable architecture search can exploit continuous relaxations of discrete design decisions, potentially improving search efficiency. However, it also raises questions about whether the relaxed optimization landscape faithfully represents the discrete circuit-design problem. 

\textbf{Outlook.} These progress from small-scale automated circuit design to increasingly structured and scalable AI-assisted discovery frameworks suggest that the field is moving beyond mere circuit optimization toward the automated discovery of reusable, hardware-aware, and fault-tolerance-compatible quantum algorithmic components. One important open question is to use AI to verify the correctness of large-scale circuits or quantum algorithms. Another promising future direction is to leverage AI and formal proof tools to discover algorithms based on entirely new mathematical structures. One example of such a link between algorithms and mathematics is the equivalence between quantum signal processing algorithms \cite{Low2017OptimalProcessing,martyn2021grand,joven2026scalable} and nonlinear Fourier transform \cite{alexis2024quantum,laneve2025generalized}. Another example is group and representation theory and the corresponding quantum algorithms such as Schur transform and generalized phase estimation \cite{bacon2006efficient}. Other mathematical structures will inspire novel quantum algorithms that are unknown before. In addition, there are more design spaces across the entire quantum computing stack where AI can be helpful to explore, including co-designing QEC codes with algorithms and architectures.

\subsection{Learning, Correction, and Control of Noisy Quantum Systems}

Correction and control are needed because realistic QC hardware has several coupled sources of error. Physical qubits are unavoidably noisy \citep{preskill2018quantum,kjaergaard2020superconducting}. These imperfections can be grouped into four broad categories: (i) decoherence, which causes stored quantum information to relax or lose phase coherence over time; (ii) control imperfections, which produce gate errors and can push the state out of the intended computational subspace; (iii) readout infidelity, meaning that the measured bit may not match the underlying qubit state; and (iv) cross-talk and temporal drift, meaning that neighboring qubits can influence one another and calibrated device parameters can slowly change over time.

Quantum error correction (QEC) addresses these limitations by encoding logical information redundantly across many physical qubits and repeatedly measuring stabilizer operators to infer and correct faults without disturbing the logical state \citep{Sivak_2023,Brock_2025}. At the abstract level, QEC is defined by a code subspace $\mathcal{C}$, a noise channel $\mathcal{N}$, and a recovery channel $\mathcal{R}$ such that $(\mathcal{R}\circ\mathcal{N})(\rho)=\rho$ for all logical states $\rho$ encoded in $\mathcal{C}$. Stabilizer-based QEC~\citep{gottesman1997stabilizer,terhal2015quantum} is a practically important realization of this framework.

Each syndrome measurement cycle produces classical data, and a classical decoder uses those data to infer the most likely fault pattern and apply a recovery operation. QEC is therefore a sequential inference and feedback-control problem: syndrome data arrive round by round, and the system must update its fault estimate and recovery action in real time. AI methods support this broader workflow through decoding, noise characterization, adaptive control, and code discovery.

\begin{table*}[t]
\centering
\caption{Representative AI tasks for learning, correction, and control of noisy quantum systems.}
\label{tab:noisy-quantum-systems}
{\scriptsize
\renewcommand{\arraystretch}{1.08}
\setlength{\tabcolsep}{4pt}
\newcommand{\taskA}[1]{\parbox[t]{0.19\textwidth}{\raggedright #1}}
\newcommand{\taskB}[1]{\parbox[t]{0.27\textwidth}{\raggedright #1}}
\newcommand{\taskC}[1]{\parbox[t]{0.43\textwidth}{\raggedright #1}}
\begin{tabular*}{0.94\textwidth}{@{\extracolsep{\fill}}lll}
\toprule
\taskA{\textbf{Task}} & \taskB{\textbf{Typical AI methods}} & \taskC{\textbf{Role in noisy quantum systems}} \\
\midrule
\taskA{Syndrome decoding} & \taskB{CNNs, recurrent models, Transformers, GNNs, differentiable message passing} & \taskC{Infer likely fault patterns from repeated syndrome measurements under strict latency constraints.} \\
\taskA{Noise-model learning and adaptation} & \taskB{Supervised learning, likelihood-based inference, adaptive decoders, surrogate models} & \taskC{Estimate drift, correlations, leakage, biased noise, and readout errors, then adapt decoders or calibration policies.} \\
\taskA{Adaptive QEC and code discovery} & \taskB{Reinforcement learning, search, differentiable optimization, generative design} & \taskC{Choose recovery actions, measurement schedules, code layouts, or protocol structures matched to hardware constraints.} \\
\taskA{Control and calibration} & \taskB{Bayesian optimization, RL, neural controllers, autonomous experimental agents} & \taskC{Tune device parameters, stabilize operation, and close the loop between characterization, correction, and hardware control.} \\
\bottomrule
\end{tabular*}}
\end{table*}

Table~\ref{tab:noisy-quantum-systems} organizes the discussion in this subsection by task and AI method family. Here, decoding belongs most directly to the standard inner loop of QEC, whereas noise adaptation, adaptive control, and code discovery are better viewed as supporting tasks in the broader workflow of making QEC effective on realistic hardware.

\subsubsection{Neural and Graph-Based Decoders}

The decoder must process syndrome data in real time, handle faulty measurements, and generalize across noise models and code families, all within the strict latency budgets imposed by qubit coherence times. A broad range of neural architectures are applied to this problem, each exploiting different structural properties of the syndrome data.

\textbf{Convolutional and feedforward neural decoders.} CNNs are among the earliest neural architectures applied to decoding, exploiting the planar lattice structure of the surface code to assign local error likelihoods~\cite{Breuckmann_2018,Ueno_2022,bordoni2023cnn}. A recent systems-oriented example is the AI-based pre-decoding framework of \citet{chamberland2026fast}, which uses local, parallel neural pre-decoders for surface-code syndrome volumes before passing residual syndromes to a downstream global decoder such as PyMatching, achieving microsecond-scale decoding runtimes on NVIDIA GPUs while reducing logical error rates relative to global decoding alone. A parallel line of work uses feedforward, or fully connected, neural networks as another decoder family. In this family, \citet{gicev2023surface} design scalable syndrome decoders for surface codes of arbitrary shape and size, incorporating noise-model flexibility for biased and spatially inhomogeneous noise, and related feedforward decoders are also benchmarked directly on IBM quantum processors~\cite{hall2024annibm}.

\textbf{Transformer and recurrent decoders.} A major direction replaces spatial convolutions with sequence models that process repeated syndrome rounds as a time series, capturing temporal correlations across measurement cycles. A recurrent transformer decoder demonstrates performance advantages over minimum-weight perfect matching (MWPM) on both simulated and experimental data from Google's Sycamore processor~\cite{bausch2023learning}. \textbf{AlphaQubit}~\cite{bausch2024alphaQubit} extends this approach with a two-stage training strategy: large-scale simulation pretraining followed by finetuning on scarce experimental data. The resulting decoder achieves improved logical error rates on distance-3 and distance-5 surface-code experiments, as well as on simulated distances up to 11 under realistic noise including cross-talk, leakage, and analog readout. This pretraining-then-finetuning paradigm is now widely adopted: \citet{Varbanov2023} apply a similar strategy and explicitly incorporate analog readout information to improve decoding accuracy. Transfer learning across code distances is further explored using Transformer decoders, reporting over an order-of-magnitude reduction in training cost~\cite{wang2023transformer}. End-to-end approaches that optimize directly against decoding performance metrics extend naturally to the practically important setting of faulty syndrome measurements~\cite{Choukroun2023deep}.

\textbf{Graph neural network decoders.} Graph neural networks (GNNs) offer a complementary approach well suited to codes with irregular sparse structure, where MWPM is inapplicable and belief propagation (BP) is impaired by short cycles in the Tanner graph. By learning message-passing schedules directly from syndrome data on the Tanner or detector graph, GNN decoders generalize across code families. \citet{Lange_2025} show that sufficiently expressive GNN decoders trained on simulated and experimental data can match matching-based approaches for stabilizer codes. The \textbf{Astra} decoder~\cite{Maan_2025} operates on Tanner graphs and reports improved thresholds and logical error rates over BP+OSD for qLDPC codes, while demonstrating extrapolation to larger code distances using models trained on smaller instances. Differentiable iterative decoders interleaving BP stages with GNN layers are proposed to suppress trapping sets arising from short Tanner-graph cycles~\cite{gong2023graphneuralnetworksenhanced}, and GNN-based message-passing decoders are benchmarked across several qLDPC designs~\cite{ninkovic2024decodingquantumldpccodes}. \textbf{GraphQEC}~\cite{hu2025efficientuniversalneuralnetworkdecoder} pushes further toward universality, proposing a code-agnostic neural decoder with linear-time complexity that generalizes across surface codes, color codes, and qLDPC codes within a single model.

\textbf{Reinforcement learning decoders.} RL frames decoding as a sequential decision-making problem in which an agent is trained by assigning rewards for successful corrections~\cite{Sweke_2021}. Deep Q-learning agents are applied to toric and surface codes under faulty syndrome measurements, achieving performance comparable to MWPM at low error rates~\cite{Andreasson_2019}.

\textbf{Future directions.} Three decoder trends now stand out. First, moving to larger code distances and more general code families will require substantially larger and more representative training sets, since direct supervised scaling quickly becomes expensive even when transfer across code distances is possible~\cite{wang2023transformer,Maan_2025,hu2025efficientuniversalneuralnetworkdecoder}. Second, higher decoding accuracy will increasingly depend on fine-tuning pretrained models on experimental QPU data, because hardware-specific effects such as leakage, analog readout, drift, and correlated faults are difficult to capture faithfully with stylized simulated noise alone~\cite{bausch2024alphaQubit,Varbanov2023}. Third, practical impact will hinge on real-time deployment: strong offline accuracy is not sufficient unless neural decoders can be integrated into low-latency classical control stacks whose inference time remains compatible with syndrome-cycle deadlines and feedback bandwidth constraints~\cite{Sivak_2023,sivak2025reinforcementlearningcontrolquantum}.

\subsubsection{Noise Model Learning and Adaptive Decoding}

Most decoders assume a known, stationary noise model, yet hardware noise drifts over time, exhibits spatial correlations, and frequently deviates from the idealized depolarizing channel used during training. The \textbf{Decoding Graph Re-weighting (DGR)} method~\cite{wang2024dgrtacklingdriftedcorrelated} estimates up-to-date edge probabilities and correlations from matching statistics accumulated across syndrome rounds, and uses these estimates to reweight the MWPM decoding graph, reporting substantial logical error rate improvements under noise mismatch and correlated noise. Bayesian inference methods are used to estimate parameters of general noise models, including time-varying channels via sequential Monte Carlo~\cite{Kobori_2025}. Recent work extends syndrome-based learning to circuit-level noise models via Fourier analysis and compressed sensing, reporting sample-complexity savings over direct logical benchmarking~\cite{zheng2026efficientlearninglogicalnoise}.

\subsubsection{Adaptive QEC and Code Discovery}

RL demonstrates broad utility across the QEC pipeline beyond decoding. At the hardware level, QEC detection events can serve directly as a learning signal for steering physical control parameters under drift: \citet{sivak2025reinforcementlearningcontrolquantum} demonstrate this paradigm experimentally, showing improved stability of the surface-code logical error rate against injected drift, with scaling simulations extending to distance 15. Model-free RL is credited as a key component in achieving coherence gains greater than unity in a 3D superconducting cavity memory~\cite{Sivak_2023}, and in optimizing GKP-encoded qudit memories~\cite{Brock_2025}. Multi-agent RL is also proposed to simultaneously discover QEC cycle components and adapt to time-varying noise channels in a model-free manner~\cite{guatto2025realtimeadaptivequantumerror}.

RL also drives the automated discovery of hardware-tailored QEC codes. \citet{Olle_2024} train a noise-aware RL agent to simultaneously discover stabilizer codes and encoding circuits under specified gate sets, connectivity graphs, and noise models, scaling to approximately 20--25 physical qubits and distance-5 codes. \citet{Su_2025} combine RL with the Quantum Lego tensor-network code framework to optimize code distance and logical error metrics under biased noise, discovering constructions including an optimal $[[17,1,3]]$-type code for asymmetric noise settings. RL optimization of tensor network code geometries has similarly been shown to identify optimal stabilizer codes more efficiently than random search~\cite{mauron2023optimizationtensornetworkcodes}.

\subsection{Autonomous Quantum Workflow Orchestration}

Recent progress in AI has made \emph{agentic} systems~\cite{wang2024survey,guo2024large,jiang2025adaptation} a major direction for large language models (LLMs). These systems pursue long-horizon goals through repeated cycles of planning, acting, observing, and revision. Foundational work such as \textbf{ReAct}~\cite{yao2022react} and \textbf{Toolformer}~\cite{schick2023toolformer} shows that language models can be embedded in closed loops that combine reasoning, tool use, and environmental interaction. More recent surveys~\cite{wang2024survey,guo2024large,jiang2025adaptation} treat autonomous and multi-agent LLM systems as a systems layer above the base model, built from memory, planning, tool invocation, coordination, and self-refinement. These developments create opportunities across quantum information science and technology: agentic systems can orchestrate research and engineering pipelines that include task decomposition, experimental execution, and iterative refinement~\cite{boiko2023autonomous,m2024augmenting}. This section surveys three main thrusts at the intersection of autonomous AI and quantum workflows: autonomous quantum programming, self-driving quantum laboratories, and formal frameworks for discovery and verification.

\subsubsection{Autonomous Quantum Programming}

A central application is the automated generation, debugging, and optimization of quantum circuits from natural language descriptions. \textbf{QAgent}~\cite{fu2025qagent} is a representative LLM-powered multi-agent system that fully automates OpenQASM programming for NISQ devices. It integrates task planning, retrieval-augmented generation (RAG) for long-term context, few-shot in-context learning, chain-of-thought (CoT) reasoning, and reflection mechanisms. QAgent decomposes quantum problems into sub-tasks, routes them to either a Dynamic-few-shot Coder or a Tools-augmented Coder, and employs a Reflection Agent for iterative debugging, achieving a 71.6\% improvement in code generation correctness over prior static LLM approaches across multiple base models.

\textbf{Agent-Q}~\cite{jern2025agent} takes a complementary approach by constructing a large-scale training dataset of over 14,000 quantum circuits (covering QAOA, VQE, and adaptive VQE) and fine-tuning LLMs to generate syntactically correct parameterized quantum circuits in OpenQASM 3.0. Agent-Q is designed to be integrated into autonomous workflows where generated circuits serve as starting points for further optimization, functioning as templates in quantum machine learning and as benchmarks for compilers and hardware. \citet{campbell2025enhancing} further develops a multi-agent framework that uses semantic analysis and multi-pass inference alongside an error correction module to enhance LLM-based quantum code generation with fault-tolerant considerations.

\textit{Benchmarks and Dedicated Quantum Code Models.} The rapid development of LLM-based quantum programming tools has created a need for standardized evaluation. \textbf{Qiskit HumanEval}~\cite{vishwakarma2024qiskit} introduces the first benchmark specifically designed for quantum code generation, with more than 100 hand-curated tasks across eight categories of quantum programming functionality. Alongside the benchmark, IBM releases \texttt{granite-8b-qiskit}, a domain-adapted code LLM that significantly outperforms general-purpose models on quantum coding tasks.

Subsequent benchmarks address complementary dimensions. \textbf{QCoder}~\cite{qcoder2025} incorporates hardware-simulator-based evaluation and human-written reference solutions from programming contests, enabling assessment of whether generated circuits execute correctly on simulated quantum devices. \textbf{QuanBench}~\cite{quanbench2025} evaluates both syntactic correctness and quantum semantic fidelity via process-fidelity metrics, testing LLMs on algorithmic reasoning beyond API compliance. These benchmarks show that current LLMs perform reasonably on basic quantum circuit construction but still struggle with deeper algorithmic reasoning and hardware-aware optimization. This motivates further work on domain-specific fine-tuning, such as the reinforcement-learning-based \textbf{QSpark}~\cite{qspark2025}, which fine-tunes Qwen2.5-Coder on Qiskit HumanEval, and on multi-agent architectures that decompose complex quantum programming tasks into manageable subtasks.

Beyond code generation, recent work asks whether LLMs can reason about quantum algorithms at a deeper symbolic level. \citet{wang2024grovergpt} train \textbf{GroverGPT}, an 8-billion-parameter language model designed for quantum search, and show that large-scale pretraining on quantum-structured data can help LLMs internalize algorithmic logic together with circuit syntax. Complementing this direction, \citet{chen2026symbolic} introduce a framework for symbolic analysis of Grover's search algorithm that combines chain-of-thought (CoT) reasoning with quantum-native tokenization, enabling step-by-step symbolic execution of quantum algorithms.

\subsubsection{Self-Driving Quantum Laboratories}

The recent integration of artificial intelligence into experimental quantum science has shifted the role of automation from running batch sequences to closing the loop on design, fabrication, control, and characterization of quantum hardware. Several recent reviews give a cross-stack picture of this shift \cite{Alexeev_2025,Krenn_2023,Krenn_2020,Ma_2025,Malashin_2026,Tom_2024}. We use ``self-driving'' broadly to describe AI systems that close at least one experimental feedback loop---proposing an action, executing or recommending it, observing the result, and updating the next action---and we use ``AI systems'' equally broadly, encompassing a wide spectrum of tools that range from classical nonlinear-optimization algorithms in specific settings, to reinforcement-learning (RL) and neural-network (NN) methods, to frontier large language models (LLMs).
Depending on the experimental platform, a typical quantum laboratory progresses through some or all of the following steps to complete an experiment: (i) design of the experiment and devices, (ii) fabrication of the devices, and (iii) control and measurement---including device tune-up or state preparation, device / Hamiltonian characterization, control-parameter optimization, and readout-parameter optimization. Although a fully self-driving quantum laboratory remains out of reach today, workflows with varying levels of AI assistance and automation have been demonstrated for each of these individual steps. In what follows, we organize the discussion according to these steps.

\textbf{Experiment design.} Quantum-optics experiments built from a finite vocabulary of building blocks, such as single-photon sources, beam splitters, mirrors, shift-parameterized holograms, and Dove prisms, are a natural target for AI design. Each block has a well-defined unitary action, so the design problem becomes a structured search over how to stack the blocks. An early demonstration was MELVIN~\cite{Krenn_2016}, which autonomously composed standard quantum-optics elements into experimental setups generating complex multi-photon states. It identified previously unknown layouts for high-dimensional GHZ states and asymmetrically entangled states, and learned from simpler instances to accelerate discovery on more complex ones. Soon after, projective-simulation-based active learning~\cite{Melnikov_2018} showed that a reinforcement-learning-style (RL) agent could generate high-dimensional entangled multi-photon states more efficiently than random search and rediscover useful quantum-optics design strategies.

The next wave of these tools emphasized interpretability. Theseus~\cite{Krenn_2021} introduced a graph-based representation of quantum-optical experiments and an inverse-design algorithm orders of magnitude faster than MELVIN, with solutions that scientists can read and reason about. Klaus~\cite{Cervera_Lierta_2022} reformulated photonic experiment design as a Boolean satisfiability problem hybridized with continuous optimization, returning exact minimal configurations and bringing logic AI into quantum experimental design for the first time. The open-source successor PyTheus~\cite{Ruiz_Gonzalez_2023} discovered one hundred diverse experimental designs spanning highly entangled states, optimized measurement schemes, communication protocols, and multi-particle gates. In 2026, language-model meta-design~\cite{Arlt_2026} took the loop a step further by training a transformer to generate Python programs that solve entire \emph{families} of quantum-optical design problems in a single forward pass. Because the generated code is human-readable, researchers can inspect the underlying design principles directly, and the trained model produced novel experimental generalizations of important condensed-matter quantum states.

\textbf{Device design.} Superconducting-qubit hardware is a useful example because it offers substantial design flexibility and relatively tight agreement between designed and fabricated devices. The design problem has two stages. The first maps a target Hamiltonian or quantum function to a lumped-element circuit model and its parameters. The second maps that circuit model to a manufacturable layout through geometry optimization and electromagnetic simulation.

The first stage has seen the most direct AI uptake. SCILLA~\cite{Menke_2021} is an automated closed-loop design tool that proposes superconducting circuit diagrams whose Hamiltonian spectrum and noise sensitivities match a user-specified target; it is demonstrated on 4-local couplers, and the same group's experimental realization of tunable three-body interactions~\cite{Menke_2022} provides hardware context for the interaction designs SCILLA targets. A differentiable framework due to Ni \emph{et al.}~\cite{Ni_2022} jointly optimizes device parameters and control parameters in a single gradient-based pass, extending GRAPE-style optimization to the device layer. A more recent automatic-differentiation framework on top of the SQcircuit package~\cite{Rajabzadeh_2024} computes eigensystem gradients of large sparse circuit Hamiltonians and applies them to qubit discovery, including metrics for decoherence and fabrication-error robustness. C\'ardenas-L\'opez \emph{et al.}~\cite{Cardenas_Lopez_2025} use genetic algorithms to design superconducting-circuit elements with target spectra, selection rules, and built-in resilience to fabrication variability.

Layout-level work connects circuit design to manufacturable geometries. Li and Jin~\cite{Li_2023} derive an analytic zero-coupling condition for the qubit-coupler-qubit (QCQ) architecture, together with a geometry-only upper bound on the effective qubit--qubit coupling. They propose an automated layout procedure that approaches that bound and validate it by electromagnetic simulation, including a millimeter-scale long-range QCQ layout. Predictive machine learning on Qiskit~Metal and ANSYS simulation data~\cite{Nugraha_2023} forward-predicts transmon engineering parameters such as frequencies, anharmonicities, and couplings, enabling fast design-space exploration.

Two complementary tools support this layout-level workflow. SQuADDS~\cite{Shanto_2024} provides an open-source database of experimentally validated superconducting device designs, each generable through Qiskit~Metal and simulated via finite-element solvers, together with a front-end that returns a ``best-guess'' design from desired circuit parameters. QDesignOptimizer~\cite{Eriksson_2025} closes the optimization loop with an open-source Python package that combines Ansys~HFSS electromagnetic simulation, pyEPR Energy-Participation-Ratio analysis, and Qiskit-Metal under a physics-guided nonlinear-optimization strategy that targets user-specified mode frequencies, decay rates, and coupling strengths across qubits, resonators, and couplers. At the chip and system level, graph neural networks (GNNs) have begun to replace heuristic layout tools. Ai and Liu~\cite{Ai_2025} introduce a GNN-based parameter-design algorithm whose ``three-stair scaling'' mechanism mitigates quantum crosstalk on circuits with up to roughly 870 qubits, reducing errors to about 51\% of the previous state of the art and cutting runtime from 90~minutes to 27~seconds. Lu \emph{et al.}~\cite{Lu_2025} develop a closed-loop neural-network surrogate plus Bayesian optimization (BO) for superconducting-chip frequency configuration, accounting for nonlinear error mechanisms such as crosstalk and decoherence that linear models cannot capture, and validate the method by randomized benchmarking and cross-entropy benchmarking. 

For integrated quantum photonics, the inverse-design pipeline spans emitter selection, device-element design, and deterministic on-chip emitter assembly, as surveyed by Kudyshev \emph{et al.}~\cite{Kudyshev_2021}.

\textbf{Fabrication and device preparation.} Once a design exists, AI plays two complementary fabrication roles. The first is to identify patterns in empirical or in-line measurements, either to fine-tune the recipe or to select optimized device regions. The second is to drive robotic actuators that physically execute the recipe. A direct demonstration of the latter is robotic chip-scale nanofabrication of Josephson-junction devices~\cite{Mayor_2026}. Automating the wet-chemical resist-development step with a robotic arm reduces chip-to-chip junction-resistance spread from roughly 7\% under human operation to roughly 2\% with the robot, pointing toward more consistent device fabrication.

For the analytics-driven role, CNN analysis of in-line scanning-electron-microscopy (SEM) micrographs has been used to optimize quantum-dot qubit nanofabrication, including lithographic proximity effects and process quality control~\cite{Mei_2021}. Atomic-precision silicon-qubit fabrication has been a particularly fruitful target. Real-time machine-learning prediction of donor number during scanning-tunneling-microscope (STM) patterning enables atomic-precision manufacturing of donor qubits in silicon~\cite{Tranter_2024}, while million-atom STM simulations combined with machine learning provide post-fabrication metrology that infers donor number and positions for atomic-scale devices~\cite{Usman_2020}. Two earlier modules from the Wolkow group make this pipeline autonomous: deep-learning identification of defects on hydrogen-terminated silicon routes patterning into defect-free regions before hydrogen lithography~\cite{Rashidi_2020}, and an earlier CNN automatically detects degraded STM tips and triggers in-situ reconditioning~\cite{Rashidi_2018}. Machine-learning-enhanced in-situ electron-beam lithography (EBL) has been used to deterministically integrate single InGaAs quantum dots into photonic nanostructures~\cite{Donges_2022}. At the materials-growth front, supervised regression and BO over diamond synthesis and post-processing parameters improve nitrogen-vacancy (NV) magnetometry sensitivity by 300\% over an average sample and 55\% over the previous champion~\cite{deQuilettes_2025}, with Shapley-importance rankings revealing the dominant growth parameters.

\textbf{Control: atomic, molecular, and optical (AMO) experiments.} The control layer is where AI in quantum experiments has accumulated the deepest experimental track record. On atomic, molecular, and optical (AMO) platforms, Wigley \emph{et al.}~\cite{Wigley_2016} established the canonical example of self-optimizing quantum experiments by reducing the number of optimization iterations needed to produce a Bose--Einstein condensate (BEC) by roughly an order of magnitude. BO has since been pushed to genuinely high-dimensional control with up to 55 simultaneous controls in BEC preparation~\cite{Vendeiro_2022}, and reinforcement learning has extended these gains: Milson \emph{et al.}~\cite{Milson_2023} demonstrate experimental RL preparation of ultracold quantum gases over many simultaneous control parameters, while Reinschmidt \emph{et al.}~\cite{Reinschmidt_2024} use RL to control a magneto-optical trap, discover robust operating modes, and successfully transfer in-silico learning to a real experiment. BO has also been used to improve robust many-body state-preparation protocols relevant to few-body fractional quantum Hall physics~\cite{Blatz_2024}, bridging AI experiment control and many-body quantum-state preparation.

\textbf{Control: quantum-dot device tune-up.} Quantum-dot devices were an early adopter of machine-learning autotuning, as reviewed by Zwolak and Taylor~\cite{Zwolak_2023}. Kalantre \emph{et al.}~\cite{Kalantre_2019} used CNNs for state recognition and automated tuning, in-situ machine-learning autotuning on hardware was demonstrated for double-dot devices~\cite{Zwolak_2020}, and Moon \emph{et al.}~\cite{Moon_2020} achieved completely automatic tuning of a quantum device across gate-voltage spaces of dimension up to eight, with median tuning times below 70~minutes and decisive speedups over random search. Schuff \emph{et al.}~\cite{Schuff_2026} have demonstrated fully autonomous tuning of a spin qubit from a grounded device through to coherent Rabi oscillations, combining deep learning, BO, and computer vision; RL has also been used to decide \emph{what to measure}, reducing measurement burden on quantum devices including automatic identification of double-dot bias triangles~\cite{Nguyen_2021}.

\textbf{Control: device or Hamiltonian characterization.} Closely related to control is the question of what model the controller is acting on. This has led to a parallel body of work on Hamiltonian learning and device characterization. B\'ejanin \emph{et al.}~\cite{Bejanin_2021} formulate resonant-coupling parameter estimation using offline data acquisition and modeling combined with online Bayesian learning. Genois \emph{et al.}~\cite{Genois_2021} combine domain knowledge with a recurrent neural network (RNN) to learn the dynamics of a superconducting qubit from experimental data; the same collaboration later extends this RNN-based characterization to detailed numerical quantum optimal control of single qubits~\cite{Genois_2025}. BO calibration with a state-parameter estimator for non-Markovian environments has been studied numerically~\cite{Qian_2022}, learning-based calibration of flux crosstalk has been demonstrated in transmon arrays~\cite{Barrett_2023}, and real-time binary-search Hamiltonian tracking achieves exponential scaling of calibration precision in measurement count~\cite{Berritta_2025}.

Hardware-aware characterization is also moving toward real-time and multimodal settings. FPGA-enabled real-time $T_1$ tracking resolves fast fluctuations of $T_1$ that previous nonadaptive protocols averaged out, improving temporal resolution by two orders of magnitude~\cite{Berritta_2026}. Deep transfer learning maps fluxonium spectra to Hamiltonian parameters, addressing the spectrum-inversion problem that is harder for fluxonium and 0-$\pi$ qubits than for transmons~\cite{Kung_2025}. A recent multimodal turn treats calibration plots themselves as inputs to vision-language models: \citet{cao2026qcaleval} introduce {QCalEval}, a benchmark for quantum-calibration plot understanding with 243 samples across 87 scenario types and 22 experiment families, and release {NVIDIA Ising Calibration 1} as an open-weight reference model for this setting.

Beyond superconducting hardware, autonomous \emph{characterization} agents have operated directly on real quantum hardware. An autonomous protocol combining unsupervised machine learning with a genetic algorithm reverse-engineers Hamiltonian models from nitrogen-vacancy-center experiments and retrieves plausible Hamiltonian models in 74\% of experimental-data instances~\cite{Gentile_2021}. The Quantum Model Learning Agent (QMLA), built around exploration-tree strategies and Elo-style objective functions, identifies the correct interaction family in 72\% of simulated Ising / Heisenberg / Hubbard cases~\cite{Flynn_2022}. On a reconfigurable integrated photonic device, an experimental gray-box agent simultaneously controls the device and reconstructs internal unitaries / Hamiltonians, accommodating fabrication imperfections~\cite{Youssry_2024}.

\textbf{Control: gate and operation-parameter optimization.} Model-free RL has become a workhorse for non-classical state preparation and gate design. Sivak \emph{et al.}~\cite{Sivak_2022} proposed and numerically demonstrated model-free RL with adaptive measurement-based feedback for non-classical-state preparation, an approach that has subsequently been adapted in Gottesman--Kitaev--Preskill (GKP) control work. Porotti \emph{et al.}~\cite{Porotti_2022} study coherent transport and feedback control with deep RL, including Fock-state preparation under weak nonlinear measurements. Li \emph{et al.}~\cite{Li_2025} introduce a sample-efficient RL-from-demonstration approach for quantum gate calibration, targeting workflows where individual experiments are expensive.

Machine learning has also been used to find pulse parameters for unconventional gates, including three-qubit Toffoli and parity-check operations in transmon systems~\cite{Daraeizadeh_2020}. On real IBM hardware, Baum \emph{et al.}~\cite{Baum_2021} show that experimental deep RL can produce error-robust gate sets. Single-qubit gate design has been demonstrated with model-free RL operating on real-time readout signals~\cite{Wright_2023}. In numerical simulation, model-free RL has designed two-qubit transmon entangling gates such as CNOT and cross-resonance~\cite{Nguyen_2024}, and an RL-parameterized ansatz has been combined with classical optimal control to design fast two-qubit gates~\cite{Sarma_2025}. An embedded / FPGA implementation pipeline for arbitrary single-qubit rotations~\cite{Bhat_2025} and a multi-objective deep-RL formulation for superconducting quantum optimal control~\cite{Liu_2025} extend this thread toward deployed hardware and multi-criterion optimization. Across gate-design work, a clear separation emerges between numerical pulse design and a growing set of hardware-validated demonstrations, with the latter increasingly using real-time feedback.

\textbf{Control: readout or state-discrimination parameter optimization.} Readout is the layer where machine learning has had a particularly cleanly measured impact. Magesan \emph{et al.}~\cite{Magesan_2015} introduced machine-learning-based discrimination of quantum-measurement trajectories using IBM superconducting-qubit data, and trapped-ion readout was likewise improved by neural networks trained on experimental data~\cite{Seif_2018}. Hidden Markov models capture temporal structure in readout signals~\cite{Martinez_2020}. Recurrent networks reconstruct full quantum dynamics from physical observations in superconducting circuits~\cite{Flurin_2020}; an LSTM estimator tracks fast superconducting-qubit dynamics~\cite{Koolstra_2022}; and deep networks improve frequency-multiplexed five-qubit superconducting readout~\cite{Lienhard_2022}.

Other work improves the signal-processing and deployment layers. Excited-state-promoted readout has been combined with feedforward classifiers~\cite{Azad_2022}, hardware-efficient machine-learning architectures have been engineered for readout scaling~\cite{Maurya_2023}, and autoencoders~\cite{Luchi_2023} and Bayesian learning~\cite{Cosco_2023} each enhance discrimination at the signal level. Time-resolved modulated networks~\cite{You_2023} move beyond binary discrimination to full state tomography, while path-signature features of the time-domain readout~\cite{Cao_2024} further improve assignment fidelity across multiplexed setups.

A growing body of work pushes inference onto field-programmable gate arrays (FPGAs) at nanosecond latency, enabling mid-circuit measurement and feedback. Machine-learning-powered FPGA-based real-time discrimination has demonstrated mid-circuit measurements for superconducting qubits~\cite{Vora_2024}, and a low-latency neural-network accelerator has been designed for multi-qubit-state discrimination~\cite{Gautam_2024}. RL has been used to co-optimize readout pulses and classifiers~\cite{Chatterjee_2025}. End-to-end deployment workflows now combine the RFSoC FPGA framework with the hls4ml (high-level-synthesis-for-machine-learning) deployment toolchain~\cite{DiGuglielmo_2025}, knowledge-distilled lightweight networks tailored for FPGA readout~\cite{Guo_2025}, and dedicated architectures for multi-level (qutrit / leakage-aware) readout at scale~\cite{Mude_2025}. Bayesian-learning measurement-error mitigation has been extended to multi-qubit experiments on near-term superconducting devices~\cite{Cosco_2025}, and reservoir computing provides a computationally lighter alternative to deep networks for crosstalk-aware multiplexed readout~\cite{Kent_2026}. Closing the feedback loop, Reuer \emph{et al.}~\cite{Reuer_2023} realized a deep-RL agent in real-time FPGA hardware demonstrated on a qubit-reset task.

\textbf{Agentic quantum control.} Agentic AI systems are beginning to close the full control-and-measurement loop by translating human instructions directly into laboratory experiments. An agent-based framework~\cite{Cao_2025} translates natural language into executable laboratory scripts and reads figures, such as plots and scope traces, to choose the next measurement step. An LLM-assisted superconducting-qubit experimentation framework~\cite{Li_2026} demonstrates two end-to-end tasks: finding readout resonators from scratch, and reading a research paper to reproduce the quantum non-demolition (QND) measurement protocol it describes.

\subsubsection{Formal Frameworks, Discovery, and Outlook}

On the conceptual side, \citet{sultanow2025quantum} provides the first formal definition of \textbf{quantum agents} and proposes architectures that integrate quantum workflows with autonomous agent-based systems. Their analysis covers both how QI may enhance AI capabilities and how autonomous AI can support quantum workflows. \citet{breen2025ax} presents \textbf{Ax-Prover}, a multi-agent theorem-proving system in Lean that combines LLM reasoning with formal verification tools. Ax-Prover outperforms frontier LLMs on the authors' \textit{QuantumTheorems} benchmark and further demonstrates assistant capabilities in quantum cryptography.

Going beyond individual theorem proving, \textbf{MerLean}~\cite{ren2026merlean} introduces a fully automated agentic framework for \emph{end-to-end autoformalization} of quantum-computing research papers. MerLean extracts mathematical statements directly from \LaTeX{} source files, formalizes them into verified Lean~4 code built on Mathlib, and translates the results back into human-readable \LaTeX{} for semantic review. Evaluated on three theoretical quantum-computing papers, MerLean produces 2,050 Lean declarations from 114 statements, reducing the verification burden to only newly introduced definitions and axioms. This pipeline offers both a practical tool for machine-verified peer review of quantum-computing research and a scalable engine for mining high-quality synthetic data to train future reasoning models. A more domain-specific formalization effort is the end-to-end Lean~4 formalization of quantum error correction by \citet{ehatamm2026end}. This highlights a complementary role for proof assistants: not only autoformalizing research papers, but also closing trust gaps in concrete fault-tolerant quantum-computing primitives such as QEC definitions, code properties, and correctness certificates.

\citet{karosas2025agentic} examines the governance and ethical implications of combining autonomous AI with QI. Beyond orchestration and verification, autonomous AI is also emerging as a creative partner in physics discovery. \citet{guevara2026single} reports a collaboration between GPT-5.2 and physicists in which the model autonomously conjectures a closed-form formula showing that single-minus tree-level gluon scattering amplitudes are nonzero in the half-collinear regime, overturning a decades-old assumption in quantum field theory, and derives a formal proof subsequently verified by the human collaborators.

Despite rapid progress, significant challenges remain: the reliability and verifiability of autonomously generated quantum protocols, the limited context windows of current LLMs for representing complex quantum systems~\cite{cao2024agents}, and the need for benchmarks that go beyond syntactic correctness to evaluate quantum semantic fidelity and hardware-aware optimization~\cite{fu2025qagent,jern2025agent,vishwakarma2024qiskit,quanbench2025}. The emerging benchmarking infrastructure (Qiskit HumanEval, QCoder, QuanBench) and autoformalization pipelines (Ax-Prover, MerLean) provide initial foundations, but substantial gaps remain in evaluating agentic systems on end-to-end quantum research workflows that span algorithm design, implementation, execution, and interpretation.

\subsection{AI for quantum sensing}
\subsubsection{Overview}
Artificial intelligence is emerging as a useful design tool for advanced quantum sensing protocols. In conventional quantum metrology, the goal is often to estimate an unknown parameter with minimum variance. Many sensing applications require a downstream decision, such as classification, hypothesis testing, anomaly detection, or extraction of a nonlinear feature. This motivates a task-oriented formulation in which the probe state, sensing interaction, quantum control, measurement, and classical decision rule are optimized end to end. This view naturally connects quantum sensing with VQAs and quantum machine learning.

\subsubsection{Entangled sensor networks}
A key example is supervised learning assisted by an entangled sensor network (SLAEN)~\cite{Zhuang2019SLAEN}. In SLAEN, distributed quantum sensors coordinate entanglement and measurement strategy to perform a classification task directly at the physical layer. The entangled network learns collective observables matched to the decision boundary. The original theory considered a support-vector-machine-inspired classification problem and trained a squeezing--linear-optics circuit to minimize classification error. An experimental demonstration later showed quantum-enhanced data classification using a variational entangled sensor network for multidimensional radio-frequency signals~\cite{Xia2021SLAEN}. These works established that entanglement can be shaped to improve parameter precision and reduce task-specific decision error.

\subsubsection{Nonlinear and computational sensing}
Gaussian SLAEN architectures are naturally suited to linear classification tasks. This limitation was addressed by controllable bosonic variational sensor networks, where universal bosonic control enables nonlinear classification by training non-Gaussian sensing and measurement operations~\cite{Liao2024BosonicVSN}. In this setting, the sensor acts as a trainable quantum feature processor: the physical channel embeds data into a quantum state, while the variational circuit maps nonlinear task-relevant features onto measurement outcomes. Related ideas appear in quantum computational sensing, where quantum processing is integrated with the sensing interaction so that the device directly outputs a property or class label~\cite{Khan2025QCSA}. Recent experimental work on quantum computational displacement sensing used a superconducting qubit--oscillator platform and trained parameterized circuits to classify displacement signals, reporting a task-specific sensing advantage over estimate-then-classify baselines~\cite{Prabhu2026QCDS}.

\subsubsection{Variational quantum state classification}
The classification formulation is also closely related to quantum state hypothesis testing. Variational circuits can be trained to minimize state-discrimination error, as in quantum convolutional neural networks for many-body phase classification~\cite{Cong2019QCNN}, universal discriminative quantum neural networks~\cite{Chen2021UDQNN}, and noisy quantum neural-network state discrimination~\cite{Patterson2021NoisyQNN}. Analyses of circuit depth and classification-error decay further clarify how variational expressivity, noise, and trainability determine sensing performance~\cite{Zhang2022FastDecay}.

\subsubsection{Adaptive quantum sensing}

Adaptive quantum sensing has roots in optical coherent-state discrimination, where the receiver is optimized to distinguish nonorthogonal quantum states with minimum error. The Kennedy receiver introduced a displacement-and-photon-counting strategy for binary coherent states~\cite{Kennedy1973Receiver}, and the Dolinar receiver showed that real-time feedback can attain the Helstrom bound for binary coherent-state discrimination~\cite{Dolinar1973Receiver}. An early experimental milestone was the closed-loop coherent-state measurement, which used real-time quantum feedback to emulate an optimal measurement for optical coherent-state discrimination~\cite{Cook2007ClosedLoop}. Conditional-nulling ideas were later extended to optical codeword demodulation, where optimized coherent-pulse nulling, photon counting, and quantum feedforward achieved error rates below direct detection~\cite{Chen2012ConditionalNulling}.

This receiver viewpoint naturally connects to adaptive and learning-based sensing. Becerra and co-workers demonstrated adaptive displacement and photon-counting receivers beating the standard quantum limit for multiple nonorthogonal coherent states~\cite{Becerra2013NatPhoton}, followed by photon-number-resolving receivers robust to realistic imperfections~\cite{Becerra2015NatPhoton}, multi-state discrimination below the quantum noise limit~\cite{Ferdinand2017NPJQI}, and robust binary coherent-state discrimination~\cite{DiMario2018PRL}. Machine-learning approaches then generalized adaptivity from hand-designed feedback to optimized policies: Hentschel and Sanders used machine learning to design adaptive phase-estimation strategies~\cite{Hentschel2010MLMeasurement}; Fiderer, Schuff, and Braun introduced neural-network heuristics for adaptive Bayesian quantum estimation~\cite{Fiderer2021NNBayesian}; QREAL used adaptive learning to optimize a feed-forward quantum receiver under different coherent-state discrimination tasks and operating conditions~\cite{Cui2022QREAL}; and Cimini \emph{et al.} applied deep reinforcement learning to quantum multiparameter estimation~\cite{Cimini2023DRLMetrology}. These works show a chronological progression from analytically designed adaptive receivers to AI-optimized sensing policies.

\subsubsection{Variational quantum metrology}
A parallel literature on variational quantum sensing optimizes probes and measurements for metrological objectives in ions and atoms. Examples include variational spin-squeezing algorithms on programmable sensors~\cite{Kaubruegger2019SpinSqueezing}, optimal metrology with trapped-ion quantum sensors~\cite{Marciniak2022OptimalMetrology}, and variational multiparameter quantum metrology for vector-field sensing~\cite{Kaubruegger2023Multiparameter}. These works often minimize estimation error and share the same design principle: use trainable, hardware-native quantum circuits to discover entangled probes and measurements adapted to the task.

\subsubsection{Outlook}
Overall, AI-enabled quantum sensing reframes the sensor as a trainable quantum information processor optimized for an operational task. The central opportunity is end-to-end design: entanglement, squeezing, nonlinear control, measurement, and classical postprocessing can be co-optimized under realistic hardware constraints. Open challenges include rigorous advantage criteria, fair resource accounting, sample-efficient training, robustness to noise and model mismatch, and scalable benchmarks. Recent theory and experiments suggest that AI-designed quantum sensors may be especially powerful when the desired output is a decision, feature, or hypothesis.

\subsection{AI for quantum networking}
\label{sec:ai_network}

Quantum networks aim to distribute entanglement between distant nodes so that the resulting shared resource can support tasks such as quantum key distribution (QKD), distributed sensing, and distributed quantum computation~\cite{azuma2023quantum}. The central obstacle is that entanglement cannot be amplified or copied the way classical signals can, so photon loss and decoherence accumulate exponentially with fiber length. The standard solution is the \emph{quantum repeater}, which divides a long channel into shorter segments, generates entangled pairs across each segment, and connects them using entanglement swapping at intermediate nodes. Connection and local operations are themselves imperfect, so each swap lowers the fidelity of the resulting pair, and repeaters therefore interleave swapping with \emph{entanglement purification}, which probabilistically distills higher-fidelity pairs from lower-fidelity ones at the cost of consuming additional pairs and additional classical communication. The resulting nested protocols deliver high-fidelity end-to-end entanglement using resources that grow only polynomially with distance~\cite{briegel1998quantum,dur1999quantum}.

Once repeater nodes are connected into networks rather than linear chains, additional control problems appear. Switches must route entanglement among many users, and grids of repeaters must support multiple simultaneous flows. The network must decide which users to serve, which paths to use, and how to share limited memory across competing requests, so the operation of every node becomes a sequence of decisions: when to attempt generation, when to swap, when to purify, when to discard a stored pair whose fidelity has decayed, and when to consume a pair for the application. These decisions are naturally modeled as a Markov decision process (MDP), with states capturing the occupancy and quality of each memory and actions corresponding to the local operations available at each node.

Several features of realistic networks make the resulting MDPs hard for classical methods. The state space grows rapidly with the number of memories, links, and pending classical messages, so exact dynamic programming becomes intractable beyond small systems; even the bipartite--tripartite switch with two memories per link is analyzed only numerically~\cite{vardoyan2020exact,vardoyan2023capacity}, and routing in 2D grids is solved through carefully designed local heuristics rather than optimal policies~\cite{Pant2019}. Operationally relevant objectives such as the secret-key rate (SKR) of a QKD protocol, or the joint rate region across competing entanglement flows, are non-linear in fidelity and generation rate, breaking the additive-reward assumption that standard reinforcement learning (RL) relies on. And classical-communication delays between nodes mean that any node's view of the network is necessarily stale, so the underlying decision problem is partially observable rather than fully Markovian. Recent work therefore turns to RL, graph neural networks, and partially observable MDP (POMDP) methods to discover network control policies, alongside machine-learning approaches to characterize the links and nodes those policies must operate on.

\textbf{RL for entanglement distribution and repeater control.} The starting point for learning-based repeater control is the finite MDP formulation of I\~{n}esta et al.~\cite{Inesta2023}, which models a homogeneous repeater chain with per-memory cutoffs that discard stored entangled links once their fidelity has decayed too far. The state encodes the ages of stored entangled links, and the action at each time step is the set of nodes that perform a swap. Value iteration and policy iteration yield globally optimal policies for end-to-end entanglement delivery time, outperforming the standard swap-as-soon-as-possible (swap-asap) heuristic. The framework assumes global instantaneous knowledge of the MDP state, an idealization that subsequent work systematically relaxes.

Rei\ss{} and van Loock~\cite{loock2023} apply deep RL to a four-segment repeater chain targeting the BB84 SKR. The generalized non-additive objective they adopt is one whose equivalence to the true SKR they cannot rigorously establish. Haldar et al.~\cite{Haldar2023-dz} use Q-learning on inhomogeneous chains to discover dynamic, state-adaptive cutoff and node-collaboration policies that improve on swap-asap in both waiting time and fidelity, identifying global knowledge of the chain state as the structural source of advantage. Classical-communication costs, however, remain outside their model.

Subsequent work targets that cost directly. Haldar et al.~\cite{Haldar2024-xm} introduce \emph{quasi-local} policies for multiplexed repeater chains, in which each node acts on information from a bounded sub-region of the network; this sharply reduces classical-communication overhead while remaining competitive with fully global strategies. Li et al.~\cite{Li2024-bw} take the complementary centralized route, folding classical-communication delays into the MDP state via an action--result history; their RL policy outperforms the natural wait-for-broadcast generalization of swap-asap in the high-success-probability regime, where partial-information decision-making outweighs the cost of acting on stale information. Mobayenjarihani et al.~\cite{Mobayenjarihani} characterize the same wait-versus-act trade-off analytically through an \emph{optimistic} purification scheme in which nodes proceed without waiting for heralding messages, trading classical-communication latency against the risk of acting on failed pairs. Casado et al.~\cite{casado2025} extend the RL framing to entanglement-distribution decisions on general network topologies.

Yau et al.~\cite{Yau2026-fx} return to the question of the objective itself, developing an RL framework that directly optimizes non-linear, differentiable application-driven objectives such as the BB84 SKR. Their MDP models two-node entanglement distribution with multi-memory configurations, multiplexing, and distillation, and includes classical-communication-induced uncertainty about distillation outcomes as part of the state. To handle the non-additive objective, they extend standard policy-gradient methods to optimize functions of multiple expected discounted returns, capturing the joint dependence on fidelity and generation time. The resulting policies improve the SKR over threshold-based baselines by up to 23\% in the parameter regimes considered.

\textbf{Routing, resource allocation, and protocol primitives.} Beyond a single chain, networks of repeaters serving multiple users introduce decisions about which path to use for each request, how to schedule competing requests against finite memory, and how to revise routing when links fluctuate. Early deep RL approaches address these directly through deep Q-networks for request scheduling on qubit-limited grids~\cite{lenguyen2022}, proximal policy optimization (PPO) agents that exploit the non-additive composition of quantum errors to find higher-fidelity routes than classical shortest-path algorithms~\cite{Roik2024}, and Q-learning policies that cache unused entangled links and proactively swap commonly used segments~\cite{islam2024}. More recent work confronts larger topologies and time-varying links: graph neural networks with local message passing yield policies competitive with global-information baselines on both terrestrial and satellite topologies~\cite{Meuser2025-kg,Meuser2026-bv}, and belief-state methods cast routing as a POMDP with formal convergence guarantees under time-varying decoherence~\cite{Taherpour2025}.

These controllers depend on resource primitives that are themselves under active study. Vardoyan and Wehner~\cite{Vardoyan_qnum} extend classical network utility maximization to quantum networks, providing the formal framework for jointly allocating entanglement rate and quality across competing users with fairness encoded through log-composition of per-route utilities. Related work develops the protocol- and design-level primitives these controllers use: entanglement buffering with closed-form availability and consumed-fidelity bounds for any purification policy~\cite{Inesta2026}, and machine-learning surrogates of expensive network simulators for tuning memory allocations and protocol parameters at a fraction of the simulation cost~\cite{Prielinger2024}.

\textbf{Network characterization.} The controllers and routing agents above presuppose channel parameters that, in practice, must be estimated from measurements without direct access to the internal links. Quantum network tomography (QNT) addresses this problem by inferring per-link channel parameters from end-to-end measurements at peripheral nodes, in analogy to its classical counterpart but contending with quantum-specific obstacles such as the multiplicative composition of channel parameters and the sign ambiguities it induces. Guedes de Andrade et al.~\cite{GuedesDeAndrade2024} formalize QNT for star topologies with single-parameter Pauli channels and characterize the regimes in which their multicast-based estimators achieve identifiability. Wang et al.~\cite{Wang2025} extend the framework to arbitrary topologies and full Pauli channels via a Mergecast protocol that entangles peripheral qubits at an intermediate node, paired with a progressive etching procedure that identifies internal channels working from the periphery inward; they further give estimators for state-preparation and measurement errors and validate the combined workflow under photon loss and memory decoherence.

Machine learning offers a complementary route that does not assume a fixed channel class. Mukherjee et al.~\cite{Mukherjee2024-mf} demonstrate a multi-stage neural-network pipeline on a Rydberg-array black box that classifies the number of network nodes, regresses their positions, and recovers the system--environment Hamiltonian and Lindblad operators from a single-time snapshot of excitation transport, with reconstruction degrading once decoherence becomes comparable to the mean dipolar coupling. At the device layer beneath such a network, Canonici et al.~\cite{canonici2024} apply supervised learning to estimate physical noise parameters of a neutral-atom processor from occupation-probability measurements. Together these directions cover targets ranging from device-level noise to network-level open-system models, supplying the channel parameters and cutoff times the controllers above treat as inputs.

\textbf{Physical-layer networking optimization.}
Beyond network-level tasks, AI also has significant potential to enhance physical-layer protocols for quantum networking and transduction. Early work by Krastanov \emph{et al.}~\cite{krastanov2019optimized} used genetic algorithms to optimize entanglement purification protocols for qubit systems. More recently, Zhao \emph{et al.}~\cite{zhao2021practical} introduced LOCCNet, a machine-learning framework for the design and optimization of distributed quantum information-processing protocols, including entanglement purification. This approach was later extended to dynamic LOCCNet~\cite{liu2026dynamic}, which improves scalability by constructing larger LOCC protocols from smaller adaptive modules. In a hardware-oriented setting, Zhang \emph{et al.}~\cite{zhang2022hybrid} developed a qubit--qumode hybrid variational quantum circuit to optimize the physical-layer distillation of entangled qubits from noisy entangled qumodes, enabling improved performance over conventional time-bin approaches. Later on, similar technique are applied to variational quantum transduction~\cite{liao2026variational}. Another related direction is variational quantum network optimization: Doolittle \emph{et al.}~\cite{doolittle2023variational} employed variational quantum circuits to optimize network-level performance measures such as nonlocality, using noisy qubit channels as the underlying network model.

\textbf{Outlook.} The picture above is largely one of single-agent learning over a fixed substrate. Several directions push beyond that frame. Distributed and multi-agent learning over quantum networks is one: DeRieux and Saad~\cite{DeRieux2024-io} propose entangled quantum multi-agent reinforcement learning, in which agents share an entangled critic over a quantum channel and coordinate without exchanging local observations. Quantum-secured federated learning over QKD networks goes the other way, with the network's QKD service underwriting classical aggregation rather than itself becoming the learning target. Both directions sharpen, rather than resolve, the open challenges that already cut across the section. Chief among them is scaling RL controllers from elementary links to long multi-hop chains under realistic partial observability, and ensuring that the operating regime of a deployed controller does not silently invalidate the channel statistics on which it was trained --- characterization and control must ultimately be co-designed rather than staged. Benchmarking AI-discovered policies against analytic baselines on standardized quantum-network simulators, and bridging the gap between elementary-link RL and continuous multi-hop network operation remain practical priorities.

\section{Quantum Computing and Quantum-Inspired Methods for Artificial Intelligence}
\label{sec:quantum4ai}

\subsection{Quantum Algorithmic Speedups for Classical ML Tasks}
\label{subsec:qc-speedup}

Many machine learning workloads rely on a small set of linear algebra primitives. Two examples are solving linear systems and extracting spectral structure through eigendecomposition or singular value decomposition. Quantum algorithms can sometimes accelerate these primitives by exploiting superposition and entanglement in exponentially large Hilbert spaces. The usual strategy is to place a quantum subroutine inside a larger ML pipeline. For instance, a quantum linear system algorithm (QLSA) can solve the normal equation in least-squares regression, or serve as a subroutine inside a larger optimization routine. Whether the speedup survives in practice depends on how the data are encoded, what access model is assumed, and how much information must be read out at the end. This subsection traces the development from the foundational HHL algorithm to modern QLSAs and then surveys applications in optimization, machine learning, and scientific computing. For a concise primer on this algorithmic family, see \citet{dervovic2018quantum}.

\textbf{HHL Algorithm: A Foundational Quantum Linear Systems Solver.} The Harrow-Hassidim-Lloyd (HHL) algorithm~\cite{harrow2009quantum} is the foundational quantum algorithm for the linear systems problem. Formally, it addresses a state-preparation version of this problem: given a matrix $A$ and an efficiently preparable input state $|b\rangle$, the goal is to prepare a state $|x\rangle$ satisfying
\begin{equation}
\left\|\,|x\rangle-\frac{A^{-1}|b\rangle}{\|A^{-1}|b\rangle\|}\right\| \le \varepsilon ,
\end{equation}
where $\varepsilon$ is the target error in the output state~\cite{harrow2009quantum,dervovic2018quantum}. This definition makes the key point explicit: HHL returns a quantum state that approximates the normalized solution. In the usual HHL input model, $A$ is an $s$-sparse Hermitian $N \times N$ matrix. This means that each row has at most $s$ nonzero entries and that the row can be queried efficiently: given a row index, one can identify the nonzero column indices and obtain their values in time $\mathcal{O}(s)$~\cite{harrow2009quantum}. The primer of \citet{dervovic2018quantum} formalizes this access model and explains why it is needed for efficient Hamiltonian simulation.


The original HHL algorithm performs matrix inversion by working in the eigenbasis of $A$. It encodes the input vector $\mathbf{b}$ as a quantum state $|b\rangle$, treats $A$ as a Hamiltonian whose evolution can be efficiently simulated, and uses quantum phase estimation to coherently estimate the eigenvalues of $A$ into an ancillary register. Conditioned on these eigenvalue estimates, the algorithm applies a controlled rotation to another ancilla state whose successful post-selection weights each eigencomponent by the inverse of its eigenvalue. Applying inverse phase estimation then uncomputes the eigenvalue register, leaving a normalized quantum state proportional to $A^{-1}|b\rangle$. Thus HHL implements matrix inversion by leaving the eigenvectors unchanged while inverting the corresponding nonzero eigenvalues, producing a quantum state from which selected properties of the solution can be estimated by measurement.

If the input state $|b\rangle$ can also be prepared efficiently, then the original HHL runtime scales as $\widetilde{\mathcal{O}}(\log(N)\, s^2 \kappa^2/\varepsilon)$, where $\kappa$ is the condition number of $A$~\cite{harrow2009quantum}. This can yield an exponential improvement in the system size $N$ when the quantum state output is itself a useful final representation. Subsequent work by \cite{childs2017quantum} improves the precision dependence, reducing the runtime to $\mathcal{O}(\log(N)\kappa \log(1/\varepsilon))$ up to the same model-dependent overheads.

This improvement uses the \emph{linear combination of unitaries} (LCU) technique. In LCU, the inverse transformation is approximated by a weighted sum of efficiently implementable unitaries, often built from short-time evolutions under $A$. By combining these unitaries coherently, the algorithm applies an approximation to $A^{-1}|b\rangle$ directly, without estimating every eigenvalue one by one. This idea foreshadows the modern QLSA methods discussed below.

\textbf{HHL and QLSAs for Machine Learning Applications.} The HHL algorithm and its successors apply to a broad range of machine learning tasks in which the computational bottleneck is a linear algebra primitive. Some of the earliest examples come from regression and classification, but recent work continues to expand these ideas to broader regression families, multi-class classification, and modern generative models. We organize these applications by task type:

\emph{Regression.} In least-squares regression, one seeks a parameter vector $\boldsymbol{\theta}$ that minimizes the squared residual norm $\|X\boldsymbol{\theta}-\mathbf{y}\|_2^2$. The corresponding first-order optimality condition is the normal equation $(X^T X)\boldsymbol{\theta} = X^T \mathbf{y}$, a linear system that HHL can address when the data matrix $X$ and target vector $\mathbf{y}$ admit efficient quantum state preparation. Building on this idea, \citet{wiebe2012quantum} gives a quantum algorithm for least-squares fitting that efficiently estimates fit quality and, in many cases, recovers a concise fitting function. More recently, \citet{liu2025accelerating} extends the discussion beyond standard linear fitting to a broader family of regression tasks, including ridge-, Lasso-, Huber-, and $\ell_p$-type objectives, and shows that quantum techniques can provide up to a quadratic improvement in the sample parameter under their structured access assumptions.

\emph{Classification.} \citet{rebentrost2014quantum} introduces a quantum least-squares support vector machine (LS-SVM) that applies HHL-style matrix inversion to an augmented system built from the training-data kernel matrix. Under strong data-access and low-rank assumptions, the method achieves logarithmic scaling in both feature dimension and training-set size. Subsequent work refines this approach. \citet{pinheiro2025ls} studies optimized HHL circuits for LS-SVM classifiers, with an emphasis on reducing circuit size and execution cost. The multi-class study of \citet{pinheiro2025quantum} compares quantum-kernel QSVMs with HHL-based LS-SVM schemes using two reductions to binary classification. The first is a \emph{one-vs-rest} (OvR) construction, in which one binary classifier is trained for each class against the union of all remaining classes, and the predicted label is chosen from the classifier with the largest decision score. The second is a \emph{two-step hierarchical} construction. In the three-class astronomical-object task built from a reduced \emph{Sloan Digital Sky Survey} (SDSS) dataset, the first classifier isolates one chosen class from the other two, and a second classifier distinguishes between the two remaining classes. That comparison highlights a practical tradeoff: the HHL-based route retains attractive asymptotic scaling, but it is more sensitive to noise and tends to underperform the kernel-based alternative on the reduced SDSS benchmark considered in that work.

\textbf{Limitations and Practical Constraints.} The applications above share two bottlenecks. First, classical data must be encoded efficiently into quantum states, often through QRAM-style assumptions. Second, the quantum output is a state, so extracting a full classical solution may require enough measurements to erase the speedup. QLSA-based advantages are therefore most relevant when expectation values or other partial functions of the solution are sufficient.

HHL and related QLSAs also face practical constraints emphasized early by \citet{aaronson2014quantum} and summarized in later QLSA surveys such as \citet{dervovic2018quantum}. The original HHL algorithm requires fault-tolerant quantum computation with long coherence times to implement quantum phase estimation and its inverse reliably~\cite{harrow2009quantum}, which makes it unsuitable for current NISQ devices~\cite{wang2024comprehensive}. Its runtime depends quadratically on the condition number $\kappa$ of the matrix $A$, scaling as $\mathcal{O}(\kappa^2/\varepsilon)$, and later QLSAs still retain condition-number dependence even when they improve this factor to linear. Data encoding remains another bottleneck: preparing classical data as quantum states using QRAM or related access models may require $\mathcal{O}(N)$ operations and can negate the speedup~\cite{aaronson2014quantum,jaques2025qram}. \citet{jaques2025qram} further argue that most asymptotic quantum linear-algebra advantages disappear with \emph{active} QRAM architectures, where every query requires external intervention and control. Scalable \emph{passive} QRAM would need memory routing to proceed after the query is initiated without further external input or energy, which rests on demanding physical assumptions.

Readout creates a second practical barrier. Measurements of the output state $|x\rangle$ reveal only partial information, and reconstructing the full solution may require $\mathcal{O}(N)$ runs of the algorithm~\cite{aaronson2014quantum,dervovic2018quantum}. Applicability is further limited by the need for an efficient quantum representation of $A$, such as sparse access or block-encoding; arbitrary matrices do not admit such representations without prohibitive overhead. Later dequantization results sharpen this issue: for low-rank linear systems and related QML pipelines, quantum-inspired classical algorithms can sometimes recover polylogarithmic dimension dependence under sampling-access assumptions analogous to state-preparation assumptions~\cite{chia2018quantuminspired,arrazola2020quantuminspired}. These limitations mean that HHL is best suited for structured problems with favorable condition numbers, efficient data encoding, and applications that require only partial information about the solution. The algorithmic developments below are best understood against these constraints.

\emph{Beyond linear algebra primitives.} Recent work extends QLSAs to more complex ML workloads. \citet{liu2024towards} develops fault-tolerant quantum algorithms for sparse, sufficiently dissipative large-scale models with small learning rates, showing that certain stochastic-gradient-descent dynamics can be simulated in time polylogarithmic in the model size. \citet{wang2025towards} gives a particularly current example on the generative-model side by formulating quantum algorithms for representative diffusion probability models~\cite{ho2020ddpm,nichol2021improved,song2021sde,song2021ddim}, and by recasting higher-order ODE solvers such as DPM-solver-$k$~\cite{lu2022dpmsolver} and UniPC~\cite{zhao2023unipc} into quantum linear-solver and Hamiltonian-simulation primitives. These extensions illustrate ongoing efforts to carry quantum algorithmic advantages beyond core linear algebra, while maintaining rigorous specification of the access models and problem structures under which speedups hold.

\textbf{Beyond HHL: Modern Quantum Linear System Algorithms.}
Modern QLSAs are built around two shifts. First, as recalled in the computational preliminaries in Sec.~\ref{subsec:qc-preliminaries}, \emph{block-encoding}~\cite{chakraborty2019block,low2019hamiltonian} gives a flexible matrix access model. A matrix $A$ is embedded as a subblock of a unitary $U_A$ such that $(\langle 0| \otimes I)\, U_A\, (|0\rangle \otimes I) = A/\alpha$ for some normalization factor $\alpha$. This replaces the sparse Hamiltonian-simulation oracle assumed in HHL with a more composable primitive.

Second, modern algorithms can approximate the matrix function $f(A)=A^{-1}$ directly as a polynomial or rational transformation~\cite{childs2017quantum,gilyen2019quantum}, thereby bypassing QPE. The LCU-based algorithm of \cite{childs2017quantum} is an early concrete example: it represents $A^{-1}$ as a linear combination of unitary operators derived from Hamiltonian simulation and achieves the first exponential improvement in $\varepsilon$ over HHL. The QSVT framework~\cite{gilyen2019quantum} later gives a general mechanism. Given a block-encoding of $A$, QSVT applies a bounded polynomial $p(x)$ to the singular values of $A$ using $\mathcal{O}(\mathrm{deg}(p))$ applications of $U_A$ and its inverse, interleaved with single-qubit signal-processing rotations. To solve $Ax=b$, one chooses $p(x)\approx 1/x$ on the spectral interval $[1/\kappa,1]$, which yields the optimal $\mathcal{O}(\kappa\log(1/\varepsilon))$ query complexity. Building on this perspective, \cite{lin2020optimal} introduces eigenstate filtering, and \cite{tong2021fast} further develops fast inversion methods and preconditioned quantum linear system solvers within the block-encoding framework.

A different line of work uses integral and adiabatic ideas. The Laplace transform identity $A^{-1} = \int_0^\infty e^{-At}\, dt$ expresses the inverse as a continuous superposition of matrix exponentials. \cite{subasi2019quantum} discretizes such an integral via quadrature rules and approximates $A^{-1}|b\rangle$ as a linear combination of time-evolved states $e^{-At_j}|b\rangle$ at quadrature nodes $t_j$. Each term can be implemented through Hamiltonian simulation. \cite{costa2022optimal} achieves the same optimal $\mathcal{O}(\kappa \log(1/\varepsilon))$ complexity through a combination of the discrete adiabatic theorem and qubitized quantum walks. \cite{wu2024efficient} shows that this approach is especially natural for structured systems such as low-rank tensor-sum linear systems arising from discretized partial differential equations.

Most recently, \cite{dalzell2024shortcut} develops kernel reflection, a conceptually simple quantum linear-system solver that extends the eigenstate-filtering idea of \cite{lin2020optimal}. When the solution norm $\|x\|$ is known, a single reflection about the kernel of an augmented linear operator suffices. When $\|x\|$ is unknown, it can be estimated with $\mathcal{O}(\log\log\kappa)$ kernel projections. These modern QLSAs achieve the same information-theoretically optimal $\mathcal{O}(\kappa \log(1/\varepsilon))$ scaling~\cite{harrow2009quantum}, but they use different strategies. LCU and QSVT treat inversion as a directly implementable matrix function, adiabatic methods follow continuous paths in solution space, and kernel reflection projects onto the solution subspace through a small number of reflections. Together, these developments turn quantum linear system solving from a single algorithm into a broader toolkit.

Modern QLSAs also serve as computational engines inside larger machine-learning and scientific-computing pipelines. In optimization, many ML training procedures reduce to structured programs whose Newton steps require solving linear systems. Quantum interior point methods use QLSAs inside each iteration: \cite{kerenidis2020quantum} initiate this direction for linear and semidefinite programs, and \cite{casares2021quantum} develop a predictor-corrector variant.

Later work handles the fact that QLSAs return approximate linear solves. Inexact-feasible formulations preserve primal feasibility despite QLSA error~\cite{mohammadisiahroudi2024efficient,mohammadisiahroudi2025improvements}. Related ideas extend to quadratic~\cite{wu2023inexact} and semidefinite~\cite{augustino2023quantum} optimization. Iterative refinement controls growing condition numbers near optimality~\cite{wu2024quantum}, and preconditioning reduces condition-number dependence in the duality gap~\cite{wu2025preconditioned}. This line culminates in an optimally scaling QIPM framework for dense large-scale linear optimization~\cite{mohammadisiahroudi2025optimal}. A different route, due to \cite{apers2026quantum}, uses Grover-accelerated leverage-score sampling to approximate the Hessian and gradient of the barrier function. Quantum SDP solvers based on matrix multiplicative weight updates form a complementary direction~\cite{brandao2017quantum,vanapeldoorn2020quantum,vanapeldoorn2019improvements}.

QLSAs also appear as subroutines in recommendation systems~\cite{kerenidis2017quantum}, quantum gradient descent for least squares~\cite{kerenidis2020gradient}, and scientific-computing tasks such as discretized differential equations~\cite{berry2017quantum}. Across these examples, the lesson is consistent: QLSAs are most compelling when the linear system has favorable structure, the oracle model is realistic, and the end-to-end pipeline still has an advantage after data loading and readout are included.

\textbf{Other Quantum Linear-Algebra Primitives for ML.}
Not all quantum speedups for ML linear-algebra tasks are rooted in solving linear systems. Two notable examples rely on fundamentally different quantum primitives.

For dimensionality reduction, \citet{lloyd2014quantum} proposes quantum principal component analysis (qPCA) for unknown low-rank density matrices. qPCA uses density matrix exponentiation to extract dominant eigenvectors and eigenvalues in quantum form with runtime polylogarithmic in the dimension under the paper's access assumptions.

For topological data analysis, \citet{lloyd2016quantum} gives quantum algorithms that estimate Betti numbers across persistent homology scales and diagonalize the combinatorial Laplacian via quantum simulation---again without reducing the problem to a linear system. Under QRAM-style access assumptions, these routines can offer exponential improvements over the best classical methods known at the time. \citet{cade2022towards} provides complexity-theoretic evidence that this TDA formulation is likely resistant to dequantization, making it a comparatively strong candidate for quantum speedup.

\subsection{Quantum Computational Advantages for Learning}
Quantum learning advantages are most plausible when the learning task matches the quantum resources used for data access, encoding, and processing. Evidence for this structure dependence appears in quantum sample complexity, data access models, and the geometry of quantum feature maps. A growing body of results indicates that the clearest learning advantages arise when the data themselves come from quantum physical systems.

\textbf{Learning Complexity and Sample Scaling Laws}
The classical PAC framework~\cite{valiant1984theory,blumer1989learnability} gives sample complexity $\Theta(d_{\mathrm{VC}}/\varepsilon)$ in the realizable setting and $\Theta(d_{\mathrm{VC}}/\varepsilon^2)$ in the agnostic setting, suppressing logarithmic dependence on the confidence parameter~$\delta$. Here \(d_{\mathrm{VC}}\) is the VC dimension, which measures the richness of the hypothesis class. Arunachalam and de Wolf~\cite{arunachalam2017guest,arunachalam2018optimal} prove that quantum and classical sample complexity coincide up to constant factors in both settings; exponential gaps appear only in query complexity. Huang, Kueng, and Preskill~\cite{huang2021information} further show that potential quantum gains are restricted in the average-case prediction setting.

Quantum state prediction gives a related lesson. The classical shadow framework~\cite{huang2020predicting} predicts $M$ linear functions with $O(\log M \cdot \max_i\|O_{i,0}\|^2_{\mathrm{shadow}}/\varepsilon^2)$ samples, where $O_{i,0}=O_i-\Tr(O_i)\mathbb{I}/2^n$. For weight-$k$ Pauli observables under local Pauli measurements, the shadow norm scales as $3^k$, so local observables can be estimated efficiently while global operators can remain exponentially costly. Generalization analyses point in the same direction. Caro et al.~\cite{caro2022generalization} bound generalization error by $\sqrt{T/N}$, tightening to $\sqrt{K/N}$ when only $K\!\ll\!T$ gates are active, while Gil-Fuster et al.~\cite{gil2306understanding} show that these uniform bounds can substantially overestimate true error because QNNs can fit random labels. Consistent with these observations, the expressivity--trainability relationship is nonmonotonic~\cite{abbas2021power}: increasing expressive capacity does not necessarily improve learning performance. These results indicate that quantum models rarely achieve advantage through reduced sample complexity alone.

\textbf{Provable Quantum Benefits in Physical Learning.}
When the learning target is a quantum system, the measurement strategy often determines the access complexity. Aaronson~\cite{aaronson2018shadow} establishes shadow tomography: $M$ binary observables require only $\tilde{O}(\varepsilon^{-4}\log^4\!M\log D)$ copies, improved from $\varepsilon^{-5}$ using the online framework of~\cite{NEURIPS2018_c1a3d347}. Huang et al.~\cite{huang2022provably} prove an information-theoretic exponential gap between separable and entangled measurements, verified experimentally on 40-qubit superconducting processors. Chen et al.~\cite{chen2022exponential} show that access complexity interpolates continuously: $k$ memory qubits move the learner between classical and fully quantum sample complexity.

Hamiltonian learning shows similar structure. Haah--Kothari--Tang~\cite{HaahKothariTang2024} achieves $O(\log N/(\beta\varepsilon)^2)$ from high-temperature Gibbs states, while Huang et al.~\cite{huang2023learning} attain Heisenberg-limited $1/\varepsilon$ scaling with entangled sensors. Tang's dequantization program~\cite{tang2019quantum,tang2021quantum} shows that quantum speedups on low-rank classical data are often classically emulable; robust advantage persists mainly for quantum-structured or high-rank data.

Very recently, \citet{zhao2026exponential} establish a provable exponential advantage on classical data by showing that a polylogarithmic-size quantum computer can perform classification and dimension reduction on massive datasets that any sub-exponential-size classical machine cannot match. Their primitive, quantum oracle sketching, accesses classical data in superposition using only random samples, bypasses QRAM entirely, and is validated on single-cell RNA-seq and sentiment classification with fewer than 60 logical qubits.

\textbf{Expressivity, Trainability, and the Limits of High-Dimensional Hilbert Spaces.}
High-dimensional Hilbert spaces alone do not guarantee learning benefits. As discussed in detail in Sec.~\ref{subsec:optimization} (see in particular the analyses of quantum kernels and barren plateaus), quantum feature maps can embed data into exponentially large spaces, yet the resulting models may suffer from kernel concentration~\cite{thanasilp2024exponential}, barren plateaus~\cite{mcclean2018barren,cerezo2021cost,ragone2024lie}, and noise-induced trainability loss~\cite{wang2021noise}. The main tension is that highly expressive circuits can be hard to train, while circuits that avoid barren plateaus often have restricted algebraic structure and may be classically simulable~\cite{cerezo2025does}. Structured architectures such as equivariant circuits~\cite{larocca2022group,PRXQuantum.4.010328} offer one route forward by constraining the dynamical Lie algebra while preserving task-relevant expressivity. In infinite-dimensional continuous-variable systems, linear optical models are efficiently trainable~\cite{volkoff2021efficient}, whereas universal discrete-continuous hybrid models exhibit a distinctive energy-dependent barren plateau phenomenon~\cite{zhang2025energy}.

\medskip\noindent
\textbf{Outlook} The structure-dependence principle emerges consistently across all three axes: quantum gains in learning arise from structural alignment, not raw dimensionality. Quantum sensing for learning is supported by rigorous information-theoretic results and hardware verification~\cite{huang2022provably}. Quantum processing for learning faces the trainability simulability dilemma~\cite{cerezo2025does,ragone2024lie}. Key open problems: (i)~existence of a simultaneously non-simulable and barren-plateau-free circuit family; (ii)~whether quantum data gains compound iteratively; (iii)~beyond-worst-case regimes for variational training~\cite{bittel2021training}; (iv)~quantum sample complexity gains under structural distributional assumptions~\cite{arunachalam2017guest}.

\subsection{Quantum Neural Networks}
\label{subsec:qc-ml}

Building on recent tutorials and surveys of quantum neural network (QNN) architectures and models, such as \cite{du2025quantum}, this subsection benchmarks QNNs against classical neural networks (NNs) by addressing three questions:
\emph{(i) What provable quantum benefits of QNNs over classical NNs are established theoretically?
(ii) Do these provable benefits translate into practical gains on common, real-world workloads?
(iii) What gaps remain between theoretical analysis and practical demonstrations?}
We survey existing benchmarking results comparing classical and quantum models on representative tasks including generative modeling, classification, and time-series prediction to assess whether, and under what conditions, such gains manifest in practice.

\textbf{Theoretical Analysis of Provable Quantum Benefits.} For theoretical frameworks of provable quantum benefits in QNNs, we present four main directions.

One line of theory links quantum benefits in QNNs to quantum contextuality \cite{anschuetz2023interpretable, anschuetz2026arbitrary, gao2022enhancing}. Contextuality is a foundational feature of quantum mechanics: the outcome of a measurement can depend on which other compatible measurements are performed jointly, and no classical hidden-variable model can reproduce this dependence~\cite{kochen2011problem,spekkens2005contextuality}. \citet{gao2022enhancing} show that a quantum-enhanced hidden Markov model (HMM) with a $D$-dimensional latent space requires a classical HMM with at least $D^{\Omega(\log D)}$ hidden states to simulate, where $\Omega(\cdot)$ denotes an asymptotic lower bound. Equivalently, an $n$-qubit quantum model can require $2^{\Omega(n^2)}$ classical states. The gap arises from quantum contextuality, demonstrated via the Mermin--Peres magic square construction, and is further validated through numerical experiments.

Building on this idea, \citet{anschuetz2023interpretable} introduce contextual recurrent neural networks (CRNNs). They prove unconditionally that CRNNs with $\mathcal{O}(n)$ qumodes can express distributions that require $\Omega(n^2)$-dimensional latent spaces in any reasonable classical sequence model. This gives a quadratic memory gap and the first unconditional expressivity benefit of a quantum neural network over a classical neural network on classical data. Considering trainability barriers, \citet{anschuetz2026arbitrary} propose $k$-hypergraph recurrent neural networks ($k$-HRNNs). On the $(l,n,k)$-hypergraph stabilizer measurement task, a $k$-HRNN with $\mathcal{O}(n)$ qumodes solves the task perfectly, while any classical network satisfying smoothness assumptions requires a latent space of dimension at least $\binom{n}{k}-1$. For constant $k$, this yields an $\Omega(n^k)$-versus-$\mathcal{O}(n)$ memory gap, with the polynomial degree controlled by the choice of~$k$.

Another line of work emphasizes entanglement as a computational resource \cite{zhao2025entanglement}. \citet{zhao2025entanglement} show that a quantum model with $\mathcal{O}(1)$ parameters and $2n$ Bell pairs achieves a perfect score on a magic-square translation task, whereas any communication-bounded classical model requires $\Omega(n)$ parameters to achieve a score above $2^{-o(n)}$. This constant-versus-linear gap is robust to depolarizing noise up to a threshold strength $p^{*}\approx 0.0064$.

A third approach realizes benefits by using QNNs to instantiate quantum algorithms that already admit provable speedups. \citet{du2021learnability} show that any concept class efficiently learnable in the quantum statistical query (QSQ) model~\cite{arunachalam2020quantum} can also be efficiently learned by parameterized quantum circuits. Because QSQ can solve parity learning, which requires exponentially many classical statistical queries, QNNs inherit this exponential benefit over classical SQ learners for such hard concept classes.

Finally, QNNs can be studied with classical learning-theoretic tools. \citet{abbas2021power} introduce the \emph{effective dimension}, a capacity measure derived from the Fisher information matrix~\cite{amari2000methods}, and prove a generalization bound in terms of this quantity. They show empirically that certain QNN architectures achieve higher effective dimension than comparable classical feedforward networks with the same parameter count, and train faster. This framework is largely task-agnostic.

Overall, existing theoretical studies of quantum benefits in QNNs make impressive progress in delivering clean, provable results with clear physical interpretations. They also exhibit a notable characteristic: most results are inherently task-specific, and they establish benefits on carefully constructed problem families often based on contextuality, entanglement-assisted nonlocal correlations, or translation-style formulations.

\textbf{Practical Benchmarking between QNNs and NNs.}  A fair comparison between QNNs and classical NNs requires careful benchmark design. As noted in \cite{bowles2024better}, QML benchmarks can be distorted by unfair baselines, inappropriate evaluation metrics, or metrics that miss quantum-specific gains. The authors therefore emphasize careful experimental design and principled evaluation protocols. To support standardized benchmarking, \cite{garcia2024lazyqml} develop LazyQML, a Python library for systematic comparison of quantum and classical learning models. These methodological contributions provide the basis for rigorous empirical evaluation of QNNs.

The theoretical results above establish quantum benefits for specific, often abstract, task families. Practical advantage is less settled. Current benchmarking studies show a more nuanced picture: quantum models can be competitive in some regimes, but the benefits depend strongly on the task, data structure, hardware assumptions, and classical baseline. We focus on three common workload categories: generative modeling, classification, and time-series prediction.

Generative modeling is one of the most actively benchmarked areas. \cite{hibat2024framework} compare quantum and classical generative models, define evaluation protocols, and identify conditions under which quantum generative models show gains. Building on this framework, \cite{riofrio2023performance} characterize performance across architectures and datasets, revealing scalability patterns and tradeoffs. \cite{gili2023quantum} study the generalization behavior of quantum circuit Born machines and find regimes in which they generalize beyond the training data. \cite{huang2021experimental} demonstrate quantum generative adversarial networks for image generation on quantum hardware, providing an early hardware implementation of quantum GANs. These works show that quantum generative models can be competitive in specific regimes, while scalability and generalization remain open challenges.
Recently, \citet{zhang2024generative} and others~\cite{parigi2025quantum} propose quantum diffusion models for generative learning, with a focus on pure-state ensemble learning. This direction has stimulated studies in machine-learning applications~\cite{de2024quantum, kwun2025mixed, huang2025continuous, zhang2026parameter} and theory~\cite{zhang2025scaling, liu2025measurement}. The possible quantum advantage of quantum diffusion models remains largely unexplored.

For classification tasks, benchmark studies report mixed results. \cite{alvarez2025benchmarking} study quantum-kernel training for classification and find advantages for certain data characteristics, with benefits that remain dataset-dependent. \cite{eslami2025reproducibility} conduct a reproducibility study of quantum machine learning methods in fundus analysis, revealing challenges in reproducing quantum ML results and highlighting the need for standardized evaluation protocols.

For time-series prediction, \cite{fellner2025quantum} compare quantum and classical approaches using variational QML. They find that quantum models can match or exceed classical performance in some scenarios, especially when the time series has favorable structure, although quantum-circuit overhead often limits practical gains. Beyond standard accuracy metrics, \cite{west2023benchmarking} benchmark adversarially robust QML at scale. They find that quantum models can have different robustness properties from classical models, but the advantage is not universal and depends on the model architecture and attack strategy.

Together, these benchmarks show that QNN gains are highly context-dependent. They appear in specific tasks, data regimes, and problem sizes, especially when quantum feature spaces, entanglement, or interference align with the problem structure. Standardized tools such as LazyQML and carefully designed evaluation protocols are therefore essential for fair comparison between quantum and classical approaches.

\textbf{Gaps between Theory and Application.} The theory and benchmarks above point to the same conclusion: QNN performance is highly context-dependent, and a gap remains between provable benefits and real-world applications. Theoretical results often rely on abstract learning settings, idealized assumptions, or restricted models such as SQ/QSQ frameworks. These results identify possible mechanisms for quantum benefit, but they do not directly guarantee gains on common practical workloads. An important open question is how to use the theory to guide the design of efficient and practically useful QNNs.

\begin{table*}
\centering
\caption{Summary of quantum-inspired approaches to classical machine learning.}
\label{tab:quantum-properties}
\resizebox{\textwidth}{!}{%
\begin{tabular}{lll}
\hline
\textbf{Reference} & \textbf{Quantum Property} & \textbf{Model} \\
\hline
\citet{deng2017quantum} & Entanglement (entanglement entropy) & RBM (Restricted Boltzmann Machine) \\
\citet{levine2019quantum} & Quantum entanglement (entanglement measures) & Deep neural networks (various architectures) \\
\citet{halverson2021neural} & Quantum field theory & Wide neural networks (infinite-width limit) \\
\citet{pomarico2025grokking} & Entanglement (entanglement transition) & MPS (Matrix Product State) tensor network \\
\citet{pomarico2025transfer} & Transfer entropy, O-information & MPS (Matrix Product State) tensor network \\
\citet{chen2026artificial} & Artificial entanglement (entanglement entropy) & LLM (Large Language Models) \\
\citet{tull2024towards} & Quantum-inspired categorical frameworks & Neural networks (general architectures) \\
\hline
\end{tabular}%
}
\end{table*}

\subsection{Quantum-Inspired Analysis of Classical Neural Networks}
\label{subsec:qc-arch}

Direct quantum algorithms for machine learning face significant practical challenges~\cite{aaronson2014quantum,liu2021efficient,tang2021quantum,wang2025towards,costa2025further,gan2025provably}. A separate line of work uses quantum concepts to analyze classical machine learning architectures. We focus on how ideas from quantum information theory, especially entanglement, can be used to study classical neural networks. We call this approach \emph{quantum-inspired analysis of classical neural networks}.

\textbf{Quantum Entanglement.} Quantum entanglement is a fundamental concept in quantum information theory that describes non-classical correlations between quantum systems. It explains why a subsystem can have nonzero entropy even when the full quantum system is in a pure state~\cite{eisert2008area}. One standard way to quantify this effect is \emph{entanglement entropy}~\cite{horodecki2009quantum,plenio2005introduction}.

\textit{Entanglement Measures and Quantification.} Entanglement in quantum systems is quantified using measures derived from quantum information theory. For a bipartite quantum system described by a density matrix $\rho_{AB}$, the entanglement entropy is defined as the von Neumann entropy of the reduced density matrix:
\begin{equation}
S(\rho_A) = -\Tr(\rho_A \log \rho_A),
\end{equation}
where $\rho_A = \Tr_B(\rho_{AB})$ is the reduced density matrix obtained by tracing out subsystem $B$. For a bipartite pure state, the Schmidt decomposition gives
\begin{equation}
|\psi\rangle_{AB} = \sum_i \lambda_i |i\rangle_A \otimes |i\rangle_B .
\end{equation}
The Schmidt coefficients $\lambda_i$ determine the entanglement, and the corresponding entanglement entropy is
\begin{equation}
S = -\sum_i \lambda_i^2 \log \lambda_i^2 .
\end{equation}
Higher entanglement entropy indicates stronger non-classical correlation between the two subsystems. For mixed states, additional measures such as logarithmic negativity or entanglement of formation are often used.

\textit{Area Law and Volume Law Scaling.} The scaling behavior of entanglement entropy with subsystem size plays a crucial role in understanding quantum many-body systems. For a subsystem $A$ with linear size $\ell_A$ in spatial dimension $d_{\mathrm{sp}}$, the \emph{area law} states that the entanglement entropy scales as:
\begin{equation}
S(\rho_A) \sim \ell_A^{d_{\mathrm{sp}}-1},
\end{equation}
where $d_{\mathrm{sp}}$ is the spatial dimension. This means that in one dimension ($d_{\mathrm{sp}}=1$), the entanglement entropy is constant (scales as $\ell_A^0$), while in two dimensions ($d_{\mathrm{sp}}=2$), it scales linearly with the boundary size. The area law holds for ground states of gapped local Hamiltonians and reflects the local nature of correlations in such systems. In contrast, the \emph{volume law} describes systems where entanglement entropy scales with the volume of the subsystem:
\begin{equation}
S(\rho_A) \sim |A|,
\end{equation}
where $|A|$ denotes the size of subsystem $A$. Volume-law scaling typically appears in highly entangled states, such as random states or thermal states at finite temperature. Critical systems at quantum phase transitions can exhibit logarithmic violations of the area law. In some systems, transitions between area-law and volume-law behavior signal changes such as many-body-localization transitions or measurement-induced phase transitions. Other quantum phase transitions involve a change from area-law scaling to logarithmic area-law violation.

Early work connects entanglement concepts from quantum information theory to machine learning. \citet{deng2017quantum} interpret the restricted Boltzmann machine (RBM) as a variational ansatz for a quantum many-body wavefunction and ask what entanglement structures this ansatz can represent. They show that short-range RBMs exhibit area-law scaling, while long-range RBMs can exhibit volume-law scaling under certain constructions. In this setting, entanglement measures describe how efficiently the neural network represents a quantum state. \citet{levine2017deep} show an equivalence between the function realized by a deep convolutional arithmetic circuit (ConvAC) and a quantum many-body wavefunction. Building on these works, \citet{levine2019quantum} apply quantum entanglement measures to deep convolutional and recurrent neural networks. Beyond entanglement, \citet{halverson2021neural} connect neural networks with quantum field theory and use field-theoretic tools to analyze training dynamics, generalization, and expressivity, especially in the infinite-width limit.

Recent work applies these ideas to specific learning phenomena in AI. \emph{Grokking}~\cite{power2022grokking} refers to cases where a neural network generalizes only after an extended period of training. \citet{pomarico2025grokking} analyze grokking in tensor-network machine learning and interpret it as an entanglement transition: the model moves from a less entangled phase to a more entangled phase that captures the structure of the problem. Complementing this work, \citet{pomarico2025transfer} use transfer entropy and O-information to detect grokking in tensor-network multi-class classification. These information-theoretic measures track how information flows and is shared across different parts of the network during training.

More recently, \citet{chen2026artificial} introduce \emph{artificial entanglement}, defined as the entanglement entropy of neural-network parameters. They use this measure to study fine-tuning dynamics in large language models and numerically identify area-law and volume-law behavior. They also derive, under suitable limits, an \emph{attention Cardy formula} that connects artificial entanglement entropy to the structure of attention patterns. The same work draws an analogy to the No-Hair Theorem in black hole physics to explain why low-rank updates can achieve strong performance with far fewer trainable parameters.

Beyond entanglement measures, \citet{tull2024towards} develop a compositional approach to explainable AI using quantum-inspired categorical frameworks, especially category theory and diagrammatic reasoning. Their goal is to analyze how neural networks compose simpler operations into complex behavior. We also treat tensor networks as a quantum-inspired framework for analyzing neural-network learning dynamics, and we review this line of work in Section~\ref{subsec:qc-tensor}.

To sum up, we organize these works in Table~\ref{tab:quantum-properties}.
These quantum-inspired perspectives offer new ways to analyze classical AI and form one thread in the broader physics-of-AI literature~\cite{liu2025physicsofairecipe,krenn2022scientific,jiao2024ai}. They also require care. Classical neural networks do not exhibit true quantum entanglement, so analogies between entanglement and classical correlations must be interpreted precisely. It also remains open when quantum-inspired measures lead to practical improvements in neural-network design or training. A useful next step is to connect these measures more directly to observable changes in optimization, generalization, or architecture design.

\subsection{Tensor Network Methods for Machine Learning}
\label{subsec:qc-tensor}

Tensor network methods are quantum-inspired machine-learning tools built from mathematical structures that originated in quantum many-body physics~\cite{white1992density,white1993density,fannes1992finitely,verstraete2004renormalization}. They can be implemented efficiently on classical computers and are useful for representing high-dimensional data with controlled correlation structure. Table~\ref{tab:tensor-network-applications} provides an overview of representative tensor network methods discussed in this section.

\begin{table*}
\centering
\caption{Summary of tensor network methods for machine learning discussed in this section.}
\label{tab:tensor-network-applications}
\resizebox{\textwidth}{!}{%
\begin{tabular}{lll}
\hline
\textbf{Reference} & \textbf{TN Structure} & \textbf{Target Model / Task} \\
\hline
\multicolumn{3}{l}{\emph{Tensorizing Neural Networks for Compression and Efficiency}} \\
\citet{novikov2015tensorizing} & TT (MPS) & FC layers (VGG) \\
\citet{novikov2016exponential} & TT & Exponential machines \\
\citet{anjum2024tensor} & TT (MPS) & LLM fine-tuning (LoRA) \\
\citet{kunwar2025tt} & TT (MPS) & LLM (MoE) \\
\citet{feng2024long} & Tensor factorization & Transformer attention \\
\citet{zhang2025tensor} & Tensor decomposition & Transformer $Q/K/V$ \\
\hline
\multicolumn{3}{l}{\emph{Tensor Networks as Standalone Learning Models}} \\
\citet{stoudenmire2016supervised} & MPS & Supervised classification \\
\citet{sun2020generative} & MPS & Generative classification \\
\citet{cheng2021supervised} & PEPS & Image classification \\
\cite{reyes2020multi,reyes2021multi} & MERA + MPS & Classification and regression \\
\citet{chen2024machine} & TTN & ML with tree structure \\
\citet{nie2025deep} & Deep TTN & Image recognition \\
\citet{selvan2020tensor} & TN & Medical image classification \\
\citet{moore2025using} & MPS & Time-series models \\
\citet{han2018unsupervised} & MPS & Unsupervised generative models \\
\citet{cheng2019tree} & TTN & Generative models \\
\citet{liu2023tensor} & TN & Unsupervised ML \\
\citet{vieijra2022generative} & PEPS & 2D generative models \\
\citet{meiburg2025generative} & TN & Continuous generative models \\
\citet{pozas2024privacy} & TN (MPS) & Privacy-preserving ML \\
\hline
\multicolumn{3}{l}{\emph{Theoretical Equivalence between TNs and NNs}} \\
\cite{cohen2016expressive,cohen2016convolutional} & Hierarchical TN & Deep CNNs \\
\cite{levine2017deep,levine2019quantum} & TN & Deep NNs \\
\citet{chen2026artificial} & MPS & LLM (PEFT) \\
\citet{chen2018equivalence} & TNS & RBM \\
\citet{glasser2020probabilistic} & Generalized TN & Probabilistic graphical models \\
\citet{wu2023tensor} & TN/MPS & RNNs \\
\citet{omranpour2024higher} & Kronecker-factorized TN & Higher-order Transformers \\
\citet{liang2024tensor} & Tensor attention & Transformers \\
\citet{atad2026tensorlens} & High-order attention tensors & Transformer stack \\
\citet{wu2024no} & TN & TN-ML models \\
\hline
\end{tabular}%
}
\end{table*}

\subsubsection{Relation between tensor networks and QC}
Important tensor network structures include matrix product states (MPS)~\cite{fannes1992finitely,klumper1993matrix}, tree tensor networks (TTN)~\cite{shi2006classical}, projected entangled pair states (PEPS)~\cite{verstraete2004renormalization}, Multiscale Entanglement Renormalization Ansatz (MERA)~\cite{vidal2007entanglement}, and Branching MERA~\cite{evenbly2014scaling}. These structures represent high-dimensional tensors compactly by factorizing them into lower-dimensional tensors connected through virtual bonds; see \citet{orus2019tensor} for a detailed introduction.

Their connection to QC comes from a shared mathematical origin. Tensor networks were developed to represent quantum many-body wavefunctions efficiently~\cite{white1992density,white1993density,fannes1992finitely,verstraete2004renormalization}, where entanglement structure controls how large the representation must be. They are now used across QC applications, including simulation of quantum computation, quantum circuit synthesis, quantum error correction and mitigation, and quantum machine learning; see \citet{berezutskii2025tensor} for a detailed introduction. We therefore treat tensor network methods as quantum-inspired approaches.

\subsubsection{Existing Surveys}
Several surveys have examined tensor networks in the context of machine learning.
\citet{wang2023tensor} provide a comprehensive overview of \underline{tensorial neural networks} (combinations of TNs and NNs) and cover how tensor decompositions are integrated into CNNs, RNNs, Transformers, GNNs, and LLMs
for model compression, as well as their use in data fusion and multi-task learning.
\citet{ran2023tensor} focus on tensor networks as quantum-inspired machine learning models,
highlighting their \underline{intrinsic interpretability} grounded in quantum information theory
and their potential for deployment on quantum hardware.
\citet{yan2025tensor} organize the field through an \underline{\emph{AI4Science}} and \underline{\emph{Science4AI}} dual perspective,
reviewing both TN-assisted scientific simulation and TN-based neural network tensorization.
\citet{berezutskii2025tensor} survey tensor networks specifically within \underline{QC and quantum machine learning}. \citet{borsoi2025low}
focuses on low-rank tensor methods for the \underline{theoretical analysis} of neural networks and connects tensor decompositions to neural-network theory. Our review places tensor network methods within the broader study of \underline{quantum information and artificial intelligence}. We organize the discussion along three axes: tensorizing neural networks for compression and efficiency, tensor networks as standalone learning models, and theoretical equivalences between tensor networks and neural networks. We emphasize how quantum information-theoretic concepts, especially entanglement structure and bond-dimension constraints, inform both the design and analysis of machine learning architectures.

\subsubsection{Mathematical Foundations of Tensor Networks}
We first introduce the fundamental concepts, notation, and operations of multilinear algebra and tensor network theory.

\textbf{Tensor Notation.}
We follow the notational conventions of~\cite{cohen2016expressive}. Vectors are denoted by bold lowercase letters, \eg, $\mathbf{v} \in \mathbb{R}^{d}$, with coordinates in regular typeface, \eg, $v_i \in \mathbb{R}$. Matrices are denoted by uppercase letters, \eg, $M \in \mathbb{R}^{d_1 \times d_2}$. Tensors (multi-dimensional arrays) of order $n \geq 3$ are denoted by calligraphic letters, \eg, $\mathcal{T} \in \mathbb{R}^{d_1 \times d_2 \times \cdots \times d_n}$, where $d_i$ is the dimension of the $i$-th mode and specific entries are referenced as $\mathcal{T}_{i_1, i_2, \ldots, i_n}$ with $i_k \in [d_k] \coloneqq \{1, \ldots, d_k\}$. The number of modes $n$ is the \emph{order} of the tensor, generalizing scalars ($n=0$), vectors ($n=1$), and matrices ($n=2$). Superscripts index elements within a collection, \eg, $A^{(k)}$ denotes the $k$-th tensor in a sequence.

\begin{figure*}[t]
\centering
\includegraphics[width=0.92\textwidth]{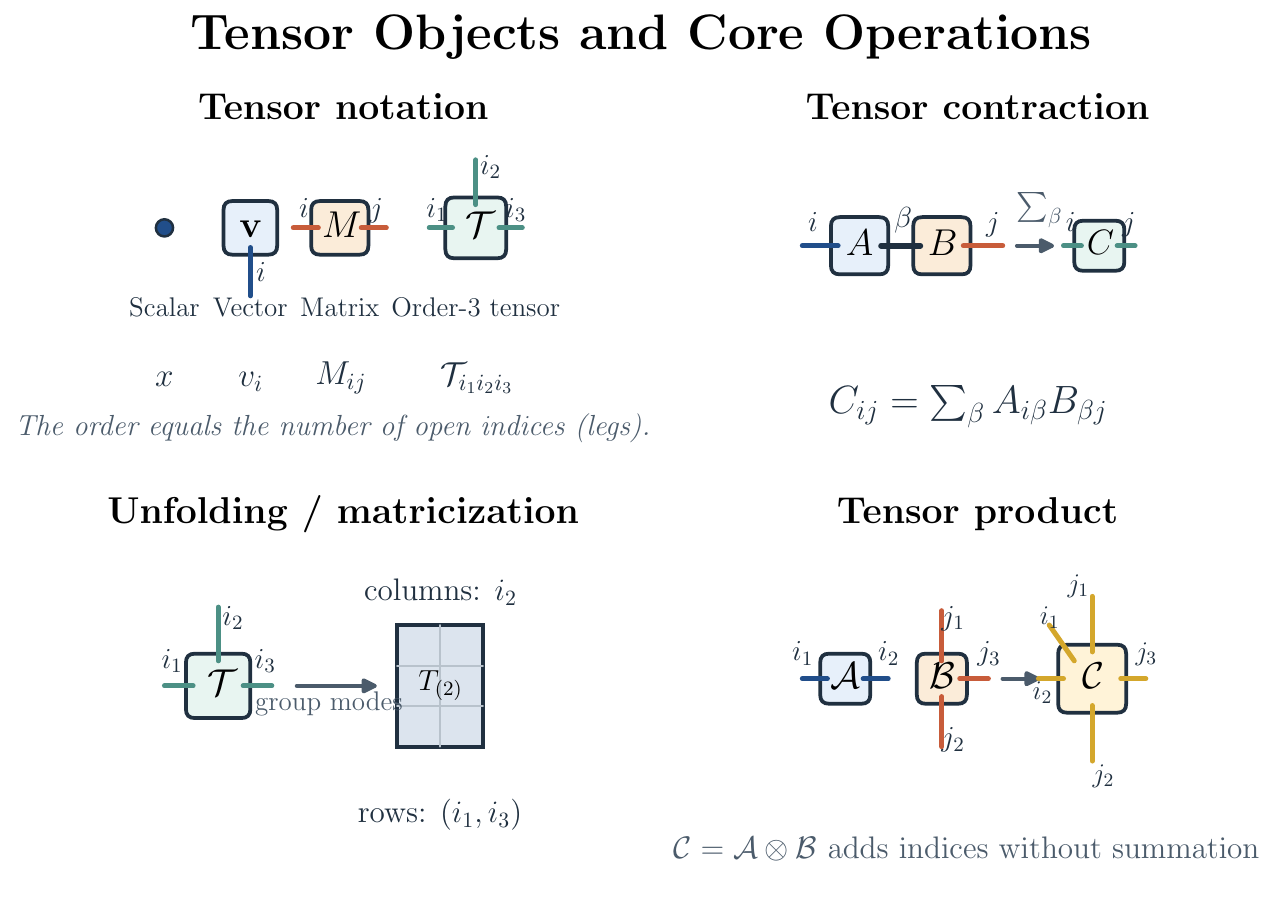}
\caption{Schematic overview of tensor notation and basic multilinear operations. Top left: scalars, vectors, matrices, and higher-order tensors are distinguished by the number of open indices (legs). Top right: tensor contraction sums over a shared index, generalizing matrix multiplication. Bottom left: unfolding or matricization groups tensor modes into a matrix so that standard linear algebra tools such as SVD can be applied. Bottom right: the tensor product combines two tensors by concatenating their index sets without summing over any shared mode.}
\label{fig:tn-basics}
\end{figure*}

\textbf{Tensor Operations.}
Fundamental tensor operations form the computational building blocks of these networks. (\textit{i})~\textbf{Tensor contraction} generalizes matrix multiplication to higher-order tensors by summing over shared indices: given $A_{i\beta}$ and $B_{\beta j}$, their contraction yields $C_{ij} = \sum_{\beta} A_{i\beta} B_{\beta j}$. (\textit{ii})~\textbf{Reshaping} (unfolding/matricization) reorganizes a tensor into a matrix by grouping modes: the mode-$k$ unfolding of $\mathcal{T} \in \mathbb{R}^{d_1 \times \cdots \times d_n}$ creates a matrix $T_{(k)}$ by arranging the mode-$k$ fibers as columns, enabling the application of standard linear algebra techniques such as SVD. (\textit{iii})~The \textbf{tensor product} $\mathcal{C} = \mathcal{A} \otimes \mathcal{B}$ combines two tensors into a higher-order tensor whose order is the sum of the orders of $\mathcal{A}$ and $\mathcal{B}$, without contraction over any shared index.
Figure~\ref{fig:tn-basics} summarizes these basic tensor objects and operations in the graphical language commonly used throughout tensor network theory.

\textbf{SVD and Low-Rank Approximation.}
The Singular Value Decomposition (SVD) is the basic tool behind tensor network compression. Any real matrix $M$ can be decomposed as $M = U \Sigma V^T$, where $U$ and $V$ are orthogonal matrices and $\Sigma$ is a diagonal matrix of non-negative singular values $\sigma_i$. In tensor networks, SVD has two main roles. First, it enables \textbf{truncation}: retaining only the largest $\chi$ singular values gives the optimal rank-$\chi$ approximation of the matrix in Frobenius norm by the Eckart--Young theorem~\cite{eckart1936approximation}. Applied to tensor matricizations at each bond, this truncation reduces the \emph{bond dimension} by discarding weak correlations. Second, SVD produces \textbf{canonical forms}: it orthogonalizes local tensors and transforms the network into left- or right-canonical form, improving numerical stability and making the representation easier to manipulate.

\begin{figure*}[t]
\centering
\includegraphics[width=0.92\textwidth]{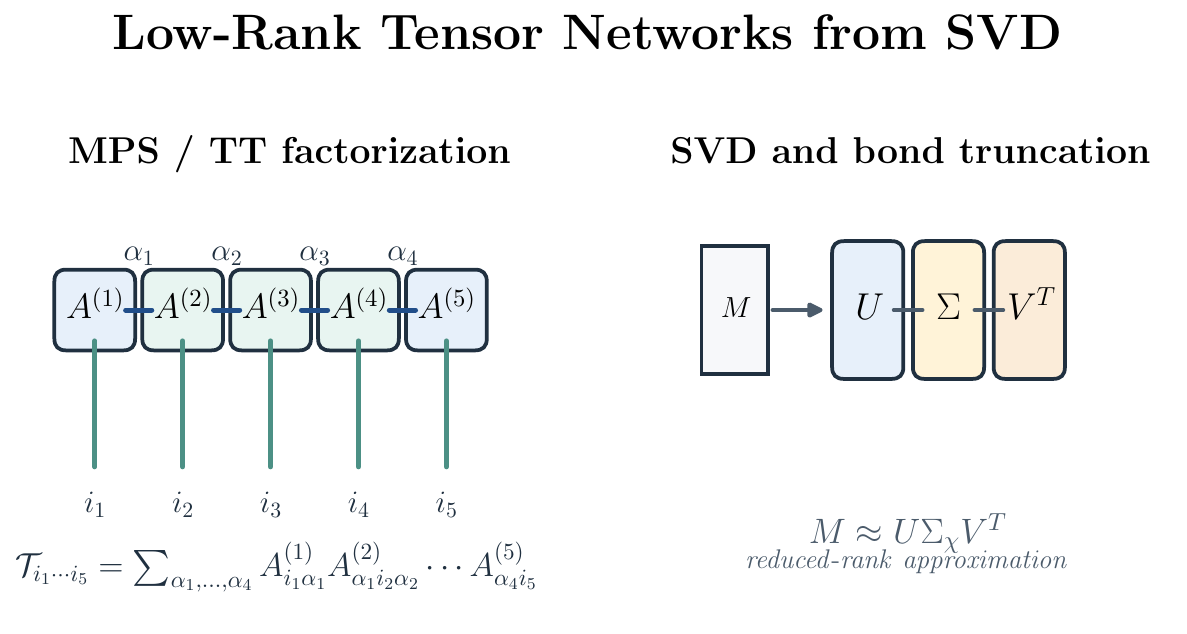}
\caption{Two core ideas behind tensor-network representations. Left: an order-$n$ tensor can be factorized into a chain of low-order cores in an MPS or TT decomposition, where the physical indices remain open and the virtual bonds $\alpha_k$ transmit correlations between neighboring cores. Right: SVD exposes the singular spectrum across a bipartition, and keeping only the largest $\chi$ singular values yields a low-rank approximation that sets the effective bond dimension and compresses the representation.}
\label{fig:mps-svd}
\end{figure*}

\textbf{Tensor Decomposition and Networks.}
A \emph{tensor network} factorizes a high-dimensional tensor into a set of low-order component tensors connected by virtual bonds. The most common structure is MPS, also known as the Tensor-Train (TT) decomposition~\cite{oseledets2011tensor}. An order-$n$ tensor $\mathcal{T}$ is decomposed as:
\begin{equation}
\mathcal{T}_{i_1, i_2, \ldots, i_n} = \sum_{\alpha_1, \dots, \alpha_{n-1}} A^{(1)}_{i_1, \alpha_1} A^{(2)}_{\alpha_1, i_2, \alpha_2} \cdots A^{(n)}_{\alpha_{n-1}, i_n},
\end{equation}
where $A^{(k)}$ are the local core tensors and $\alpha_k$ are the auxiliary \emph{virtual bond indices} of dimension $r_k$ (the bond dimension). The boundary tensors $A^{(1)} \in \mathbb{R}^{d_1 \times r_1}$ and $A^{(n)} \in \mathbb{R}^{r_{n-1} \times d_n}$ are matrices, while the interior cores $A^{(k)} \in \mathbb{R}^{r_{k-1} \times d_k \times r_k}$ ($2 \leq k \leq n-1$) are order-3 tensors. The total number of parameters scales as $O(n d r^2)$, compared to the exponential scaling $O(d^n)$ of the full tensor. The bond dimension $r = \max_k r_k$ controls the expressivity: a larger $r$ captures more correlations (entanglement) between variables, while a smaller $r$ imposes a tighter information bottleneck.
Figure~\ref{fig:mps-svd} illustrates how SVD truncation induces low-rank tensor-network factorizations such as MPS, with the retained singular values directly controlling the bond dimension.

\subsubsection{Novel Applications of Tensor Networks}
Tensor network methods have traditionally been applied to calculate ground- and low-lying excited states~\cite{schollwock2011ann,stoudenmire2012annu,catarina2023epjb}, thermal states~\cite{verstraete2004prl,zwolak2004prl,Stoudenmire2010}, non-equilibrium steady states~\cite{gabriel2021rmp,bertini2021rmp,weimer2021rmp}, and time evolution~\cite{paeckel2019ann}, of low-dimensional many body quantum systems with short-range interactions. Examples include strongly-correlated electron systems in condensed matter~\cite{Jiang2019,qin2020prx,wietek2021prx,tang2023nat,mortier2025scipost,zhehao2025prl}, chemical and molecular systems~\cite{baiardi2020jvp,Werner2026}, and cold atoms in optical lattices~\cite{ronzheimer2013prl,vidmar2015prl,kshetrimayum2019prl,baldelli2024prl}. The general formalism of tensor networks, the different types of networks, their connection to entanglement in many-body quantum systems, and examples of applications are covered in several reviews~\cite{orus2014ann,cirac2021rmp,banuls2023annu,orus2019tensor,Montangero2018,Ran2020tensor,Xiang2024}.

Many recent applications of tensor networks successfully demonstrate the efficient simulation of strongly correlated quantum systems beyond the low-dimensional and short-range interaction limits~\cite{zalatel2015prb}. First, entanglement remains upper-bounded by a constant value in non-critical one-dimensional systems with long-range interactions with locally-bounded Hamiltonians~\cite{Kuwahara2020nat}, such as chains with dipolar interactions~\cite{rosson2020pra,hughes2022pra,aramthottil2023prb}. In addition, systems that feature effective long-range interactions when represented as one-dimensional lattices have been investigated with MPS representations of their states. Examples include impurity models~\cite{wolf2014prb,we_dmft,linden2020prb,erpenbeck2023prb,yu2025prb}, nonequilibrium systems connected to discretized reservoirs~\cite{rams2020prl,brenes2020prx,tamascelli2019prl}, and lattices embedded in optical cavities~\cite{gammelmark2012pra,fabio2020prr,passetti2023prl,baccioni2023scipost,halati2020prl,halati2025prl,halati2025prl2}. Crucially, some of the most remarkable demonstrations of the power of tensor networks have been achieved in simulations of high-dimensional interacting quantum systems. Notable examples are bilayer systems~\cite{soejima2020prb,parker2021prl,faulstich2023prb}, as well as three-dimensional lattices when combined with mean-field approximations~\cite{bollmark2023prx,bollmark2025prb}. Furthermore, efficient simulations of two- and three-dimensional systems incorporating contractions based on belief propagation have challenged the most recent claims of quantum advantage~\cite{joey2023prxq,joey2026science}.

A novel multidisciplinary avenue of tensor network applications, mostly based on MPS structures, corresponds to the simulation of high-dimensional partial differential equations (PDEs). Accurate numerical analysis of multiscale dynamical systems, such as turbulent flows~\cite{pope2000}, suffers from a complexity similar to that of exact diagonalization of many-body quantum systems. Namely, direct numerical simulation (DNS) of turbulence relies on resolving all relevant length and time scales of the dynamics, leading to an exponential complexity with the Reynolds number (the ratio of inertial and viscous forces). This complexity motivates the application of tensor network structures to compress the high-order tensors that encode the transport variables of turbulent flows. In particular, representing such variables as MPS~\cite{Lubasch2018} (also known as quantics MPS or tensor trains~\cite{oseledets2011tensor,Oseledets2012,Osel2013,oseledets2021front,lindsey2024multiscale,bohun2026}), and performing operations such as derivatives, element-wise multiplication~\cite{Michailidis_2024,Meng_2026,ritter2026}, and Fourier transform within the MPS manifold~\cite{dolgov2012fourier,Ripoll2021quantum,chen2023prxq,niedermeier2024prr}, the nonlinear governing PDEs such as the Navier-Stokes equations can be efficiently solved. This \textit{quantum-inspired} application was first demonstrated for incompressible turbulent flows in Ref.~\cite{Gourianov_NatComputSci2022}, where a 2D temporally-developing jet and a 3D Taylor-Green vortex were accurately simulated while requiring one order of magnitude fewer parameters compared to DNS. Such a significant compression was enabled by the notion of locality among length scales resulting from the energy cascade mechanism in turbulent flows. Further applications of MPS for PDEs and continuous functions include simulations of various types of classical fluid dynamics systems~\cite{Gourianov_SciA2025,kiffner2023tensor,kornev2023numerical,holscher2024,peddinti2024commun,truong2024jcp,bobby2025,vanhulst2026cpc,adak2024mathematics,adak2025jsci,gross2026,vanhulst2026}, quantum turbulence~\cite{Connor2026TensorNetworkGPE,felipe2025}, plasma physics~\cite{ye2022pre,ye2024quantized,ye2026}, phonon transport in crystals~\cite{sangyeop2026}, wave equations~\cite{lively2025}, and calculation of Green's functions in quantum field theory~\cite{Shinaoka_PRX2023,takahashi2025scipost}. Efforts in these directions have also been made using tree tensor networks, also referred to as Tucker tensors~\cite{Tindall_2024,Ghahremani_CMAME2024}. Whether other types of network geometries enable an efficient simulation of high-dimensional multiscale PDEs remains an open question.

To take advantage of emerging quantum computing resources, a natural strategy consists of performing classical tensor network simulations first and continuing them on quantum hardware via shallow circuits~\cite{anselme2024pra}. Recent works based on efficient MPS representation of flows at high Reynolds numbers~\cite{siegl2026prr,mello2025,pisoni2026} suggest that such states can be efficiently uploaded to quantum hardware via diverse strategies~\cite{schon2007pra,ran2020pra,rudolph2023qst,lhc2022pra,malz2024prl,smith2024prxq,robertson2025acm,ballarin2025,szodra2026}. Moreover, the recent demonstration that matrix product operators can be well represented as quantum circuits\cite{dolgov2024njp,siegl2026prr} indicates that tensor-network time evolution algorithms can be performed on the uploaded quantum MPS. Similarly to classical simulations~\cite{Michailidis_2024,Meng_2026,ritter2026}, the main bottleneck of the quantum simulation of PDEs is the implementation of nonlinearities. However, dealing with nonlinear terms is much more complex in the quantum computing regime \cite{Succi2023epl,Tennie_NatRP2025}. This complexity comes mainly from the no cloning theorem and the exponential cost of measuring variables encoded in a large number of qubits. A proposal to overcome these limitations, built on MPS encoding, is based on variational algorithms using quantum nonlinear processing units~\cite{lubasch2020pra,dieter2023aiaa,pool2024nonlinear,umer2024nonlinear}. Alternatively, variational quantum time evolution algorithms~\cite{McArdle_npjQI2019, Yuan_Quantum2019, Cerezo_NatRP2021} could be used to solve PDEs~\cite{hirad2025,nguyen2024solving} based on an MPS Ansatz~\cite{li2024jcp}.

\subsubsection{Tensorizing Neural Networks for Compression and Efficiency}

A fundamental application of tensor networks in machine learning is neural-network compression. Early work uses tensor-train (TT) decompositions to compress fully connected layers and to model higher-order feature interactions with far fewer parameters \cite{novikov2015tensorizing,novikov2016exponential,simonyan2014very}. The same basic idea later extends to convolutional, recurrent, transformer, and graph models~\cite{wang2018wide,lebedev2014speeding,kim2015compression,garipov2016ultimate,yang2017tensor,pan2019compressing,tjandra2020recurrent,wu2020hybrid,su2020convolutional,tangpanitanon2022explainable,ma2019tensorized,liu2021enabling,li2022hypoformer,zhao4502093tensor,yin2022nimble}.

More recently, TT-style factorizations move into LLMs, where they are used for both model compression and parameter-efficient fine-tuning (PEFT)~\cite{xu2023tensorgpt,chekalina2023efficient}. Representative examples include TT-LoRA, which compresses LoRA-style updates~\cite{anjum2024tensor}, and TT-LoRA MoE, which combines TT structure with mixture-of-experts design~\cite{kunwar2025tt}. This direction now sits within a broader family of tensor-network-inspired PEFT methods~\cite{chen2024superlora,koike2025quantum,he2024lora,si2024flora,bershatsky2024lotr,xu2024geometry,chen2024quanta,shibo2023fact,hu2025dota,yang2024loretta,hounie2024lorta}.

Tensor decompositions also help at inference time, for example through KV-cache compression~\cite{liu2024unlocking} and production-oriented TN compression pipelines~\cite{kozyrev2026practical}. A closely related line of work tensorizes attention itself in addition to the weights: recent methods factorize long-context attention, compress the $Q/K/V$ maps, or redesign attention kernels around tensor structure~\cite{feng2024long,zhang2025tensor,xu2026flashattention}. For a more comprehensive survey, see \citet{wang2023tensor}.

The predominance of MPS (or equivalently, TT) in neural-network compression is mainly due to its favorable balance between expressivity and computational tractability. Other topologies, such as tensor rings (TR), PEPS, and TTN, encode different structural biases and may be better matched to some architectures. Their use in large-scale language models nevertheless remains limited. The main obstacles are more expensive contractions, weaker alignment with common network geometries, and the lack of optimization pipelines as mature as those available for TT. Even so, these alternative geometries remain promising for targeted components such as hierarchical feedforward blocks or attention modules with structured correlation patterns.

\subsubsection{Tensor Networks as Standalone Learning Models}
Beyond compression, tensor networks can also serve as standalone learning models. Existing work spans supervised classification, generative modeling, and privacy-preserving learning~\cite{stoudenmire2016supervised,sun2020generative,cheng2021supervised,reyes2020multi,selvan2020tensor,chen2024machine,nie2025deep,cheng2019tree,han2018unsupervised,liu2023tensor,meiburg2025generative,pozas2024privacy}.

\citet{stoudenmire2016supervised} show that MPS can act as effective classifiers, especially when the data have sequential or locally correlated structure. This basic supervised-learning idea now extends in several directions. \citet{sun2020generative} combine generative and discriminative objectives in a tensor-network classifier. \citet{cheng2021supervised} use PEPS to handle two-dimensional spatial correlations, while \citet{reyes2020multi,reyes2021multi,chen2024machine,nie2025deep} develop hierarchical and tree-based architectures that better capture multiscale structure. Domain-specific applications now include medical imaging~\cite{selvan2020tensor} and time-series learning~\cite{moore2025using}, where the network geometry aligns naturally with the data.

Tensor networks also form a useful family of generative models. MPS-based approaches learn high-dimensional distributions efficiently~\cite{han2018unsupervised,liu2023tensor}, TTN variants capture multiscale correlations~\cite{cheng2019tree}, and PEPS-based models extend the same logic to higher-dimensional data~\cite{vieijra2022generative}. More recent work pushes these models to continuous distributions~\cite{meiburg2025generative}. Beyond prediction and generation, tensor factorizations are also used to support privacy-preserving learning frameworks~\cite{pozas2024privacy}.

\textbf{Challenges.} Despite these advances, standalone tensor network models face several persistent
challenges that limit their broader adoption. A fundamental limitation shared across
nearly all tensor network architectures is the \emph{sensitivity to input ordering}:
MPS-based models require imposing a one-dimensional ordering on inherently
multi-dimensional data such as images, and the choice of this ordering can
significantly impact performance~\citep{stoudenmire2016supervised, han2018unsupervised}.
While higher-dimensional tensor network structures such as
PEPS~\citep{cheng2021supervised, vieijra2022generative} and
MERA-based architectures~\citep{reyes2020multi} alleviate this issue by natively accommodating
spatial correlations, they introduce a different bottleneck: their contraction costs
scale exponentially on classical hardware, severely limiting
scalability~\citep{cheng2021supervised}.
The \emph{bond dimension} remains a critical hyperparameter governing the
expressiveness--efficiency tradeoff across all architectures; increasing it improves
model capacity but incurs rapidly growing computational and memory
costs~\citep{stoudenmire2016supervised, moore2025using}.
In generative modeling, early MPS-based Born machines suffer from exponentially
decaying correlations that restrict their ability to model complex, long-range
dependencies~\citep{han2018unsupervised, cheng2019tree}, and the overall
performance of tensor network generative models has historically lagged behind
state-of-the-art neural network approaches, with prior work acknowledging that
existing tensor network models ``only work as a proof of
principle''~\citep{liu2023tensor}.
Although recent developments such as autoregressive MPS~\citep{liu2023tensor},
deep tree tensor networks~\citep{nie2025deep}, and continuous Born
machines~\citep{meiburg2025generative} have substantially narrowed this gap,
\citet{nie2025deep} still note that ``current tensor networks are predominantly
suited for simpler tasks and face limitations in computational efficiency and
expressive power.''
For time-series analysis, the MPS ansatz naturally aligns with sequential data
structures~\citep{moore2025using}, but its applicability to very long or
high-dimensional time series remains constrained by bond dimension scaling.
In privacy-preserving settings, the canonical form-based protection proposed
by~\citet{pozas2024privacy} currently applies rigorously only to MPS architectures,
and extending such guarantees to more general tensor network families remains an
open challenge.
More broadly, tensor network models are inherently \emph{multilinear}, lacking
the nonlinear activation functions that endow deep neural networks with their
expressive power, which may fundamentally limit their capacity for certain
learning tasks, though this same property provides advantages in interpretability
and theoretical analysis.

Collectively, these limitations suggest that the path forward for standalone
tensor network models lies in \emph{exploiting their unique structural
advantages}: exact tractability of probabilistic quantities (\eg, partition
functions and marginals), principled interpretability through entanglement
measures, intrinsic privacy-preserving properties via gauge symmetry, and
natural compatibility with quantum hardware. We argue that future research might
focus on (i)~developing hybrid architectures that combine the interpretability
and structural guarantees of tensor networks with the nonlinear expressiveness
of neural networks, (ii)~designing adaptive tensor network topologies that
automatically discover data-dependent optimal structures, and (iii)~extending the theoretical privacy and generalization
guarantees currently established for MPS to broader tensor network families,
thereby enabling principled deployment in safety-critical and privacy-sensitive
applications.

\subsubsection{Theoretical Equivalence between TNs and NNs}
\label{sec:tn_nn_equivalence}
Recent theory connects TNs and NNs by treating both as parameterizations of high-order multilinear objects.
TNs represent functions or distributions through explicit tensor contraction graphs, where topology and bond dimension impose
structured low-rank constraints. In quantum language, these are also low-entanglement constraints.
Many classical ML models can also be recast as generalized tensor factorizations once their computations are expressed
over suitable feature maps. This gives a direct way to compare architecture, expressivity, and inductive bias
\cite{wang2023tensor,yan2025tensor,borsoi2025low}.

\textbf{Expressivity and depth efficiency via tensor decompositions.}
A seminal line of work establishes that the expressive power of deep architectures can be analyzed using tensor
decomposition theory. In particular, deep convolutional-style computations correspond to \emph{hierarchical} tensor decompositions,
while shallower counterparts correspond to \emph{flatter} decompositions, yielding formal depth-efficiency statements:
for broad families of functions, hierarchical factorizations can achieve compact representations that would require exponentially
many parameters in shallow forms \cite{cohen2016expressive,cohen2016convolutional}.
This ``tensorization'' perspective makes architectural motifs such as locality, pooling, and weight sharing amenable to rigorous rank-based analysis.

\textbf{Entanglement-inspired measures of representational capacity.}
Another influential bridge interprets neural architectures as representations of many-body wavefunctions.
This makes it possible to use entanglement-related quantities to characterize the complexity of functions they can represent.
In this formulation, network design choices determine entanglement capacity, often through graph cuts or effective bond dimensions
in the corresponding TN picture. This gives a principled account of how architectural inductive bias arises
\cite{levine2017deep,levine2019quantum}.
Such ideas have recently resurfaced in the context of parameter-efficient fine-tuning, where effective rank/entanglement measures
help explain why low-dimensional updates can still induce rich task-specific behavior \cite{chen2026artificial}.

\textbf{Constructive correspondences: RBMs, graphical models, and sequence models.}
Beyond expressivity analysis, several works provide \emph{constructive} translations between model classes.
Restricted Boltzmann machines (RBMs) admit explicit equivalences with tensor network states (TNS), clarifying when and how RBMs can
emulate particular TN topologies and how entanglement/rank constraints bound their representational power \cite{chen2018equivalence}.
Probabilistic graphical models can likewise be embedded into generalized TN formalisms (often using copy/replication tensors),
yielding a shared language for inference and supervised learning \cite{glasser2020probabilistic}.
For sequential data, TN representations of quantum states can be mapped to recurrent-style computations, resulting in tensorial RNNs
that inherit TN training intuitions and regularization mechanisms \cite{wu2023tensor}.

\textbf{Transformers through tensor-network methods.}
Recent work has further enriched the theoretical toolkit connecting TNs and Transformers.
\emph{Higher-order Transformers} generalize attention to tensor-structured inputs using Kronecker-factorized interactions,
offering favorable scaling for multi-axis data \cite{omranpour2024higher}.
Learning-theoretic advances show that certain higher-order/tensor attention mechanisms admit provably efficient training,
including near-linear-time gradient computation under suitable assumptions \cite{liang2024tensor}.
End-to-end analysis tools reformulate the Transformer stack via high-order attention tensors to unify attention, MLPs,
normalization, and residual pathways into a single tensor-centric description \cite{atad2026tensorlens}.

\textbf{Limits, assumptions, and when equivalence helps.}
Equivalence should not be interpreted as universality.
Recent no-free-lunch results characterize regimes in which TN-based ML models cannot guarantee generalization without assumptions on
data structure and encoding \cite{wu2024no}, and analyses of classical data distributions highlight cases where strict low-entanglement TN biases
may be insufficient for efficient description \cite{lu2025tensor}.
Taken together, these developments suggest a nuanced conclusion for this survey:
TNs provide a mathematically transparent language for understanding and designing neural architectures, but their benefits depend critically
on matching TN topology and rank constraints to the compositional structure of the target data and task
\cite{wang2023tensor,borsoi2025low}.

\section{Cross-Cutting Challenges and Outlook}
\label{sec:crosscutting}

Recent surveys in this area typically close with the challenges and outlooks that cut across subfields \citep{alexeev2025artificial,klusch2024qai,cerezo2022challenges}. However we identify that \emph{AI for QI} and \emph{QI for AI} share several hard problem structures: \underline{(1)} Both rely on indirect or noisy access to the object of interest; \underline{(2)} Both operate under severe resource constraints; \underline{(3)}  Both still face a large gap between elegant theory and robust large-scale deployment. Concretely, three cross-cutting challenges appear as detailed below:

\textbf{Benchmarking and Baselines.}
A first challenge is evaluation. In \emph{AI for QI}, many methods are validated on simulators, simplified noise models, or narrowly defined hardware tasks. This makes it difficult to compare performance across platforms, code distances, control settings, sensing tasks, networking protocols, and experimental pipelines. In \emph{QI for AI}, performance claims are often benchmarked against weak classical baselines, incomplete end-to-end resource accounting, or problem instances that already encode favorable structure. Reported improvements can therefore reflect a mismatch in evaluation protocol and may obscure whether a genuine algorithmic or scientific advantage has been demonstrated.
Therefore, more standardized datasets, careful ablations, transparent classical baselines, and clearer resource reporting are needed. Just as importantly, the field would benefit from more negative results and stress tests. Failures under noise drift, distribution shift, limited training data, or larger system size are scientifically valuable and should become part of the evidence base.

\textbf{Scalability, Data, and Hardware Realism.}
A second challenge is the transition from proof-of-principle settings to realistic scales. Across both directions, many appealing results are still established on small systems, synthetic datasets, idealized noise models, or asymptotic regimes far from current hardware. In \emph{AI for QI}, the main bottlenecks are larger training requirements, online adaptation to drift, and real-time deployment limits set by classical feedback latency and experimental overhead. In \emph{QI for AI}, the bottlenecks are data loading, fault-tolerance overhead, trainability, and the question of whether asymptotic speedups survive realistic constants and error rates. Bridging this gap will require tighter interaction between theory, simulation, and experiment. We expect increasing importance of methods fine-tuned on experimental QPU data, benchmark suites derived from real hardware traces, and analyses that report both asymptotic complexity and the regime in which a proposed advantage could plausibly appear. Future surveys will be most useful when they clearly separate mathematically interesting results, near-term deployable techniques, and genuinely fault-tolerant long-term opportunities.

\textbf{Toward Co-Designed Quantum--AI Systems.}
The third challenge is co-designing quantum--AI system. The strongest results in this area emerge when the learning model, the physical constraints, the optimization loop, and the computational architecture are designed together. Examples already appear throughout this survey: geometry-aware training for variational circuits, adaptive decoders tuned to hardware noise, AI-assisted experimental control, AI-designed sensing and networking protocols, tensor-network methods that exploit problem structure, and quantum-inspired models that feed back into classical architecture design.

Looking ahead, we expect the most productive direction to be hybrid and hierarchical. In the near term, AI will likely continue to deliver practical gains in calibration, control, error mitigation, decoding, compilation, sensing, networking, and autonomous experimentation. In parallel, quantum computing may contribute more selectively to AI through specialized subroutines, new representational principles, and sharper theoretical insights into learning and expressivity. Over a longer horizon, the field may evolve toward AI-native quantum stacks in which learning algorithms, compilers, controllers, simulators, sensors, networked devices, and hardware are coupled in closed loop. At the same time, responsible progress will require restraint in claims of quantum speedup and greater clarity about uncertainty, deployment assumptions, and energy or infrastructure costs.

\section{Conclusion}
\label{sec:conclusion}

This survey reviews the interface between AI and QI from both directions. On one side, AI is increasingly central to the practical development of QI, from learning and characterizing quantum systems to optimizing variational algorithms, designing sensing and networking protocols, stabilizing noisy hardware, and automating laboratory and programming workflows. On the other side, quantum computing and quantum-inspired methods offer new algorithmic primitives, representational tools, and theoretical perspectives for machine learning and artificial intelligence.

Our main message is that these two directions are best understood together. They are linked by recurring questions about what can be learned from limited data, how high-dimensional models can be represented efficiently, which optimization signals remain usable at scale, how robustness is maintained under noise and finite resources, and when a claimed advantage is scientifically meaningful. In that sense, the interface between AI and QI is an emerging common language for learning, representation, and control under quantum constraints. We expect future progress to require rigorous benchmarking, hardware-aware modeling, and the co-design of hybrid quantum--classical systems.

\section*{Acknowledgement}
MC, YG, XJ, YL, JW, ZW, and JL are supported in part by the University of Pittsburgh, School of Computing and Information, Department of Computer Science, Pitt Cyber, Pitt Momentum fund, PQI Community Collaboration Awards, John C. Mascaro Faculty Scholar in Sustainability, Switzerland NSF award 2000-1-243053, NSF award 2535915, Thinking Machines Lab and Cisco Research. This research used resources of the Oak Ridge Leadership Computing Facility, which is a DOE Office of Science User Facility supported under Contract DE-AC05-00OR22725.
BZ and QZ acknowledges support from NSF (CCF-2240641, 2350153, OMA-2326746), AFOSR MURI FA9550-24-1-0349, ONR MURI N000142612102 and DOE ARPA-E Grant No. DE-AR0002067.
YLiu acknowledges support from DOE Advanced Scientific Computing Research under contract number DE-SC0025384, and NSF OSI-2531350 (with a sub-contract from Duke University). XZ acknowledges the support from the AWS Center for Quantum Computing, Samsung Global Research Outreach program, and Columbia University. KPS thanks the U.S. Department of Energy, Office of Science, Advanced Scientific Computing Research (ASCR) program, for support under Award Number DE-SC0026264, and PQI Community Collaboration Awards.

\bibliography{references,XZ_refs,QZ,kps,yl,RefsJJ}

@techreport{Kennedy1973Receiver,
  title = {A Near-Optimum Receiver for the Binary Coherent State Quantum Channel},
  author = {Kennedy, R. S.},
  institution = {Research Laboratory of Electronics, Massachusetts Institute of Technology},
  type = {Quarterly Progress Report},
  number = {108},
  pages = {219--225},
  year = {1973},
  url = {https://dspace.mit.edu/handle/1721.1/56346}
}

@techreport{Dolinar1973Receiver,
  title = {An Optimum Receiver for the Binary Coherent State Quantum Channel},
  author = {Dolinar, S. J.},
  institution = {Research Laboratory of Electronics, Massachusetts Institute of Technology},
  type = {Quarterly Progress Report},
  number = {111},
  pages = {115--120},
  year = {1973},
  url = {https://dspace.mit.edu/handle/1721.1/56414}
}

@article{Hentschel2010MLMeasurement,
  title = {Machine Learning for Precise Quantum Measurement},
  author = {Hentschel, Alexander and Sanders, Barry C.},
  journal = {Physical Review Letters},
  volume = {104},
  number = {6},
  pages = {063603},
  year = {2010},
  publisher = {American Physical Society},
  doi = {10.1103/PhysRevLett.104.063603},
  url = {https://doi.org/10.1103/PhysRevLett.104.063603}
}

@article{Fiderer2021NNBayesian,
  title = {Neural-Network Heuristics for Adaptive Bayesian Quantum Estimation},
  author = {Fiderer, Lukas J. and Schuff, Jonas and Braun, Daniel},
  journal = {PRX Quantum},
  volume = {2},
  number = {2},
  pages = {020303},
  year = {2021},
  publisher = {American Physical Society},
  doi = {10.1103/PRXQuantum.2.020303},
  url = {https://doi.org/10.1103/PRXQuantum.2.020303}
}

@article{Cimini2023DRLMetrology,
  title = {Deep Reinforcement Learning for Quantum Multiparameter Estimation},
  author = {Cimini, Valeria and Valeri, Mauro and Polino, Emanuele and Piacentini, Simone and Ceccarelli, Francesco and Corrielli, Giacomo and Spagnolo, Nicol{\`o} and Osellame, Roberto and Sciarrino, Fabio},
  journal = {Advanced Photonics},
  volume = {5},
  number = {1},
  pages = {016005},
  year = {2023},
  doi = {10.1117/1.AP.5.1.016005},
  url = {https://doi.org/10.1117/1.AP.5.1.016005}
}

@article{Cook2007ClosedLoop,
  title = {Optical Coherent State Discrimination Using a Closed-Loop Quantum Measurement},
  author = {Cook, R. L. and Martin, P. J. and Geremia, J. M.},
  journal = {Nature},
  volume = {446},
  pages = {774--777},
  year = {2007},
  doi = {10.1038/nature05655},
  url = {https://doi.org/10.1038/nature05655}
}

@article{Chen2012ConditionalNulling,
  title = {Optical Codeword Demodulation with Error Rates Below the Standard Quantum Limit Using a Conditional Nulling Receiver},
  author = {Chen, Jian and Habif, Jonathan L. and Dutton, Zachary and Lazarus, Richard and Guha, Saikat},
  journal = {Nature Photonics},
  volume = {6},
  pages = {374--379},
  year = {2012},
  doi = {10.1038/nphoton.2012.113},
  url = {https://doi.org/10.1038/nphoton.2012.113}
}

@article{Cui2022QREAL,
  title = {Quantum Receiver Enhanced by Adaptive Learning},
  author = {Cui, Chaohan and Horrocks, William and Hao, Shuhong and Guha, Saikat and Peyghambarian, Nasser and Zhuang, Quntao and Zhang, Zheshen},
  journal = {Light: Science \& Applications},
  volume = {11},
  pages = {344},
  year = {2022},
  doi = {10.1038/s41377-022-01039-5},
  url = {https://doi.org/10.1038/s41377-022-01039-5}
}

@article{Becerra2013NatPhoton,
  title = {Experimental Demonstration of a Receiver Beating the Standard Quantum Limit for Multiple Nonorthogonal State Discrimination},
  author = {Becerra, F. E. and Fan, J. and Baumgartner, G. and Goldhar, J. and Kosloski, J. T. and Migdall, A.},
  journal = {Nature Photonics},
  volume = {7},
  number = {2},
  pages = {147--152},
  year = {2013},
  doi = {10.1038/nphoton.2012.316},
  url = {https://doi.org/10.1038/nphoton.2012.316}
}

@article{Becerra2015NatPhoton,
  title = {Photon Number Resolution Enables Quantum Receiver for Realistic Coherent Optical Communications},
  author = {Becerra, F. E. and Fan, J. and Migdall, A.},
  journal = {Nature Photonics},
  volume = {9},
  number = {1},
  pages = {48--53},
  year = {2015},
  doi = {10.1038/nphoton.2014.280},
  url = {https://doi.org/10.1038/nphoton.2014.280}
}

@article{Ferdinand2017NPJQI,
  title = {Multi-State Discrimination Below the Quantum Noise Limit at the Single-Photon Level},
  author = {Ferdinand, A. R. and DiMario, M. T. and Becerra, F. E.},
  journal = {npj Quantum Information},
  volume = {3},
  pages = {43},
  year = {2017},
  doi = {10.1038/s41534-017-0042-2},
  url = {https://doi.org/10.1038/s41534-017-0042-2}
}

@article{DiMario2018PRL,
  title = {Robust Measurement for the Discrimination of Binary Coherent States},
  author = {DiMario, M. T. and Becerra, F. E.},
  journal = {Physical Review Letters},
  volume = {121},
  number = {2},
  pages = {023603},
  year = {2018},
  doi = {10.1103/PhysRevLett.121.023603},
  url = {https://doi.org/10.1103/PhysRevLett.121.023603}
}

@article{Zhuang2019SLAEN,
  title = {Physical-Layer Supervised Learning Assisted by an Entangled Sensor Network},
  author = {Zhuang, Quntao and Zhang, Zheshen},
  journal = {Physical Review X},
  volume = {9},
  number = {4},
  pages = {041023},
  year = {2019},
  publisher = {American Physical Society},
  doi = {10.1103/PhysRevX.9.041023},
  url = {https://doi.org/10.1103/PhysRevX.9.041023}
}

@article{Xia2021SLAEN,
  title = {Quantum-Enhanced Data Classification with a Variational Entangled Sensor Network},
  author = {Xia, Yi and Li, Wei and Zhuang, Quntao and Zhang, Zheshen},
  journal = {Physical Review X},
  volume = {11},
  number = {2},
  pages = {021047},
  year = {2021},
  publisher = {American Physical Society},
  doi = {10.1103/PhysRevX.11.021047},
  url = {https://doi.org/10.1103/PhysRevX.11.021047}
}

@article{Liao2024BosonicVSN,
  title = {Quantum-Enhanced Learning with a Controllable Bosonic Variational Sensor Network},
  author = {Liao, Pengcheng and Zhang, Bingzhi and Zhuang, Quntao},
  journal = {Quantum Science and Technology},
  volume = {9},
  number = {4},
  pages = {045040},
  year = {2024},
  doi = {10.1088/2058-9565/ad752d},
  url = {https://doi.org/10.1088/2058-9565/ad752d}
}

@article{Khan2025QCSA,
  title = {Quantum Computational-Sensing Advantage},
  author = {Khan, Saeed A. and Prabhu, Sridhar and Wright, Logan G. and McMahon, Peter L.},
  journal = {arXiv preprint arXiv:2507.16918},
  year = {2025},
  eprint = {2507.16918},
  archivePrefix = {arXiv},
  primaryClass = {quant-ph},
  url = {https://arxiv.org/abs/2507.16918}
}

@article{Prabhu2026QCDS,
  title = {Quantum Computational Displacement Sensing},
  author = {Prabhu, Sridhar and Khan, Saeed A. and Song, Xingrui and Ouellet, Mathieu and Yanagimoto, Ryotatsu and Roy, Saswata and Senanian, Alen and Wright, Logan G. and Fatemi, Valla and McMahon, Peter L.},
  journal = {arXiv preprint arXiv:2604.13177},
  year = {2026},
  eprint = {2604.13177},
  archivePrefix = {arXiv},
  primaryClass = {quant-ph},
  url = {https://arxiv.org/abs/2604.13177}
}

@article{Kaubruegger2019SpinSqueezing,
  title = {Variational Spin-Squeezing Algorithms on Programmable Quantum Sensors},
  author = {Kaubruegger, Raphael and Silvi, Pietro and Kokail, Christian and van Bijnen, Rick and Rey, Ana Maria and Ye, Jun and Kaufman, Adam M. and Zoller, Peter},
  journal = {Physical Review Letters},
  volume = {123},
  number = {26},
  pages = {260505},
  year = {2019},
  publisher = {American Physical Society},
  doi = {10.1103/PhysRevLett.123.260505},
  url = {https://doi.org/10.1103/PhysRevLett.123.260505}
}

@article{Marciniak2022OptimalMetrology,
  title = {Optimal Metrology with Programmable Quantum Sensors},
  author = {Marciniak, Christian D. and Feldker, Thomas and Pogorelov, Ivan and Kaubruegger, Raphael and Vasilyev, Denis V. and van Bijnen, Rick and Schindler, Philipp and Zoller, Peter and Blatt, Rainer and Monz, Thomas},
  journal = {Nature},
  volume = {603},
  number = {7902},
  pages = {604--609},
  year = {2022},
  doi = {10.1038/s41586-022-04435-4},
  url = {https://doi.org/10.1038/s41586-022-04435-4}
}

@article{Kaubruegger2023Multiparameter,
  title = {Optimal and Variational Multiparameter Quantum Metrology and Vector-Field Sensing},
  author = {Kaubruegger, Raphael and Shankar, Athreya and Vasilyev, Denis V. and Zoller, Peter},
  journal = {PRX Quantum},
  volume = {4},
  number = {2},
  pages = {020333},
  year = {2023},
  publisher = {American Physical Society},
  doi = {10.1103/PRXQuantum.4.020333},
  url = {https://doi.org/10.1103/PRXQuantum.4.020333}
}

@article{Cong2019QCNN,
  title = {Quantum Convolutional Neural Networks},
  author = {Cong, Iris and Choi, Soonwon and Lukin, Mikhail D.},
  journal = {Nature Physics},
  volume = {15},
  number = {12},
  pages = {1273--1278},
  year = {2019},
  doi = {10.1038/s41567-019-0648-8},
  url = {https://doi.org/10.1038/s41567-019-0648-8}
}

@article{Chen2021UDQNN,
  title = {Universal Discriminative Quantum Neural Networks},
  author = {Chen, Hongxiang and Wossnig, Leonard and Severini, Simone and Neven, Hartmut and Mohseni, Masoud},
  journal = {Quantum Machine Intelligence},
  volume = {3},
  pages = {1},
  year = {2021},
  doi = {10.1007/s42484-020-00025-7},
  url = {https://doi.org/10.1007/s42484-020-00025-7}
}

@article{Patterson2021NoisyQNN,
  title = {Quantum State Discrimination Using Noisy Quantum Neural Networks},
  author = {Patterson, Andrew and Chen, Hongxiang and Wossnig, Leonard and Severini, Simone and Browne, Dan and Rungger, Ivan},
  journal = {Physical Review Research},
  volume = {3},
  number = {1},
  pages = {013063},
  year = {2021},
  publisher = {American Physical Society},
  doi = {10.1103/PhysRevResearch.3.013063},
  url = {https://doi.org/10.1103/PhysRevResearch.3.013063}
}

@article{Zhang2022FastDecay,
  title = {Fast Decay of Classification Error in Variational Quantum Circuits},
  author = {Zhang, Bingzhi and Zhuang, Quntao},
  journal = {Quantum Science and Technology},
  volume = {7},
  number = {3},
  pages = {035017},
  year = {2022},
  doi = {10.1088/2058-9565/ac70f5},
  url = {https://doi.org/10.1088/2058-9565/ac70f5}
}

@article{zhang2022hybrid,
  title={Hybrid entanglement distribution between remote microwave quantum computers empowered by machine learning},
  author={Zhang, Bingzhi and Wu, Jing and Fan, Linran and Zhuang, Quntao},
  journal={Physical Review Applied},
  volume={18},
  number={6},
  pages={064016},
  year={2022},
  publisher={APS}
}

@article{doolittle2023variational,
  title={Variational quantum optimization of nonlocality in noisy quantum networks},
  author={Doolittle, Brian and Bromley, R Thomas and Killoran, Nathan and Chitambar, Eric},
  journal={IEEE Transactions on Quantum Engineering},
  volume={4},
  pages={1--27},
  year={2023},
  publisher={IEEE}
}

@article{krastanov2019optimized,
  title={Optimized entanglement purification},
  author={Krastanov, Stefan and Albert, Victor V and Jiang, Liang},
  journal={Quantum},
  volume={3},
  pages={123},
  year={2019},
  publisher={Verein zur F{\"o}rderung des Open Access Publizierens in den Quantenwissenschaften}
}

@article{zhao2021practical,
  title={Practical distributed quantum information processing with LOCCNet},
  author={Zhao, Xuanqiang and Zhao, Benchi and Wang, Zihe and Song, Zhixin and Wang, Xin},
  journal={npj Quantum Information},
  volume={7},
  number={1},
  pages={159},
  year={2021},
  publisher={Nature Publishing Group UK London}
}

@article{liu2026dynamic,
  title={Dynamic local operations and classical communication for automated entanglement manipulation},
  author={Liu, Xia and Zhao, Jiayi and Zhao, Benchi and Wang, Xin},
  journal={Communications Physics},
  volume={9},
  number={1},
  pages={113},
  year={2026},
  publisher={Nature Publishing Group UK London}
}

@article{liao2026variational,
  title={Variational Quantum Transduction},
  author={Liao, Pengcheng and Shi, Haowei and Zhuang, Quntao},
  journal={arXiv preprint arXiv:2603.03642},
  year={2026}
}

@book{pope2000,
  	title={{Turbulent Flows}},
  	author={Stephen B. Pope},
  	year={2000},
  	publisher={Cambridge University Press},
  	address={Cambridge}
}

@article{schollwock2011ann,
  title = {The density-matrix renormalization group in the age of matrix product states},
  author = {Schollw\"ock, U.},
  journal = {Ann. Phys.},
  volume = {326},
  pages = {96},
  year = {2011},
  doi = {https://doi.org/10.1016/j.aop.2010.09.012},
  URL = {https://www.sciencedirect.com/science/article/abs/pii/S0003491610001752},
}

@article{cirac2021rmp,
  title = {Matrix product states and projected entangled pair states: Concepts, symmetries, theorems},
  author = {Cirac, J. Ignacio and P\'erez-Garc\'{\i}a, David and Schuch, Norbert and Verstraete, Frank},
  journal = {Rev. Mod. Phys.},
  volume = {93},
  pages = {045003},
  year = {2021},
  publisher = {American Physical Society},
  doi = {10.1103/RevModPhys.93.045003},
  url = {https://link.aps.org/doi/10.1103/RevModPhys.93.045003}
}

@article{catarina2023epjb,
  title = {{Density-matrix renormalization group: a pedagogical introduction}},
  author = {Catarina, G. and Murta, B.},
  journal = {Eur. Phys. J. B},
  volume = {96},
  pages = {111},
  year = {2023},
  publisher = {Springer},
  doi = {10.1140/epjb/s10051-023-00575-2},
  url = {https://doi.org/10.1140/epjb/s10051-023-00575-2}
}

@article{banuls2023annu,
   author = "BaÃ±uls, Mari Carmen",
   title = "Tensor Network Algorithms: A Route Map", 
   journal= "Annu. Rev. Condens. Matter Phys.",
   year = "2023",
   volume = "14",
   pages = "173",
   doi = "https://doi.org/10.1146/annurev-conmatphys-040721-022705",
   url = "https://www.annualreviews.org/content/journals/10.1146/annurev-conmatphys-040721-022705",
   publisher = "Annual Reviews",
   type = "Journal Article"
  }

@article{paeckel2019ann,
title = {Time-evolution methods for matrix-product states},
journal = {Annals of Physics},
volume = {411},
pages = {167998},
year = {2019},
doi = {https://doi.org/10.1016/j.aop.2019.167998},
url = {https://www.sciencedirect.com/science/article/pii/S0003491619302532},
author = {Sebastian Paeckel and Thomas K\"ohler and Andreas Swoboda and Salvatore R. Manmana and Ulrich Schollw\"ock and Claudius Hubig},
}

@Article{mortier2025scipost,
	title={{Fermionic tensor network methods}},
	author={Quinten Mortier and Lukas Devos and Lander Burgelman and Bram Vanhecke and Nick Bultinck and Frank Verstraete and Jutho Haegeman and Laurens Vanderstraeten},
	journal={SciPost Phys.},
	volume={18},
	pages={012},
	year={2025},
	publisher={SciPost},
	doi={10.21468/SciPostPhys.18.1.012},
	url={https://scipost.org/10.21468/SciPostPhys.18.1.012},
}

@article{gabriel2021rmp,
  title = {Nonequilibrium boundary-driven quantum systems: Models, methods, and properties},
  author = {Landi, Gabriel T. and Poletti, Dario and Schaller, Gernot},
  journal = {Rev. Mod. Phys.},
  volume = {94},
  pages = {045006},
  numpages = {58},
  year = {2022},
  publisher = {American Physical Society},
  doi = {10.1103/RevModPhys.94.045006},
  url = {https://link.aps.org/doi/10.1103/RevModPhys.94.045006}
}

@article{bertini2021rmp,
  title = {Finite-temperature transport in one-dimensional quantum lattice models},
  author = {Bertini, B. and Heidrich-Meisner, F. and Karrasch, C. and Prosen, T. and Steinigeweg, R. and \ifmmode \check{Z}\else \v{Z}\fi{}nidari\ifmmode \check{c}\else \v{c}\fi{}, M.},
  journal = {Rev. Mod. Phys.},
  volume = {93},
  pages = {025003},
  numpages = {71},
  year = {2021},
  publisher = {American Physical Society},
  doi = {10.1103/RevModPhys.93.025003},
  url = {https://link.aps.org/doi/10.1103/RevModPhys.93.025003}
}

@article{weimer2021rmp,
  title = {Simulation methods for open quantum many-body systems},
  author = {Weimer, Hendrik and Kshetrimayum, Augustine and Or\'us, Rom\'an},
  journal = {Rev. Mod. Phys.},
  volume = {93},
  pages = {015008},
  numpages = {24},
  year = {2021},
  publisher = {American Physical Society},
  doi = {10.1103/RevModPhys.93.015008},
  url = {https://link.aps.org/doi/10.1103/RevModPhys.93.015008}
}

@article{stoudenmire2012annu,
  author    = {E. M. Stoudenmire and Steven R. White},
  title     = {Studying Two-Dimensional Systems with the Density Matrix Renormalization Group},
  journal   = {Annu. Rev. Condens. Matter Phys.},
  volume    = {3},
  number    = {1},
  pages     = {111--128},
  year      = {2012},
  doi       = {10.1146/annurev-conmatphys-020911-125018},
  url       = {https://doi.org/10.1146/annurev-conmatphys-020911-125018}
}

@incollection{Werner2026,
  author    = {Mikl{\'o}s Antal Werner and Andor Menczer and {\"O}rs Legeza},
  title     = {Chapter Five - Tensor Network State Methods and Quantum Information Theory for Strongly Correlated Molecular Systems},
  booktitle = {Hungarian Quantum Chemistry: Part B - Contemporary Research},
  series    = {Adv. Quantum Chem.},
  volume    = {94},
  pages     = {115--151},
  year      = {2026},
  publisher = {Academic Press},
  doi       = {10.1016/bs.aiq.2025.02.001}
}

@article{baiardi2020jvp,
    author = {Baiardi, Alberto and Reiher, Markus},
    title = {The density matrix renormalization group in chemistry and molecular physics: Recent developments and new challenges},
    journal = {J. Chem. Phys.},
    volume = {152},
    number = {4},
    pages = {040903},
    year = {2020},
    issn = {0021-9606},
    doi = {10.1063/1.5129672},
    url = {https://doi.org/10.1063/1.5129672}
}

@article{orus2014ann,
title = {A practical introduction to tensor networks: Matrix product states and projected entangled pair states},
journal = {Ann. Phys.},
volume = {349},
pages = {117-158},
year = {2014},
issn = {0003-4916},
doi = {https://doi.org/10.1016/j.aop.2014.06.013},
url = {https://www.sciencedirect.com/science/article/pii/S0003491614001596},
author = {Rom\'a' Or\'us'},
}

@book{Montangero2018,
  author    = {Simone Montangero},
  title     = {Introduction to Tensor Network Methods: Numerical Simulations of Low-Dimensional Many-Body Quantum Systems},
  series    = {Lecture Notes in Physics},
  volume    = {944},
  year      = {2018},
  publisher = {Springer},
  address   = {Cham},
  doi       = {10.1007/978-3-319-91479-1},
  isbn      = {978-3-319-91478-4}
}

@book{Ran2020tensor,
  author    = {Shi-Ju Ran and Emanuele Tirrito and Cheng Peng and Xi Chen and Luca Tagliacozzo and Gang Su and Maciej Lewenstein},
  title     = {Tensor Network Contractions},
  series    = {Lecture Notes in Physics},
  volume    = {964},
  year      = {2020},
  publisher = {Springer},
  address   = {Cham},
  doi       = {10.1007/978-3-030-34489-4},
  isbn      = {978-3-030-34488-7}
}

@book{Xiang2024,
  author    = {Tao Xiang},
  title     = {Density Matrix and Tensor Network Renormalization},
  year      = {2024},
  publisher = {Cambridge University Press},
  address   = {Cambridge},
  doi       = {10.1017/9781009299994},
  isbn      = {9781009299970}
}

@article{verstraete2004prl,
  title = {Matrix Product Density Operators: Simulation of Finite-Temperature and Dissipative Systems},
  author = {Verstraete, F. and Garc\'{\i}a-Ripoll, J. J. and Cirac, J. I.},
  journal = {Phys. Rev. Lett.},
  volume = {93},
  pages = {207204},
  numpages = {4},
  year = {2004},
  publisher = {American Physical Society},
  doi = {10.1103/PhysRevLett.93.207204},
  url = {https://link.aps.org/doi/10.1103/PhysRevLett.93.207204}
}

@article{zwolak2004prl,
  title = {Mixed-State Dynamics in One-Dimensional Quantum Lattice Systems: A Time-Dependent Superoperator Renormalization Algorithm},
  author = {Zwolak, Michael and Vidal, Guifr\'e},
  journal = {Phys. Rev. Lett.},
  volume = {93},
  issue = {20},
  pages = {207205},
  year = {2004},
  publisher = {American Physical Society},
  doi = {10.1103/PhysRevLett.93.207205},
  url = {https://link.aps.org/doi/10.1103/PhysRevLett.93.207205}
}

@article{Stoudenmire2010,
  author    = {E. M. Stoudenmire and Steven R. White},
  title     = {Minimally Entangled Typical Thermal State Algorithms},
  journal   = {New J. Phys.},
  volume    = {12},
  number    = {5},
  pages     = {055026},
  year      = {2010},
  doi       = {10.1088/1367-2630/12/5/055026},
  url       = {https://doi.org/10.1088/1367-2630/12/5/055026}
}

@article{zalatel2015prb,
  title = {Time-evolving a matrix product state with long-ranged interactions},
  author = {Zaletel, Michael P. and Mong, Roger S. K. and Karrasch, Christoph and Moore, Joel E. and Pollmann, Frank},
  journal = {Phys. Rev. B},
  volume = {91},
  pages = {165112},
  year = {2015},
  publisher = {American Physical Society},
  doi = {10.1103/PhysRevB.91.165112},
  url = {https://link.aps.org/doi/10.1103/PhysRevB.91.165112}
}

@article{Kuwahara2020nat,
   title={Area law of noncritical ground states in 1D long-range interacting systems},
   volume={11},
   ISSN={2041-1723},
   url={http://dx.doi.org/10.1038/s41467-020-18055-x},
   DOI={10.1038/s41467-020-18055-x},
   number={1},
   journal={Nat. Commun.},
   publisher={Springer Science and Business Media LLC},
   author={Kuwahara, Tomotaka and Saito, Keiji},
   year={2020},
}

@misc{Meng_2026,
  title = {Recursive {{Sketched Interpolation}}: {{Efficient Hadamard Products}} of {{Tensor Trains}}},
  author = {Meng, Zhaonan and Khoo, Yuehaw and Li, Jiajia and Stoudenmire, E. Miles},
  year = 2026,
  month = feb,
  number = {arXiv:2602.17974},
  doi = {10.48550/arXiv.2602.17974},
  urldate = {2026-04-30}
}

@misc{Michailidis_2024,
  title = {Tensor {{Train Multiplication}}},
  author = {Michailidis, Alexios A. and Fenton, Christian and Kiffner, Martin},
  year = 2024,
  month = oct,
  number = {arXiv:2410.19747},
  eprint = {2410.19747},
  publisher = {arXiv},
  doi = {10.48550/arXiv.2410.19747},
  urldate = {2024-10-30}
}

@misc{Tindall_2024,
  title = {Compressing Multivariate Functions with Tree Tensor Networks},
  author = {Tindall, Joseph and Stoudenmire, Miles and Levy, Ryan},
  year = 2024,
  month = oct,
  number = {arXiv:2410.03572},
  eprint = {2410.03572},
  publisher = {arXiv},
  urldate = {2024-10-17}
}

@misc{ritter2026,
      title={Fast elementwise operations on tensor trains with alternating cross interpolation}, 
      author={Marc K. Ritter},
      year={2026},
      eprint={2604.00037},
      archivePrefix={arXiv},
      primaryClass={math.NA},
      url={https://arxiv.org/abs/2604.00037}, 
}

@article{Jiang2019,
  author    = {Hong-Chen Jiang and Thomas P. Devereaux},
  title     = {Superconductivity in the Doped Hubbard Model and Its Interplay with Next-Nearest Hopping $t'$},
  journal   = {Science},
  volume    = {365},
  number    = {6460},
  pages     = {1424--1428},
  year      = {2019},
  doi       = {10.1126/science.aal5304}
}

@article{tang2023nat,
  title = {{Traces of electron-phonon coupling in one-dimensional cuprates}},
  author = {Tang, Ta and Moritz, Brian and Peng, Cheng and Shen, Zhi-Xun and Devereaux, Thomas P.},
  journal = {Nat. Commun.},
  volume = {14},
  pages = {3129},
  year = {2023},
  publisher = {Nature},
  doi = {https://www.nature.com/articles/s41467-023-38408-6}
}

@article{ronzheimer2013prl,
  title = {Expansion Dynamics of Interacting Bosons in Homogeneous Lattices in One and Two Dimensions},
  author = {Ronzheimer, J. P. and Schreiber, M. and Braun, S. and Hodgman, S. S. and Langer, S. and McCulloch, I. P. and Heidrich-Meisner, F. and Bloch, I. and Schneider, U.},
  journal = {Phys. Rev. Lett.},
  volume = {110},
  pages = {205301},
  year = {2013},
  publisher = {American Physical Society},
  doi = {10.1103/PhysRevLett.110.205301},
  url = {https://link.aps.org/doi/10.1103/PhysRevLett.110.205301}
}

@article{vidmar2015prl,
  title = {{Dynamical Quasicondensation of Hard-Core Bosons at Finite Momenta}},
  author = {Vidmar, L. and Ronzheimer, J. P. and Schreiber, M. and Braun, S. and Hodgman, S. S. and Langer, S. and Heidrich-Meisner, F. and Bloch, I. and Schneider, U.},
  journal = {Phys. Rev. Lett.},
  volume = {115},
  pages = {175301},
  year = {2015},
  publisher = {American Physical Society},
  doi = {10.1103/PhysRevLett.115.175301},
  url = {https://link.aps.org/doi/10.1103/PhysRevLett.115.175301}
}

@article{kshetrimayum2019prl,
  title = {{Tensor Network Annealing Algorithm for Two-Dimensional Thermal States}},
  author = {Kshetrimayum, A. and Rizzi, M. and Eisert, J. and Or\'us, R.},
  journal = {Phys. Rev. Lett.},
  volume = {122},
  issue = {7},
  pages = {070502},
  numpages = {6},
  year = {2019},
  month = {Feb},
  publisher = {American Physical Society},
  doi = {10.1103/PhysRevLett.122.070502},
  url = {https://link.aps.org/doi/10.1103/PhysRevLett.122.070502}
}

@article{baldelli2024prl,
  title = {{Frustrated Extended Bose-Hubbard Model and Deconfined Quantum Critical Points with Optical Lattices at the Antimagic Wavelength}},
  author = {Baldelli, Niccol\`o and Cabrera, Cesar R. and Juli\`a-Farr\'e, Sergi and Aidelsburger, Monika and Barbiero, Luca},
  journal = {Phys. Rev. Lett.},
  volume = {132},
  pages = {153401},
  year = {2024},
  publisher = {American Physical Society},
  doi = {10.1103/PhysRevLett.132.153401},
  url = {https://link.aps.org/doi/10.1103/PhysRevLett.132.153401}
}

@article{rosson2020pra,
  title = {Characterizing the phase diagram of finite-size dipolar Bose-Hubbard systems},
  author = {Rosson, Paolo and Kiffner, Martin and Mur-Petit, Jordi and Jaksch, Dieter},
  journal = {Phys. Rev. A},
  volume = {101},
  pages = {013616},
  year = {2020},
  publisher = {American Physical Society},
  doi = {10.1103/PhysRevA.101.013616},
  url = {https://link.aps.org/doi/10.1103/PhysRevA.101.013616}
}

@article{hughes2022pra,
  title = {Dipolar Bose-Hubbard model in finite-size real-space cylindrical lattices},
  author = {Hughes, Michael and Jaksch, Dieter},
  journal = {Phys. Rev. A},
  volume = {105},
  pages = {053301},
  year = {2022},
  publisher = {American Physical Society},
  doi = {10.1103/PhysRevA.105.053301},
  url = {https://link.aps.org/doi/10.1103/PhysRevA.105.053301}
}

@article{aramthottil2023prb,
  title = {Role of interaction-induced tunneling in the dynamics of polar lattice bosons},
  author = {Aramthottil, Adith Sai and \L{}acki, Mateusz and Santos, Luis and Zakrzewski, Jakub},
  journal = {Phys. Rev. B},
  volume = {107},
  pages = {104305},
  year = {2023},
  publisher = {American Physical Society},
  doi = {10.1103/PhysRevB.107.104305},
  url = {https://link.aps.org/doi/10.1103/PhysRevB.107.104305}
}

@article{Lubasch2018,
title = {Multigrid renormalization},
journal = {J. Comput. Phys.},
volume = {372},
pages = {587-602},
year = {2018},
author = {Lubasch, M. and Moinier, P. and Jaksch, D.},
issn = {0021-9991},
doi = {https://doi.org/10.1016/j.jcp.2018.06.065},
url = {https://www.sciencedirect.com/science/article/pii/S0021999118304431}
}

@article{Gourianov_NatComputSci2022,
  title = {A Quantum-Inspired Approach to Exploit Turbulence Structures},
  author = {Gourianov, Nikita and Lubasch, Michael and Dolgov, Sergey and {van den Berg}, Quincy Y. and Babaee, Hessam and Givi, Peyman and Kiffner, Martin and Jaksch, Dieter},
  year = 2022,
  journal = {Nat. Comput. Sci.},
  volume = {2},
  pages = {30--37},
  publisher = {Nature Publishing Group},
  doi = {10.1038/s43588-021-00181-1},
  urldate = {2024-03-14},
}

@article{Gourianov_SciA2025,
  title = {Tensor Networks Enable the Calculation of Turbulence Probability Distributions},
  author = {Gourianov, Nikita and Givi, Peyman and Jaksch, Dieter and Pope, Stephen B.},
  year = 2025,
  journal = {Sci. Adv.},
  volume = {11},
  pages = {eads5990},
  publisher = {American Association for the Advancement of Science},
  doi = {10.1126/sciadv.ads5990},
  urldate = {2025-03-25},
}

@article{felipe2025,
  title = {Simulating quantum turbulence with matrix-product states},
  author = {G\'omez-Lozada, Felipe and Perico-Garc\'{\i}a, Nicolas and Gourianov, Nikita and Salman, Hayder and Mendoza-Arenas, Juan Jos\'e},
  journal = {Phys. Rev. Appl.},
  volume = {25},
  pages = {064069},
  year = {2026},
  publisher = {American Physical Society},
  doi = {10.1103/x5xg-whh1},
  url = {https://link.aps.org/doi/10.1103/x5xg-whh1}
}

@article{bobby2025,
      title={Matrix Product State Simulation of Reacting Shear Flows}, 
      author={Robert Pinkston and Nikita Gourianov and Hirad Alipanah and Peyman Givi and Dieter Jaksch and Juan Jos\'e Mendoza-Arenas},
      year={2025},
      journal={arXiv:2512.13661},
      url={https://arxiv.org/abs/2512.13661}, 
      doi={https://doi.org/10.48550/arXiv.2512.13661}
}

@article{sangyeop2026,
  title         = {Solving the {Peierls--Boltzmann} transport equation with matrix product states},
  author        = {Lee, Sangyeop and Alipanah, Hirad and Mendoza-Arenas, Juan Jos{\'e}},
  year          = {2026},
  journal       = {arXiv:2604.06153},
  url           = {https://arxiv.org/abs/2604.06153},
  doi           = {https://doi.org/10.48550/arXiv.2604.06153}
}

@article{Connor2026TensorNetworkGPE,
  author    = {Ryan J. J. Connor and Callum W. Duncan and Andrew J. Daley},
  title     = {{Tensor network methods for the Gross--Pitaevskii equation on fine grids}},
  journal   = {New J. Phys.},
  volume    = {28},
  number    = {2},
  pages     = {023203},
  year      = {2026},
  doi       = {10.1088/1367-2630/ae2b05},
  url       = {https://doi.org/10.1088/1367-2630/ae2b05}
}

@article{ye2022pre,
  title = {{Quantum-inspired method for solving the Vlasov-Poisson equations}},
  author = {Ye, Erika and Loureiro, Nuno F. G.},
  journal = {Phys. Rev. E},
  volume = {106},
  pages = {035208},
  year = {2022},
  publisher = {American Physical Society},
  doi = {10.1103/PhysRevE.106.035208},
  url = {https://link.aps.org/doi/10.1103/PhysRevE.106.035208}
}

@article{ye2024quantized,
      title={{Quantized tensor networks for solving the Vlasov-Maxwell equations}}, 
      author={Erika Ye and Nuno Loureiro},
      year={2024},
      journal={J. Plasma Phys.},
      volume={90},
      pages={805900301},
      doi = {10.1017/S0022377824000503},
      url = {https://doi.org/10.1017/S0022377824000503}
}

@article{Oseledets2012,
author = {Oseledets, I. V. and Dolgov, S. V.},
title = {Solution of linear systems and matrix inversion in the {TT}-format},
volume = {34},
year = {2012},
pages = {A2718},
journal = {SIAM J. Sci. Comput.},
doi = {10.1137/110833142},
URL = {https://doi.org/10.1137/110833142}
}

@article{Osel2013,
author={Oseledets, I.},
title={Constructive representation of functions in low-rank tensor formats},
doi = {10.1007/s00365-012-9175-x},
year = {2013},
journal = {Constr. Approx.},
volume = {37},
pages ={1},
database={scopus},
}

@article{oseledets2021front,
  title = {{Solution of the Fokker-Planck Equation by Cross Approximation Method in the Tensor Train Format}},
  author = {Chertkov, A. and Oseledets, I.},
  journal = {Front. Artif. Intell.},
  volume = {4},
  pages = {668215},
  year = {2021},
  doi = {10.3389/frai.2021.668215},
  url = {https://www.frontiersin.org/articles/10.3389/frai.2021.668215/full}
}

@article{kornev2023numerical,
      title={{Numerical solution of the incompressible Navier-Stokes equations for chemical mixers via quantum-inspired Tensor Train Finite Element Method}}, 
      author={Egor Kornev and Sergey Dolgov and Karan Pinto and Markus Pflitsch and Michael Perelshtein and Artem Melnikov},
      year={2023},
      journal = {arXiv:2305.10784v2},
      doi = {https://doi.org/10.48550/arXiv.2305.10784},
      url = {https://arxiv.org/abs/2305.10784}
}

@article{kiffner2023tensor,
  title = {Tensor network reduced order models for wall-bounded flows},
  author = {Kiffner, Martin and Jaksch, Dieter},
  journal = {Phys. Rev. Fluids},
  volume = {8},
  pages = {124101},
  year = {2023},
  publisher = {American Physical Society},
  doi = {10.1103/PhysRevFluids.8.124101},
  url = {https://link.aps.org/doi/10.1103/PhysRevFluids.8.124101}
}

@article{peddinti2024commun,
  title = {Quantum-inspired framework for computational fluid dynamics},
  author = {Peddinti, R.D. and Pisoni, S. and Marini, A. and Lott, P. and Argentieri, H. and Tiunov, E. and Aolita, L.},
  journal = {Commun. Phys.},
  volume = {7},
  pages = {135},
  year = {2024},
  publisher = {Nature},
  doi = {10.1038/s42005-024-01623-8},
  url = {https://doi.org/10.1038/s42005-024-01623-8}
}

@article{holscher2024,
  title = {{Quantum-inspired fluid simulation of two-dimensional turbulence with GPU acceleration}},
  author = {H\"olscher, Leonhard and Rao, Pooja and M\"uller, Lukas and Klepsch, Johannes and Luckow, Andre and Stollenwerk, Tobias and Wilhelm, Frank K.},
  journal = {Phys. Rev. Res.},
  volume = {7},
  pages = {013112},
  year = {2025},
  publisher = {American Physical Society},
  doi = {10.1103/PhysRevResearch.7.013112},
  url = {https://link.aps.org/doi/10.1103/PhysRevResearch.7.013112}
}

@article{truong2024jcp,
title = {{Tensor networks for solving the time-independent Boltzmann neutron transport equation}},
journal = {J. Comput. Phys.},
volume = {507},
pages = {112943},
year = {2024},
doi = {https://doi.org/10.1016/j.jcp.2024.112943},
url = {https://www.sciencedirect.com/science/article/pii/S002199912400192X},
author = {Duc P. Truong and Mario I. Ortega and Ismael Boureima and Gianmarco Manzini and Kim A. Rasmussen and Boian S. Alexandrov},
}

@article{vanhulst2026cpc,
title = {Quantum-inspired tensor-network fractional-step method for incompressible flow in curvilinear coordinates},
journal = {Comput. Phys. Commun.},
volume = {325},
pages = {110169},
year = {2026},
issn = {0010-4655},
doi = {https://doi.org/10.1016/j.cpc.2026.110169},
url = {https://www.sciencedirect.com/science/article/pii/S0010465526001517},
author = {Nis-Luca van H\"ulst and Pia Siegl and Paul Over and Sergio Bengoechea and Tomohiro Hashizume and Mario Guillaume Cecile and Thomas Rung and Dieter Jaksch},
}

@article{Ghahremani_CMAME2024,
  title = {Cross Interpolation for Solving High-Dimensional Dynamical Systems on Low-Rank {{Tucker}} and Tensor Train Manifolds},
  author = {Ghahremani, Behzad and Babaee, Hessam},
  year = 2024,
  month = dec,
  journal = {Comput. Methods Appl. Mech. Eng.},
  volume = {432},
  pages = {117385},
  issn = {0045-7825},
  doi = {10.1016/j.cma.2024.117385},
  urldate = {2024-10-03}
}

@misc{pisoni2026,
      title={Compression, simulation, and synthesis of turbulent flows with tensor trains}, 
      author={Stefano Pisoni and Raghavendra Dheeraj Peddinti and Egor Tiunov and Siddhartha E. Guzman and Leandro Aolita},
      year={2026},
      eprint={2506.05477},
      archivePrefix={arXiv},
      primaryClass={physics.flu-dyn},
      url={https://arxiv.org/abs/2506.05477}, 
}

@article{adak2024mathematics,
  author  = {Adak, Dibyendu and Truong, Duc P. and Manzini, Gianmarco and Rasmussen, Kim {\O}. and Alexandrov, Boian S.},
  title   = {Tensor Network Space-Time Spectral Collocation Method for Time-Dependent Convection-Diffusion-Reaction Equations},
  journal = {Mathematics},
  volume  = {12},
  number  = {19},
  pages   = {2988},
  year    = {2024},
  doi     = {10.3390/math12192988}
}

@article{adak2025jsci,
  author  = {Adak, Dibyendu and Danis, M. Engin and Truong, Duc P. and Rasmussen, Kim {\O}. and Alexandrov, Boian S.},
  title   = {Tensor Network Space-Time Spectral Collocation Method for Solving the Nonlinear Convection Diffusion Equation},
  journal = {J. Sci. Comput.},
  volume  = {103},
  number  = {2},
  pages   = {46},
  year    = {2025},
  doi     = {10.1007/s10915-025-02860-x},
  url     = {https://doi.org/10.1007/s10915-025-02860-x}
}

@INPROCEEDINGS{lively2025,
  author={Lively, Kevin and Pagni, Vittorio and Camacho, Gonzalo},
  booktitle={2025 IEEE International Conference on Quantum Computing and Engineering (QCE)}, 
  title={A Quantum-Inspired Algorithm for Wave Simulation Using Tensor Networks}, 
  year={2025},
  volume={01},
  number={},
  pages={1926-1937},
  doi={10.1109/QCE65121.2025.00210}
}

@misc{gross2026,
      title={Tensor Network Lattice Boltzmann Method for Data-Compressed Fluid Simulations}, 
      author={Lukas Gross and Elie Mounzer and David M. Wawrzyniak and Josef M. Winter and Nikolaus A. Adams},
      year={2026},
      eprint={2512.07615},
      archivePrefix={arXiv},
      primaryClass={physics.flu-dyn},
      url={https://arxiv.org/abs/2512.07615}, 
}

@misc{ye2026,
      title={A practical investigation on time integration in the quantized tensor train format}, 
      author={Erika Ye},
      year={2026},
      eprint={2605.12833},
      archivePrefix={arXiv},
      primaryClass={physics.comp-ph},
      url={https://arxiv.org/abs/2605.12833}, 
}

@misc{vanhulst2026,
      title={Quantum-Inspired Simulation of 2D Turbulent Rayleigh-B\'enard Convection}, 
      author={Nis-Luca van Hülst and Mario Guillaume Cecile and Hai-Yen Van and Tomohiro Hashizume and Eugene de Villiers and Dieter Jaksch},
      year={2026},
      eprint={2604.16179},
      archivePrefix={arXiv},
      primaryClass={physics.flu-dyn},
      url={https://arxiv.org/abs/2604.16179}, 
}

@article{dolgov2012fourier,
title = {Superfast Fourier Transform Using QTT Approximation},
journal = {J. Fourier Anal. Appl.},
volume = {18},
pages = {915},
year = {2012},
doi = {https://doi.org/10.1016/j.jcp.2024.112943},
url = {https://www.sciencedirect.com/science/article/pii/S002199912400192X},
author = {Dolgov, S. and Khoromskij, B. and Savostyanov, D.},
}

@article{Ripoll2021quantum,
  doi = {10.22331/q-2021-04-15-431},
  url = {https://doi.org/10.22331/q-2021-04-15-431},
  title = {Quantum-inspired algorithms for multivariate analysis: from interpolation to partial differential equations},
  author = {Garc{\'{i}}a-Ripoll, J. J.},
  journal = {{Quantum}},
  publisher = {{Verein zur F{\"{o}}rderung des Open Access Publizierens in den Quantenwissenschaften}},
  volume = {5},
  pages = {431},
  year = {2021}
}

@article{chen2023prxq,
  title = {{Quantum Fourier Transform Has Small Entanglement}},
  author = {Chen, Jielun and Stoudenmire, E.M. and White, Steven R.},
  journal = {PRX Quantum},
  volume = {4},
  pages = {040318},
  year = {2023},
  publisher = {American Physical Society},
  doi = {10.1103/PRXQuantum.4.040318},
  url = {https://link.aps.org/doi/10.1103/PRXQuantum.4.040318}
}

@misc{ballarin2025,
      title={Efficient quantum state preparation of multivariate functions using tensor networks}, 
      author={Marco Ballarin and Juan José García-Ripoll and David Hayes and Michael Lubasch},
      year={2025},
      eprint={2511.15674},
      archivePrefix={arXiv},
      primaryClass={quant-ph},
      url={https://arxiv.org/abs/2511.15674}, 
}

@misc{szodra2026,
      title={Scalable Preparation of Matrix Product States with Sequential and Brick Wall Quantum Circuits}, 
      author={Tomasz Szoldra and Rick Mukherjee and Peter Schmelcher},
      year={2026},
      eprint={2602.12042},
      archivePrefix={arXiv},
      primaryClass={quant-ph},
      url={https://arxiv.org/abs/2602.12042}, 
}

@article{lindsey2024multiscale,
      title={Multiscale interpolative construction of quantized tensor trains}, 
      author={Michael Lindsey},
      year={2024},
      journal = {arXiv:2311.12554},
      doi = {https://doi.org/10.48550/arXiv.2311.12554},
      url = {https://arxiv.org/abs/2403.12826v1}
}

@misc{bohun2026,
      title={Entanglement scaling in matrix product state representation of smooth functions and their shallow quantum circuit approximations}, 
      author={Vladyslav Bohun and Illia Lukin and Mykola Luhanko and Georgios Korpas and Philippe J. S. De Brouwer and Mykola Maksymenko and Maciej Koch-Janusz},
      year={2026},
      eprint={2412.05202},
      archivePrefix={arXiv},
      primaryClass={quant-ph},
      url={https://arxiv.org/abs/2412.05202}, 
}

@article{niedermeier2024prr,
  title = {{Simulating the quantum Fourier transform, Grover's algorithm, and the quantum counting algorithm with limited entanglement using tensor networks}},
  author = {Niedermeier, Marcel and Lado, Jose L. and Flindt, Christian},
  journal = {Phys. Rev. Res.},
  volume = {6},
  pages = {033325},
  year = {2024},
  publisher = {American Physical Society},
  doi = {10.1103/PhysRevResearch.6.033325},
  url = {https://link.aps.org/doi/10.1103/PhysRevResearch.6.033325}
}

@article{joey2023prxq,
  title = {{Efficient Tensor Network Simulation of IBM's Eagle Kicked Ising Experiment}},
  author = {Tindall, Joseph and Fishman, Matthew and Stoudenmire, E. Miles and Sels, Dries},
  journal = {PRX Quantum},
  volume = {5},
  pages = {010308},
  year = {2024},
  publisher = {American Physical Society},
  doi = {10.1103/PRXQuantum.5.010308},
  url = {https://link.aps.org/doi/10.1103/PRXQuantum.5.010308}
}

@article{joey2026science,
author = {Joseph Tindall  and Antonio Francesco Mello  and Matthew Fishman  and E. Miles Stoudenmire  and Dries Sels },
title = {Dynamics of disordered quantum systems with two- and three-dimensional tensor networks},
journal = {Science},
volume = {392},
number = {6800},
pages = {868-872},
year = {2026},
doi = {10.1126/science.adx2728},
URL = {https://www.science.org/doi/abs/10.1126/science.adx2728}
}

@article{lubasch2020pra,
author = {Lubasch, M. and Joo, J. and Moinier, P. and Kiffner, M. and Jaksch, D.},
title = {Variational quantum algorithms for nonlinear problems},
volume = {101},
year = {2020},
pages = {010301(R)},
journal = {Phys. Rev. A},
doi={https://doi.org/10.1103/PhysRevA.101.010301},
url={https://journals.aps.org/pra/abstract/10.1103/PhysRevA.101.010301}
}

@article{umer2024nonlinear,
  title = {Probing the limits of variational quantum algorithms for nonlinear ground states on real quantum hardware: The effects of noise},
  author = {Umer, Muhammad and Mastorakis, Eleftherios and Evangelou, Sofia and Angelakis, Dimitris G.},
  journal = {Phys. Rev. A},
  volume = {111},
  pages = {012626},
  year = {2025},
  publisher = {American Physical Society},
  doi = {10.1103/PhysRevA.111.012626},
  url = {https://link.aps.org/doi/10.1103/PhysRevA.111.012626}
}

@article{dieter2023aiaa,
author = {Jaksch, Dieter and Givi, Peyman and Daley, Andrew J. and Rung, Thomas},
title = {{Variational Quantum Algorithms for Computational Fluid Dynamics}},
journal = {AIAA Journal},
volume = {61},
pages = {1885},
year = {2023},
doi = {10.2514/1.J062426},
URL = {https://doi.org/10.2514/1.J062426}
}

@article{pool2024nonlinear,
  title = {Nonlinear dynamics as a ground-state solution on quantum computers},
  author = {Pool, Albert J. and Somoza, Alejandro D. and Mc Keever, Conor and Lubasch, Michael and Horstmann, Birger},
  journal = {Phys. Rev. Res.},
  volume = {6},
  pages = {033257},
  year = {2024},
  publisher = {American Physical Society},
  doi = {10.1103/PhysRevResearch.6.033257},
  url = {https://link.aps.org/doi/10.1103/PhysRevResearch.6.033257}
}

@article{dolgov2024njp,
  title = {Tensor quantum programming},
  author = {Termanova, A. and Melnikov, A. and Mamenchikov, E. and Belokonev, N. and Dolgov, S. and Berezutskii, A. and Ellerbrock, R. and Mansell, C. and Perelshtein, M. R.},
  journal = {New J. Phys.},
  volume = {26},
  pages = {123019},
  year = {2024},
  publisher = {IOP},
  doi = {10.1088/1367-2630/ad985b},
  url = {https://iopscience.iop.org/article/10.1088/1367-2630/ad985b}
}

@article{siegl2026prr,
  title = {Tensor-programmable quantum circuits for solving differential equations},
  author = {Siegl, Pia and Reese, Greta Sophie and Hashizume, Tomohiro and van H\"ulst, Nis-Luca and Jaksch, Dieter},
  journal = {Phys. Rev. Res.},
  volume = {8},
  pages = {013052},
  year = {2026},
  publisher = {American Physical Society},
  doi = {10.1103/2qzh-yf49},
  url = {https://link.aps.org/doi/10.1103/2qzh-yf49}
}

@misc{mello2025,
      title={Magic of the Well: assessing quantum resources of fluid dynamics data}, 
      author={Antonio Francesco Mello and Mario Collura and E. Miles Stoudenmire and Ryan Levy},
      year={2025},
      eprint={2512.03177},
      archivePrefix={arXiv},
      primaryClass={quant-ph},
      url={https://arxiv.org/abs/2512.03177}, 
}

@article{Succi2023epl,
    url = {https://dx.doi.org/10.1209/0295-5075/acfdc7},
    year = {2023},
    month = {oct},
    volume = {144},
    number = {1},
    pages = {10001},
    author = {Sauro Succi and W. Itani and K. Sreenivasan and R. Steijl},
    title = {Quantum computing for fluids: Where do we stand?},
    journal = {Europhys. Lett.}
}

@article{Tennie_NatRP2025,
  title = {Quantum Computing for Nonlinear Differential Equations and Turbulence},
  author = {Tennie, Felix and Laizet, Sylvain and Lloyd, Seth and Magri, Luca},
  year = 2025,
  month = apr,
  journal = {Nat. Rev. Phys.},
  volume = {7},
  number = {4},
  pages = {220--230},
  publisher = {Nature Publishing Group},
  issn = {2522-5820},
  doi = {10.1038/s42254-024-00799-w},
  urldate = {2025-08-08}
}

@article{rudolph2023qst,
doi = {10.1088/2058-9565/ad04e6},
url = {https://dx.doi.org/10.1088/2058-9565/ad04e6},
year = {2023},
publisher = {IOP Publishing},
volume = {9},
pages = {015012},
author = {Manuel S. Rudolph and Jing Chen and Jacob Miller and Atithi Acharya and Alejandro Perdomo-Ortiz},
title = {Decomposition of matrix product states into shallow quantum circuits},
journal = {Quantum Sci. Technol.}
}

@article{ran2020pra,
  title = {Encoding of matrix product states into quantum circuits of one- and two-qubit gates},
  author = {Ran, Shi-Ju},
  journal = {Phys. Rev. A},
  volume = {101},
  pages = {032310},
  year = {2020},
  publisher = {American Physical Society},
  doi = {10.1103/PhysRevA.101.032310},
  url = {https://link.aps.org/doi/10.1103/PhysRevA.101.032310}
}

@article{lhc2022pra,
  title = {{Classical versus quantum: Comparing tensor-network-based quantum circuits on Large Hadron Collider data}},
  author = {Araz, Jack Y. and Spannowsky, Michael},
  journal = {Phys. Rev. A},
  volume = {106},
  pages = {062423},
  year = {2022},
  publisher = {American Physical Society},
  doi = {10.1103/PhysRevA.106.062423},
  url = {https://link.aps.org/doi/10.1103/PhysRevA.106.062423}
}

@article{malz2024prl,
  title = {{Preparation of Matrix Product States with Log-Depth Quantum Circuits}},
  author = {Malz, Daniel and Styliaris, Georgios and Wei, Zhi-Yuan and Cirac, J. Ignacio},
  journal = {Phys. Rev. Lett.},
  volume = {132},
  pages = {040404},
  year = {2024},
  publisher = {American Physical Society},
  doi = {10.1103/PhysRevLett.132.040404},
  url = {https://link.aps.org/doi/10.1103/PhysRevLett.132.040404}
}

@article{schon2007pra,
  title = {{Sequential generation of matrix-product states in cavity QED}},
  author = {Sch\"on, C. and Hammerer, K. and Wolf, M. M. and Cirac, J. I. and Solano, E.},
  journal = {Phys. Rev. A},
  volume = {75},
  pages = {032311},
  year = {2007},
  publisher = {American Physical Society},
  doi = {10.1103/PhysRevA.75.032311},
  url = {https://link.aps.org/doi/10.1103/PhysRevA.75.032311}
}

@article{anselme2024pra,
  title = {Combining matrix product states and noisy quantum computers for quantum simulation},
  author = {Anselme Martin, Baptiste and Ayral, Thomas and Jamet, Fran\ifmmode \mbox{\c{c}}\else \c{c}\fi{}ois and Ran\ifmmode \check{c}\else \v{c}\fi{}i\ifmmode \acute{c}\else \'{c}\fi{}, Marko J. and Simon, Pascal},
  journal = {Phys. Rev. A},
  volume = {109},
  pages = {062437},
  year = {2024},
  publisher = {American Physical Society},
  doi = {10.1103/PhysRevA.109.062437},
  url = {https://link.aps.org/doi/10.1103/PhysRevA.109.062437}
}

@article{robertson2025acm,
   title={{Approximate Quantum Compiling for Quantum Simulation: A Tensor Network Based Approach}},
   volume={6},
   ISSN={2643-6817},
   url={http://dx.doi.org/10.1145/3731251},
   DOI={10.1145/3731251},
   number={3},
   journal={ACM Trans. Quantum Comput.},
   publisher={Association for Computing Machinery (ACM)},
   author={Robertson, Niall and Akhriev, Albert and Vala, Jiri and Zhuk, Sergiy},
   year={2025},
   pages={1-15} 
   }

@article{smith2024prxq,
  title = {Constant-{{Depth Preparation}} of {{Matrix Product States}} with {{Adaptive Quantum Circuits}}},
  author = {Smith, Kevin C. and Khan, Abid and Clark, Bryan K. and Girvin, S.M. and Wei, Tzu-Chieh},
  year = 2024,
  month = sep,
  journal = {PRX Quantum},
  volume = {5},
  number = {3},
  pages = {030344},
  publisher = {American Physical Society},
  doi = {10.1103/PRXQuantum.5.030344},
  urldate = {2026-02-18},
}

@article{li2024jcp,
    author = {Li, Hao-En and Li, Xiang and Huang, Jia-Cheng and Zhang, Guang-Ze and Shen, Zhu-Ping and Zhao, Chen and Li, Jun and Hu, Han-Shi},
    title = {{Variational quantum imaginary time evolution for matrix product state Ansatz with tests on transcorrelated Hamiltonians}},
    journal = {J. Chem. Phys.},
    volume = {161},
    number = {14},
    pages = {144104},
    year = {2024},
    issn = {0021-9606},
    doi = {10.1063/5.0228731},
    url = {https://doi.org/10.1063/5.0228731}
}

@article{Shinaoka_PRX2023,
  title = {Multiscale {{Space-Time Ansatz}} for {{Correlation Functions}} of {{Quantum Systems Based}} on {{Quantics Tensor Trains}}},
  author = {Shinaoka, Hiroshi and Wallerberger, Markus and Murakami, Yuta and Nogaki, Kosuke and Sakurai, Rihito and Werner, Philipp and Kauch, Anna},
  year = 2023,
  journal = {Phys. Rev. X},
  volume = {13},
  pages = {021015},
  publisher = {American Physical Society},
  doi = {10.1103/PhysRevX.13.021015},
  urldate = {2025-04-30},
}

@Article{takahashi2025scipost,
	title={{Compactness of quantics tensor train representations of local imaginary-time propagators}},
	author={Haruto Takahashi and Rihito Sakurai and Hiroshi Shinaoka},
	journal={SciPost Phys.},
	volume={18},
	pages={007},
	year={2025},
	publisher={SciPost},
	doi={10.21468/SciPostPhys.18.1.007},
	url={https://scipost.org/10.21468/SciPostPhys.18.1.007},
}

@article{wietek2021prx,
  title = {{Stripes, Antiferromagnetism, and the Pseudogap in the Doped Hubbard Model at Finite Temperature}},
  author = {Wietek, Alexander and He, Yuan-Yao and White, Steven R. and Georges, Antoine and Stoudenmire, E. Miles},
  journal = {Phys. Rev. X},
  volume = {11},
  pages = {031007},
  year = {2021},
  publisher = {American Physical Society},
  doi = {10.1103/PhysRevX.11.031007},
  url = {https://link.aps.org/doi/10.1103/PhysRevX.11.031007}
}

@article{qin2020prx,
  title = {Absence of Superconductivity in the Pure Two-Dimensional Hubbard Model},
  author = {Qin, Mingpu and Chung, Chia-Min and Shi, Hao and Vitali, Ettore and Hubig, Claudius and Schollw\"ock, Ulrich and White, Steven R. and Zhang, Shiwei},
  collaboration = {Simons Collaboration on the Many-Electron Problem},
  journal = {Phys. Rev. X},
  volume = {10},
  pages = {031016},
  year = {2020},
  publisher = {American Physical Society},
  doi = {10.1103/PhysRevX.10.031016},
  url = {https://link.aps.org/doi/10.1103/PhysRevX.10.031016}
}

@article{zhehao2025prl,
  title = {{Fermionic Isometric Tensor Network States in Two Dimensions}},
  author = {Dai, Zhehao and Wu, Yantao and Wang, Taige and Zaletel, Michael P.},
  journal = {Phys. Rev. Lett.},
  volume = {134},
  pages = {026502},
  year = {2025},
  publisher = {American Physical Society},
  doi = {10.1103/PhysRevLett.134.026502},
  url = {https://link.aps.org/doi/10.1103/PhysRevLett.134.026502}
}

@article{bollmark2023prx,
  title = {Solving 2D and 3D Lattice Models of Correlated Fermions---Combining Matrix Product States with Mean-Field Theory},
  author = {Bollmark, Gunnar and K\"ohler, Thomas and Pizzino, Lorenzo and Yang, Yiqi and Hofmann, Johannes S. and Shi, Hao and Zhang, Shiwei and Giamarchi, Thierry and Kantian, Adrian},
  journal = {Phys. Rev. X},
  volume = {13},
  pages = {011039},
  year = {2023},
  publisher = {American Physical Society},
  doi = {10.1103/PhysRevX.13.011039},
  url = {https://link.aps.org/doi/10.1103/PhysRevX.13.011039}
}

@article{bollmark2025prb,
  title = {Resolving competition of charge density wave and superconducting phases using the matrix product state plus mean field algorithm},
  author = {Bollmark, Gunnar and K\"ohler, Thomas and Kantian, Adrian},
  journal = {Phys. Rev. B},
  volume = {111},
  pages = {125141},
  year = {2025},
  publisher = {American Physical Society},
  doi = {10.1103/PhysRevB.111.125141},
  url = {https://link.aps.org/doi/10.1103/PhysRevB.111.125141}
}

@article{soejima2020prb,
  title = {Efficient simulation of moir\'e materials using the density matrix renormalization group},
  author = {Soejima, Tomohiro and Parker, Daniel E. and Bultinck, Nick and Hauschild, Johannes and Zaletel, Michael P.},
  journal = {Phys. Rev. B},
  volume = {102},
  pages = {205111},
  year = {2020},
  publisher = {American Physical Society},
  doi = {10.1103/PhysRevB.102.205111},
  url = {https://link.aps.org/doi/10.1103/PhysRevB.102.205111}
}

@article{faulstich2023prb,
  title = {Interacting models for twisted bilayer graphene: A quantum chemistry approach},
  author = {Faulstich, Fabian M. and Stubbs, Kevin D. and Zhu, Qinyi and Soejima, Tomohiro and Dilip, Rohit and Zhai, Huanchen and Kim, Raehyun and Zaletel, Michael P. and Chan, Garnet Kin-Lic and Lin, Lin},
  journal = {Phys. Rev. B},
  volume = {107},
  pages = {235123},
  year = {2023},
  publisher = {American Physical Society},
  doi = {10.1103/PhysRevB.107.235123},
  url = {https://link.aps.org/doi/10.1103/PhysRevB.107.235123}
}

@article{parker2021prl,
  title = {Strain-Induced Quantum Phase Transitions in Magic-Angle Graphene},
  author = {Parker, Daniel E. and Soejima, Tomohiro and Hauschild, Johannes and Zaletel, Michael P. and Bultinck, Nick},
  journal = {Phys. Rev. Lett.},
  volume = {127},
  issue = {2},
  pages = {027601},
  numpages = {7},
  year = {2021},
  month = {Jul},
  publisher = {American Physical Society},
  doi = {10.1103/PhysRevLett.127.027601},
  url = {https://link.aps.org/doi/10.1103/PhysRevLett.127.027601}
}

@article{we_dmft,
  title = {{Ultra-Fast Control of Magnetic Relaxation in a Periodically Driven Hubbard Model}},
  author = {Mendoza-Arenas, J. J. and G\'omez-Ruiz, F. J. and Eckstein, M. and Jaksch, D. and Clark, S. R.},
  journal = {Ann. Phys. (Berlin)},
  volume = {529},
  pages = {1700024},
  year = {2017},
  doi = {10.1002/andp.201700024},
  publisher = {Wiley}
}

@article{fabio2020prr,
  title = {{R\'enyi entropy singularities as signatures of topological criticality in coupled photon-fermion systems}},
  author = {M\'endez-C\'ordoba, F. P. M. and Mendoza-Arenas, J. J. and G\'omez-Ruiz, F. J. and Rodr\'{\i}guez, F. J. and Tejedor, C. and Quiroga, L.},
  journal = {Phys. Rev. Res.},
  volume = {2},
  pages = {043264},
  year = {2020},
  publisher = {American Physical Society},
  doi = {10.1103/PhysRevResearch.2.043264},
  url = {https://link.aps.org/doi/10.1103/PhysRevResearch.2.043264}
}

@article{brenes2020prx,
  title = {{Tensor-Network Method to Simulate Strongly Interacting Quantum Thermal Machines}},
  author = {Brenes, Marlon and Mendoza-Arenas, Juan Jos\'e and Purkayastha, Archak and Mitchison, Mark T. and Clark, Stephen R. and Goold, John},
  journal = {Phys. Rev. X},
  volume = {10},
  pages = {031040},
  year = {2020},
  publisher = {American Physical Society},
  doi = {10.1103/PhysRevX.10.031040},
  url = {https://link.aps.org/doi/10.1103/PhysRevX.10.031040}
}

@article{wolf2014prb,
  title = {Solving nonequilibrium dynamical mean-field theory using matrix product states},
  author = {Wolf, F. Alexander and McCulloch, Ian P. and Schollw\"ock, Ulrich},
  journal = {Phys. Rev. B},
  volume = {90},
  pages = {235131},
  year = {2014},
  publisher = {American Physical Society},
  doi = {10.1103/PhysRevB.90.235131},
  url = {https://link.aps.org/doi/10.1103/PhysRevB.90.235131}
}

@article{linden2020prb,
  title = {Imaginary-time matrix product state impurity solver in a real material calculation: Spin-orbit coupling in $\mathrm{Sr}{}_{2}\mathrm{RuO}{}_{4}$},
  author = {Linden, Nils-Oliver and Zingl, Manuel and Hubig, Claudius and Parcollet, Olivier and Schollw\"ock, Ulrich},
  journal = {Phys. Rev. B},
  volume = {101},
  pages = {041101},
  year = {2020},
  publisher = {American Physical Society},
  doi = {10.1103/PhysRevB.101.041101},
  url = {https://link.aps.org/doi/10.1103/PhysRevB.101.041101}
}

@article{erpenbeck2023prb,
  title = {Tensor train continuous time solver for quantum impurity models},
  author = {Erpenbeck, A. and Lin, W.-T. and Blommel, T. and Zhang, L. and Iskakov, S. and Bernheimer, L. and N\'u\~nez-Fern\'andez, Y. and Cohen, G. and Parcollet, O. and Waintal, X. and Gull, E.},
  journal = {Phys. Rev. B},
  volume = {107},
  pages = {245135},
  year = {2023},
  publisher = {American Physical Society},
  doi = {10.1103/PhysRevB.107.245135},
  url = {https://link.aps.org/doi/10.1103/PhysRevB.107.245135}
}

@article{yu2025prb,
  title = {Inchworm tensor train hybridization expansion quantum impurity solver},
  author = {Yu, Yang and Erpenbeck, Andr\'e and Zgid, Dominika and Cohen, Guy and Parcollet, Olivier and Gull, Emanuel},
  journal = {Phys. Rev. B},
  volume = {112},
  pages = {085120},
  year = {2025},
  publisher = {American Physical Society},
  doi = {10.1103/yt8p-vr1v},
  url = {https://link.aps.org/doi/10.1103/yt8p-vr1v}
}

@article{rams2020prl,
  title = {Breaking the Entanglement Barrier: Tensor Network Simulation of Quantum Transport},
  author = {Rams, Marek M. and Zwolak, Michael},
  journal = {Phys. Rev. Lett.},
  volume = {124},
  pages = {137701},
  year = {2020},
  publisher = {American Physical Society},
  doi = {10.1103/PhysRevLett.124.137701},
  url = {https://link.aps.org/doi/10.1103/PhysRevLett.124.137701}
}

@article{tamascelli2019prl,
  title = {Efficient Simulation of Finite-Temperature Open Quantum Systems},
  author = {Tamascelli, D. and Smirne, A. and Lim, J. and Huelga, S. F. and Plenio, M. B.},
  journal = {Phys. Rev. Lett.},
  volume = {123},
  pages = {090402},
  year = {2019},
  publisher = {American Physical Society},
  doi = {10.1103/PhysRevLett.123.090402},
  url = {https://link.aps.org/doi/10.1103/PhysRevLett.123.090402}
}

@article{gammelmark2012pra,
  title = {Interacting spins in a cavity: Finite-size effects and symmetry-breaking dynamics},
  author = {Gammelmark, S\o{}ren and M\o{}lmer, Klaus},
  journal = {Phys. Rev. A},
  volume = {85},
  pages = {042114},
  year = {2012},
  publisher = {American Physical Society},
  doi = {10.1103/PhysRevA.85.042114},
  url = {https://link.aps.org/doi/10.1103/PhysRevA.85.042114}
}

@article{halati2020prl,
  title = {{Numerically Exact Treatment of Many-Body Self-Organization in a Cavity}},
  author = {Halati, Catalin-Mihai and Sheikhan, Ameneh and Ritsch, Helmut and Kollath, Corinna},
  journal = {Phys. Rev. Lett.},
  volume = {125},
  pages = {093604},
  year = {2020},
  publisher = {American Physical Society},
  doi = {10.1103/PhysRevLett.125.093604},
  url = {https://link.aps.org/doi/10.1103/PhysRevLett.125.093604}
}

@article{halati2025prl,
  title = {Controlling the Dynamics of Atomic Correlations via the Coupling to a Dissipative Cavity},
  author = {Halati, Catalin-Mihai and Sheikhan, Ameneh and Morigi, Giovanna and Kollath, Corinna},
  journal = {Phys. Rev. Lett.},
  volume = {134},
  pages = {073604},
  year = {2025},
  publisher = {American Physical Society},
  doi = {10.1103/PhysRevLett.134.073604},
  url = {https://link.aps.org/doi/10.1103/PhysRevLett.134.073604}
}

@article{halati2025prl2,
  title = {From Light-Cone to Supersonic Propagation of Correlations by Competing Short- and Long-Range Couplings},
  author = {Halati, Catalin-Mihai and Sheikhan, Ameneh and Morigi, Giovanna and Kollath, Corinna and J\"ager, Simon B.},
  journal = {Phys. Rev. Lett.},
  volume = {135},
  pages = {190402},
  year = {2025},
  publisher = {American Physical Society},
  doi = {10.1103/tt11-vcpr},
  url = {https://link.aps.org/doi/10.1103/tt11-vcpr}
}

@Article{baccioni2023scipost,
	title={{First-order photon condensation in magnetic cavities: A two-leg ladder model}},
	author={Zeno Bacciconi and Gian M. Andolina and Titas Chanda and Giuliano ChiriacÃ² and Marco SchirÃ² and Marcello Dalmonte},
	journal={SciPost Phys.},
	volume={15},
	pages={113},
	year={2023},
	publisher={SciPost},
	doi={10.21468/SciPostPhys.15.3.113},
	url={https://scipost.org/10.21468/SciPostPhys.15.3.113},
}

@article{passetti2023prl,
  title = {{Cavity Light-Matter Entanglement through Quantum Fluctuations}},
  author = {Passetti, Giacomo and Eckhardt, Christian J. and Sentef, Michael A. and Kennes, Dante M.},
  journal = {Phys. Rev. Lett.},
  volume = {131},
  pages = {023601},
  year = {2023},
  publisher = {American Physical Society},
  doi = {10.1103/PhysRevLett.131.023601},
  url = {https://link.aps.org/doi/10.1103/PhysRevLett.131.023601}
}

@article{hirad2025,
  title = {Quantum dynamics simulation of the advection-diffusion equation},
  author = {Alipanah, Hirad and Zhang, Feng and Yao, Yong-Xin and Thompson, Richard and Nguyen, Nam and Liu, Junyu and Givi, Peyman and McDermott, Brian J. and Mendoza-Arenas, Juan Jos\'e},
  journal = {Phys. Rev. Res.},
  volume = {7},
  pages = {043318},
  year = {2025},
  publisher = {American Physical Society},
  doi = {10.1103/ndc3-bdwt},
  url = {https://link.aps.org/doi/10.1103/ndc3-bdwt}
}

@article{nguyen2024solving,
  title = {{Solving Maxwells Equations using Variational Quantum Imaginary Time Evolution}},
  author = {Nam Nguyen and Richard Thompson},
  journal = {arXiv:2402.14156},
  year = {2024},
  publisher = {arXiv},
  url = {https://arxiv.org/abs/2402.14156},
  doi = {https://doi.org/10.48550/arXiv.2402.14156}
}

@article{McArdle_npjQI2019,
  title = {Variational Ansatz-Based Quantum Simulation of Imaginary Time Evolution},
  author = {McArdle, Sam and Jones, Tyson and Endo, Suguru and Li, Ying and Benjamin, Simon C. and Yuan, Xiao},
  year = 2019,
  month = sep,
  journal = {npj Quantum Inf},
  volume = {5},
  number = {1},
  pages = {75},
  publisher = {Nature Publishing Group},
  issn = {2056-6387},
  doi = {10.1038/s41534-019-0187-2},
  urldate = {2026-02-18}
}

@article{Yuan_Quantum2019,
  title = {Theory of Variational Quantum Simulation},
  author = {Yuan, Xiao and Endo, Suguru and Zhao, Qi and Li, Ying and Benjamin, Simon C.},
  year = 2019,
  month = oct,
  journal = {Quantum},
  volume = {3},
  pages = {191},
  publisher = {Verein zur F\"orderung des Open Access Publizierens in den Quantenwissenschaften},
  doi = {10.22331/q-2019-10-07-191},
  urldate = {2026-02-18}
}

@article{Cerezo_NatRP2021,
  title = {Variational Quantum Algorithms},
  author = {Cerezo, M. and Arrasmith, Andrew and Babbush, Ryan and Benjamin, Simon C. and Endo, Suguru and Fujii, Keisuke and McClean, Jarrod R. and Mitarai, Kosuke and Yuan, Xiao and Cincio, Lukasz and Coles, Patrick J.},
  year = 2021,
  month = sep,
  journal = {Nat. Rev. Phys.},
  volume = {3},
  number = {9},
  pages = {625--644},
  publisher = {Nature Publishing Group},
  issn = {2522-5820},
  doi = {10.1038/s42254-021-00348-9},
  urldate = {2026-02-18}
}

@article{Krenn_2020,
  title   = {Computer-inspired quantum experiments},
  author  = {Krenn, Mario and Erhard, Manuel and Zeilinger, Anton},
  journal = {Nature Reviews Physics},
  volume  = {2},
  number  = {11},
  pages   = {649--661},
  year    = {2020},
  month   = sep,
  doi     = {10.1038/s42254-020-0230-4},
  url     = {http://dx.doi.org/10.1038/s42254-020-0230-4},
  issn    = {2522-5820},
  publisher = {Springer Science and Business Media LLC}
}

@article{Krenn_2023,
  title   = {Artificial intelligence and machine learning for quantum technologies},
  author  = {Krenn, Mario and Landgraf, Jonas and Foesel, Thomas and Marquardt, Florian},
  journal = {Physical Review A},
  volume  = {107},
  number  = {1},
  pages   = {010101},
  year    = {2023},
  month   = jan,
  doi     = {10.1103/PhysRevA.107.010101},
  url     = {http://dx.doi.org/10.1103/PhysRevA.107.010101},
  issn    = {2469-9934},
  publisher = {American Physical Society (APS)}
}

@article{Tom_2024,
  title   = {Self-Driving Laboratories for Chemistry and Materials Science},
  author  = {Tom, Gary and Schmid, Stefan P. and Baird, Sterling G. and Cao, Yang and Darvish, Kourosh and Hao, Han and Lo, Stanley and Pablo-Garc{\'\i}a, Sergio and Rajaonson, Ella M. and Skreta, Marta and Yoshikawa, Naruki and Corapi, Samantha and Akkoc, Gun Deniz and Strieth-Kalthoff, Felix and Seifrid, Martin and Aspuru-Guzik, Al{\'a}n},
  journal = {Chemical Reviews},
  volume  = {124},
  number  = {16},
  pages   = {9633--9732},
  year    = {2024},
  month   = aug,
  doi     = {10.1021/acs.chemrev.4c00055},
  url     = {http://dx.doi.org/10.1021/acs.chemrev.4c00055},
  issn    = {1520-6890},
  publisher = {American Chemical Society (ACS)}
}

@article{Ma_2025,
  title   = {Machine learning for estimation and control of quantum systems},
  author  = {Ma, Hailan and Qi, Bo and Petersen, Ian R. and Wu, Re-Bing and Rabitz, Herschel and Dong, Daoyi},
  journal = {National Science Review},
  volume  = {12},
  number  = {8},
  year    = {2025},
  month   = jul,
  doi     = {10.1093/nsr/nwaf269},
  url     = {http://dx.doi.org/10.1093/nsr/nwaf269},
  issn    = {2053-714X},
  publisher = {Oxford University Press (OUP)}
}

@article{Malashin_2026,
  title   = {A review of applications of machine learning in quantum dots research},
  author  = {Malashin, Ivan and Martysyuk, Dmitry and Nelyub, Vladimir and Borodulin, Aleksei and Gantimurov, Andrei and Tynchenko, Vadim},
  journal = {Discover Nano},
  volume  = {21},
  number  = {1},
  year    = {2026},
  month   = feb,
  doi     = {10.1186/s11671-026-04466-0},
  url     = {http://dx.doi.org/10.1186/s11671-026-04466-0},
  issn    = {2731-9229},
  publisher = {Springer Science and Business Media LLC}
}

@article{Alexeev_2025,
  title   = {Artificial intelligence for quantum computing},
  author  = {Alexeev, Yuri and Farag, Marwa H. and Patti, Taylor L. and Wolf, Mark E. and Ares, Natalia and Aspuru-Guzik, Al{\'a}n and Benjamin, Simon C. and Cai, Zhenyu and Cao, Shuxiang and Chamberland, Christopher and Chandani, Zohim and Fedele, Federico and Hamamura, Ikko and Harrigan, Nicholas and Kim, Jin-Sung and Kyoseva, Elica and Lietz, Justin G. and Lubowe, Tom and McCaskey, Alexander and Melko, Roger G. and Nakaji, Kouhei and Peruzzo, Alberto and Rao, Pooja and Schmitt, Bruno and Stanwyck, Sam and Tubman, Norm M. and Wang, Hanrui and Costa, Timothy},
  journal = {Nature Communications},
  volume  = {16},
  number  = {1},
  year    = {2025},
  month   = dec,
  doi     = {10.1038/s41467-025-65836-3},
  url     = {http://dx.doi.org/10.1038/s41467-025-65836-3},
  issn    = {2041-1723},
  publisher = {Springer Science and Business Media LLC}
}

@article{Krenn_2016,
  title   = {Automated Search for new Quantum Experiments},
  author  = {Krenn, Mario and Malik, Mehul and Fickler, Robert and Lapkiewicz, Radek and Zeilinger, Anton},
  journal = {Physical Review Letters},
  volume  = {116},
  number  = {9},
  pages   = {090405},
  year    = {2016},
  month   = mar,
  doi     = {10.1103/PhysRevLett.116.090405},
  url     = {http://dx.doi.org/10.1103/PhysRevLett.116.090405},
  issn    = {1079-7114},
  publisher = {American Physical Society (APS)}
}

@article{Melnikov_2018,
  title   = {Active learning machine learns to create new quantum experiments},
  author  = {Melnikov, Alexey A. and Poulsen Nautrup, Hendrik and Krenn, Mario and Dunjko, Vedran and Tiersch, Markus and Zeilinger, Anton and Briegel, Hans J.},
  journal = {Proceedings of the National Academy of Sciences},
  volume  = {115},
  number  = {6},
  pages   = {1221--1226},
  year    = {2018},
  month   = jan,
  doi     = {10.1073/pnas.1714936115},
  url     = {http://dx.doi.org/10.1073/pnas.1714936115},
  issn    = {1091-6490},
  publisher = {Proceedings of the National Academy of Sciences}
}

@article{Krenn_2021,
  title   = {Conceptual Understanding through Efficient Automated Design of Quantum Optical Experiments},
  author  = {Krenn, Mario and Kottmann, Jakob S. and Tischler, Nora and Aspuru-Guzik, Al{\'a}n},
  journal = {Physical Review X},
  volume  = {11},
  number  = {3},
  pages   = {031044},
  year    = {2021},
  month   = aug,
  doi     = {10.1103/PhysRevX.11.031044},
  url     = {http://dx.doi.org/10.1103/PhysRevX.11.031044},
  issn    = {2160-3308},
  publisher = {American Physical Society (APS)}
}

@article{Cervera_Lierta_2022,
  title   = {Design of quantum optical experiments with logic artificial intelligence},
  author  = {Cervera-Lierta, Alba and Krenn, Mario and Aspuru-Guzik, Al{\'a}n},
  journal = {Quantum},
  volume  = {6},
  pages   = {836},
  year    = {2022},
  month   = oct,
  doi     = {10.22331/q-2022-10-13-836},
  url     = {http://dx.doi.org/10.22331/q-2022-10-13-836},
  issn    = {2521-327X},
  publisher = {Verein zur F{\"o}rderung des Open Access Publizierens in den Quantenwissenschaften}
}

@article{Ruiz_Gonzalez_2023,
  title   = {Digital Discovery of 100 diverse Quantum Experiments with PyTheus},
  author  = {Ruiz-Gonzalez, Carlos and Arlt, S{\"o}ren and Petermann, Jan and Sayyad, Sharareh and Jaouni, Tareq and Karimi, Ebrahim and Tischler, Nora and Gu, Xuemei and Krenn, Mario},
  journal = {Quantum},
  volume  = {7},
  pages   = {1204},
  year    = {2023},
  month   = dec,
  doi     = {10.22331/q-2023-12-12-1204},
  url     = {http://dx.doi.org/10.22331/q-2023-12-12-1204},
  issn    = {2521-327X},
  publisher = {Verein zur F{\"o}rderung des Open Access Publizierens in den Quantenwissenschaften}
}

@article{Arlt_2026,
  title   = {Meta-designing quantum experiments with language models},
  author  = {Arlt, S{\"o}ren and Duan, Haonan and Li, Felix and Xie, Sang Michael and Wu, Yuhuai and Krenn, Mario},
  journal = {Nature Machine Intelligence},
  volume  = {8},
  number  = {2},
  pages   = {148--157},
  year    = {2026},
  month   = feb,
  doi     = {10.1038/s42256-025-01153-0},
  url     = {http://dx.doi.org/10.1038/s42256-025-01153-0},
  issn    = {2522-5839},
  publisher = {Springer Science and Business Media LLC}
}

@article{Menke_2021,
  title   = {Automated design of superconducting circuits and its application to 4-local couplers},
  author  = {Menke, Tim and H{\"a}se, Florian and Gustavsson, Simon and Kerman, Andrew J. and Oliver, William D. and Aspuru-Guzik, Al{\'a}n},
  journal = {npj Quantum Information},
  volume  = {7},
  number  = {1},
  year    = {2021},
  month   = mar,
  doi     = {10.1038/s41534-021-00382-6},
  url     = {http://dx.doi.org/10.1038/s41534-021-00382-6},
  issn    = {2056-6387},
  publisher = {Springer Science and Business Media LLC}
}

@article{Menke_2022,
  title   = {Demonstration of Tunable Three-Body Interactions between Superconducting Qubits},
  author  = {Menke, Tim and Banner, William P. and Bergamaschi, Thomas R. and Di Paolo, Agustin and Veps{\"a}l{\"a}inen, Antti and Weber, Steven J. and Winik, Roni and Melville, Alexander and Niedzielski, Bethany M. and Rosenberg, Danna and Serniak, Kyle and Schwartz, Mollie E. and Yoder, Jonilyn L. and Aspuru-Guzik, Al{\'a}n and Gustavsson, Simon and Grover, Jeffrey A. and Hirjibehedin, Cyrus F. and Kerman, Andrew J. and Oliver, William D.},
  journal = {Physical Review Letters},
  volume  = {129},
  number  = {22},
  pages   = {220501},
  year    = {2022},
  month   = nov,
  doi     = {10.1103/PhysRevLett.129.220501},
  url     = {http://dx.doi.org/10.1103/PhysRevLett.129.220501},
  issn    = {1079-7114},
  publisher = {American Physical Society (APS)}
}

@article{Ni_2022,
  title   = {Integrating quantum processor device and control optimization in a gradient-based framework},
  author  = {Ni, Xiaotong and Zhao, Hui-Hai and Wang, Lei and Wu, Feng and Chen, Jianxin},
  journal = {npj Quantum Information},
  volume  = {8},
  number  = {1},
  year    = {2022},
  month   = sep,
  doi     = {10.1038/s41534-022-00614-3},
  url     = {http://dx.doi.org/10.1038/s41534-022-00614-3},
  issn    = {2056-6387},
  publisher = {Springer Science and Business Media LLC}
}

@misc{Rajabzadeh_2024,
  title         = {A General Framework for Gradient-Based Optimization of Superconducting Quantum Circuits using Qubit Discovery as a Case Study},
  author        = {Rajabzadeh, Taha and Boulton-McKeehan, Alex and Bonkowsky, Sam and Schuster, David I. and Safavi-Naeini, Amir H.},
  year          = {2024},
  eprint        = {2408.12704},
  archivePrefix = {arXiv},
  primaryClass  = {quant-ph},
  url           = {https://arxiv.org/abs/2408.12704}
}

@article{Cardenas_Lopez_2025,
  title   = {Resilient superconducting-element design with genetic algorithms},
  author  = {C{\'a}rdenas-L{\'o}pez, F. A. and Retamal, J. C. and Chen, Xi and Romero, G. and Sanz, M.},
  journal = {Physical Review Applied},
  volume  = {23},
  number  = {5},
  pages   = {054068},
  year    = {2025},
  month   = may,
  doi     = {10.1103/PhysRevApplied.23.054068},
  url     = {http://dx.doi.org/10.1103/PhysRevApplied.23.054068},
  issn    = {2331-7019},
  publisher = {American Physical Society (APS)}
}

@article{Li_2023,
  title   = {Quantum chip design optimization and automation in superconducting coupler architecture},
  author  = {Li, Fei-Yu and Jin, Li-Jing},
  journal = {Quantum Science and Technology},
  volume  = {8},
  number  = {4},
  pages   = {045015},
  year    = {2023},
  month   = aug,
  doi     = {10.1088/2058-9565/ace8b6},
  url     = {http://dx.doi.org/10.1088/2058-9565/ace8b6},
  issn    = {2058-9565},
  publisher = {IOP Publishing}
}

@inproceedings{Nugraha_2023,
  title     = {Machine Learning-Based Predictive Modeling for Designing Transmon Superconducting Qubits},
  author    = {Nugraha, Ferris Prima and Shao, Qiming},
  booktitle = {2023 IEEE International Conference on Quantum Computing and Engineering (QCE)},
  publisher = {IEEE},
  pages     = {1360--1368},
  year      = {2023},
  month     = sep,
  doi       = {10.1109/QCE57702.2023.00154},
  url       = {http://dx.doi.org/10.1109/QCE57702.2023.00154}
}

@article{Shanto_2024,
  title   = {{SQuADDS}: A validated design database and simulation workflow for superconducting qubit design},
  author  = {Shanto, Sadman and Kuo, Andre and Miyamoto, Clark and Zhang, Haimeng and Maurya, Vivek and Vlachos, Evangelos and Hecht, Malida and Shum, Chung Wa and Levenson-Falk, Eli},
  journal = {Quantum},
  volume  = {8},
  pages   = {1465},
  year    = {2024},
  month   = sep,
  doi     = {10.22331/q-2024-09-09-1465},
  url     = {http://dx.doi.org/10.22331/q-2024-09-09-1465},
  issn    = {2521-327X},
  publisher = {Verein zur F{\"o}rderung des Open Access Publizierens in den Quantenwissenschaften}
}

@misc{Eriksson_2025,
  title         = {Automated, physics-guided, multi-parameter design optimization for superconducting quantum devices},
  author        = {Eriksson, Axel M. and Splitthoff, Lukas J. and Upadhyay, Harsh Vardhan and Campana, Pietro and Pittan Narendiran, Niranjan and Helambe, Kunal and Andersson, Linus and Gasparinetti, Simone},
  year          = {2025},
  eprint        = {2508.18027},
  archivePrefix = {arXiv},
  primaryClass  = {quant-ph},
  url           = {https://arxiv.org/abs/2508.18027}
}

@article{Ai_2025,
  title   = {Scalable Parameter Design for Superconducting Quantum Circuits with Graph Neural Networks},
  author  = {Ai, Hao and Liu, Yu-xi},
  journal = {Physical Review Letters},
  volume  = {135},
  number  = {4},
  pages   = {040601},
  year    = {2025},
  month   = jul,
  doi     = {10.1103/yr9d-7z8k},
  url     = {http://dx.doi.org/10.1103/yr9d-7z8k},
  issn    = {1079-7114},
  publisher = {American Physical Society (APS)}
}

@article{Lu_2025,
  title   = {Neural Network-Based Frequency Optimization for Superconducting Quantum Chips},
  author  = {Lu, Bin-Han and Li, Qing-Song and Wang, Peng and Chen, Zhao-Yun and Wu, Yu-Chun and Guo, Guo-Ping},
  journal = {Chinese Physics Letters},
  volume  = {42},
  number  = {3},
  pages   = {030204},
  year    = {2025},
  month   = mar,
  doi     = {10.1088/0256-307X/42/3/030204},
  url     = {http://dx.doi.org/10.1088/0256-307X/42/3/030204},
  issn    = {1741-3540},
  publisher = {IOP Publishing}
}

@article{Kudyshev_2021,
  title   = {Machine Learning for Integrated Quantum Photonics},
  author  = {Kudyshev, Zhaxylyk A. and Shalaev, Vladimir M. and Boltasseva, Alexandra},
  journal = {ACS Photonics},
  volume  = {8},
  number  = {1},
  pages   = {34--46},
  year    = {2021},
  month   = jan,
  doi     = {10.1021/acsphotonics.0c00960},
  url     = {http://dx.doi.org/10.1021/acsphotonics.0c00960},
  issn    = {2330-4022},
  publisher = {American Chemical Society (ACS)}
}

@article{Mayor_2026,
  title   = {Robotic chip-scale nanofabrication for superior consistency},
  author  = {Mayor, Felix M. and Guan, Wenyan and Szakiel, Erik and Safavi-Naeini, Amir H. and Gyger, Samuel},
  journal = {Applied Physics Letters},
  volume  = {128},
  number  = {14},
  pages   = {144101},
  year    = {2026},
  month   = apr,
  doi     = {10.1063/5.0313577},
  url     = {http://dx.doi.org/10.1063/5.0313577},
  issn    = {1077-3118},
  publisher = {AIP Publishing}
}

@article{Mei_2021,
  title   = {Optimization of quantum-dot qubit fabrication via machine learning},
  author  = {Mei, Antonio B. and Milosavljevic, Ivan and Simpson, Amanda L. and Smetanka, Valerie A. and Feeney, Colin P. and Seguin, Shay M. and Ha, Sieu D. and Ha, Wonill and Reed, Matthew D.},
  journal = {Applied Physics Letters},
  volume  = {118},
  number  = {20},
  pages   = {204001},
  year    = {2021},
  month   = may,
  doi     = {10.1063/5.0040967},
  url     = {http://dx.doi.org/10.1063/5.0040967},
  issn    = {1077-3118},
  publisher = {AIP Publishing}
}

@article{Tranter_2024,
  title   = {Machine Learning-Assisted Precision Manufacturing of Atom Qubits in Silicon},
  author  = {Tranter, Aaron D. and Kranz, Ludwik and Sutherland, Sam and Keizer, Joris G. and Gorman, Samuel K. and Buchler, Benjamin C. and Simmons, Michelle Y.},
  journal = {ACS Nano},
  year    = {2024},
  month   = jul,
  doi     = {10.1021/acsnano.4c00080},
  url     = {http://dx.doi.org/10.1021/acsnano.4c00080},
  issn    = {1936-086X},
  publisher = {American Chemical Society (ACS)}
}

@article{Usman_2020,
  title   = {Framework for atomic-level characterisation of quantum computer arrays by machine learning},
  author  = {Usman, Muhammad and Wong, Yi Zheng and Hill, Charles D. and Hollenberg, Lloyd C. L.},
  journal = {npj Computational Materials},
  volume  = {6},
  number  = {1},
  year    = {2020},
  month   = mar,
  doi     = {10.1038/s41524-020-0282-0},
  url     = {http://dx.doi.org/10.1038/s41524-020-0282-0},
  issn    = {2057-3960},
  publisher = {Springer Science and Business Media LLC}
}

@article{Rashidi_2020,
  title   = {Deep learning-guided surface characterization for autonomous hydrogen lithography},
  author  = {Rashidi, Mohammad and Croshaw, Jeremiah and Mastel, Kieran and Tamura, Marcus and Hosseinzadeh, Hedieh and Wolkow, Robert A.},
  journal = {Machine Learning: Science and Technology},
  volume  = {1},
  number  = {2},
  pages   = {025001},
  year    = {2020},
  month   = mar,
  doi     = {10.1088/2632-2153/ab6d5e},
  url     = {http://dx.doi.org/10.1088/2632-2153/ab6d5e},
  issn    = {2632-2153},
  publisher = {IOP Publishing}
}

@article{Rashidi_2018,
  title   = {Autonomous Scanning Probe Microscopy in Situ Tip Conditioning through Machine Learning},
  author  = {Rashidi, Mohammad and Wolkow, Robert A.},
  journal = {ACS Nano},
  volume  = {12},
  number  = {6},
  pages   = {5185--5189},
  year    = {2018},
  month   = may,
  doi     = {10.1021/acsnano.8b02208},
  url     = {http://dx.doi.org/10.1021/acsnano.8b02208},
  issn    = {1936-086X},
  publisher = {American Chemical Society (ACS)}
}

@article{Donges_2022,
  title   = {Machine learning enhanced in situ electron beam lithography of photonic nanostructures},
  author  = {Donges, Jan and Schlischka, Marvin and Shih, Ching-Wen and Pengerla, Monica and Limame, Imad and Schall, Johannes and Bremer, Lucas and Rodt, Sven and Reitzenstein, Stephan},
  journal = {Nanoscale},
  volume  = {14},
  number  = {39},
  pages   = {14529--14536},
  year    = {2022},
  doi     = {10.1039/D2NR03696G},
  url     = {http://dx.doi.org/10.1039/D2NR03696G},
  issn    = {2040-3372},
  publisher = {Royal Society of Chemistry (RSC)}
}

@misc{deQuilettes_2025,
  title         = {Machine Learning Enables Optimization of Diamond for Quantum Applications},
  author        = {deQuilettes, Dane W. and Price, Eden and Pham, Linh M. and Kurlej, Arthur and Vattam, Swaroop and Melville, Alexander and Osadchy, Tom and Li, Boning and Wang, Guoqing and Muniz, Collin N. and Cappellaro, Paola and Schloss, Jennifer M. and Mallek, Justin L. and Braje, Danielle A.},
  year          = {2025},
  eprint        = {2510.22121},
  archivePrefix = {arXiv},
  primaryClass  = {quant-ph},
  url           = {https://arxiv.org/abs/2510.22121}
}

@article{Wigley_2016,
  title   = {Fast machine-learning online optimization of ultra-cold-atom experiments},
  author  = {Wigley, P. B. and Everitt, P. J. and van den Hengel, A. and Bastian, J. W. and Sooriyabandara, M. A. and McDonald, G. D. and Hardman, K. S. and Quinlivan, C. D. and Manju, P. and Kuhn, C. C. N. and Petersen, I. R. and Luiten, A. N. and Hope, J. J. and Robins, N. P. and Hush, M. R.},
  journal = {Scientific Reports},
  volume  = {6},
  number  = {1},
  pages   = {25890},
  year    = {2016},
  month   = may,
  doi     = {10.1038/srep25890},
  url     = {http://dx.doi.org/10.1038/srep25890},
  issn    = {2045-2322},
  publisher = {Springer Science and Business Media LLC}
}

@article{Vendeiro_2022,
  title   = {Machine-learning-accelerated Bose-Einstein condensation},
  author  = {Vendeiro, Zachary and Ramette, Joshua and Rudelis, Alyssa and Chong, Michelle and Sinclair, Josiah and Stewart, Luke and Urvoy, Alban and Vuleti{\'c}, Vladan},
  journal = {Physical Review Research},
  volume  = {4},
  number  = {4},
  pages   = {043216},
  year    = {2022},
  month   = dec,
  doi     = {10.1103/PhysRevResearch.4.043216},
  url     = {http://dx.doi.org/10.1103/PhysRevResearch.4.043216},
  issn    = {2643-1564},
  publisher = {American Physical Society (APS)}
}

@article{Milson_2023,
  title   = {High-dimensional reinforcement learning for optimization and control of ultracold quantum gases},
  author  = {Milson, N. and Tashchilina, A. and Ooi, T. and Czarnecka, A. and Ahmad, Z. F. and LeBlanc, L. J.},
  journal = {Machine Learning: Science and Technology},
  volume  = {4},
  number  = {4},
  pages   = {045057},
  year    = {2023},
  month   = dec,
  doi     = {10.1088/2632-2153/ad1437},
  url     = {http://dx.doi.org/10.1088/2632-2153/ad1437},
  issn    = {2632-2153},
  publisher = {IOP Publishing}
}

@article{Reinschmidt_2024,
  title   = {Reinforcement learning in cold atom experiments},
  author  = {Reinschmidt, Malte and Fort{\'a}gh, J{\'o}zsef and G{\"u}nther, Andreas and Volchkov, Valentin V.},
  journal = {Nature Communications},
  volume  = {15},
  number  = {1},
  year    = {2024},
  month   = oct,
  doi     = {10.1038/s41467-024-52775-8},
  url     = {http://dx.doi.org/10.1038/s41467-024-52775-8},
  issn    = {2041-1723},
  publisher = {Springer Science and Business Media LLC}
}

@article{Blatz_2024,
  title   = {Bayesian Optimization for Robust State Preparation in Quantum Many-Body Systems},
  author  = {Blatz, Tizian and Kwan, Joyce and L{\'e}onard, Julian and Bohrdt, Annabelle},
  journal = {Quantum},
  volume  = {8},
  pages   = {1388},
  year    = {2024},
  month   = jun,
  doi     = {10.22331/q-2024-06-27-1388},
  url     = {http://dx.doi.org/10.22331/q-2024-06-27-1388},
  issn    = {2521-327X},
  publisher = {Verein zur F{\"o}rderung des Open Access Publizierens in den Quantenwissenschaften}
}

@article{Kalantre_2019,
  title   = {Machine learning techniques for state recognition and auto-tuning in quantum dots},
  author  = {Kalantre, Sandesh S. and Zwolak, Justyna P. and Ragole, Stephen and Wu, Xingyao and Zimmerman, Neil M. and Stewart, M. D. and Taylor, Jacob M.},
  journal = {npj Quantum Information},
  volume  = {5},
  number  = {1},
  year    = {2019},
  month   = jan,
  doi     = {10.1038/s41534-018-0118-7},
  url     = {http://dx.doi.org/10.1038/s41534-018-0118-7},
  issn    = {2056-6387},
  publisher = {Springer Science and Business Media LLC}
}

@article{Zwolak_2020,
  title   = {Autotuning of Double-Dot Devices In Situ with Machine Learning},
  author  = {Zwolak, Justyna P. and McJunkin, Thomas and Kalantre, Sandesh S. and Dodson, J. P. and MacQuarrie, E. R. and Savage, D. E. and Lagally, M. G. and Coppersmith, S. N. and Eriksson, Mark A. and Taylor, Jacob M.},
  journal = {Physical Review Applied},
  volume  = {13},
  number  = {3},
  pages   = {034075},
  year    = {2020},
  month   = mar,
  doi     = {10.1103/PhysRevApplied.13.034075},
  url     = {http://dx.doi.org/10.1103/PhysRevApplied.13.034075},
  issn    = {2331-7019},
  publisher = {American Physical Society (APS)}
}

@article{Moon_2020,
  title   = {Machine learning enables completely automatic tuning of a quantum device faster than human experts},
  author  = {Moon, H. and Lennon, D. T. and Kirkpatrick, J. and van Esbroeck, N. M. and Camenzind, L. C. and Yu, Liuqi and Vigneau, F. and Zumb{\"u}hl, D. M. and Briggs, G. A. D. and Osborne, M. A. and Sejdinovic, D. and Laird, E. A. and Ares, N.},
  journal = {Nature Communications},
  volume  = {11},
  number  = {1},
  year    = {2020},
  month   = aug,
  doi     = {10.1038/s41467-020-17835-9},
  url     = {http://dx.doi.org/10.1038/s41467-020-17835-9},
  issn    = {2041-1723},
  publisher = {Springer Science and Business Media LLC}
}

@article{Zwolak_2023,
  title   = {Colloquium: Advances in automation of quantum dot devices control},
  author  = {Zwolak, Justyna P. and Taylor, Jacob M.},
  journal = {Reviews of Modern Physics},
  volume  = {95},
  number  = {1},
  pages   = {011006},
  year    = {2023},
  month   = feb,
  doi     = {10.1103/RevModPhys.95.011006},
  url     = {http://dx.doi.org/10.1103/RevModPhys.95.011006},
  issn    = {1539-0756},
  publisher = {American Physical Society (APS)}
}

@article{Schuff_2026,
  title   = {Fully autonomous tuning of a spin qubit},
  author  = {Schuff, Jonas and Carballido, Miguel J. and Kotzagiannidis, Madeleine and Calvo, Juan Carlos and Caselli, Marco and Rawling, Jacob and Craig, David L. and van Straaten, Barnaby and Severin, Brandon and Fedele, Federico and Svab, Simon and Chevalier Kwon, Pierre and Eggli, Rafael S. and Patlatiuk, Taras and Korda, Nathan and Zumb{\"u}hl, Dominik M. and Ares, Natalia},
  journal = {Nature Electronics},
  volume  = {9},
  number  = {3},
  pages   = {304--313},
  year    = {2026},
  month   = feb,
  doi     = {10.1038/s41928-025-01562-4},
  url     = {http://dx.doi.org/10.1038/s41928-025-01562-4},
  issn    = {2520-1131},
  publisher = {Springer Science and Business Media LLC}
}

@article{Nguyen_2021,
  title   = {Deep reinforcement learning for efficient measurement of quantum devices},
  author  = {Nguyen, V. and Orbell, S. B. and Lennon, D. T. and Moon, H. and Vigneau, F. and Camenzind, L. C. and Yu, L. and Zumb{\"u}hl, D. M. and Briggs, G. A. D. and Osborne, M. A. and Sejdinovic, D. and Ares, N.},
  journal = {npj Quantum Information},
  volume  = {7},
  number  = {1},
  year    = {2021},
  month   = jun,
  doi     = {10.1038/s41534-021-00434-x},
  url     = {http://dx.doi.org/10.1038/s41534-021-00434-x},
  issn    = {2056-6387},
  publisher = {Springer Science and Business Media LLC}
}

@article{Sivak_2022,
  title   = {Model-Free Quantum Control with Reinforcement Learning},
  author  = {Sivak, V. V. and Eickbusch, A. and Liu, H. and Royer, B. and Tsioutsios, I. and Devoret, M. H.},
  journal = {Physical Review X},
  volume  = {12},
  number  = {1},
  pages   = {011059},
  year    = {2022},
  month   = mar,
  doi     = {10.1103/PhysRevX.12.011059},
  url     = {http://dx.doi.org/10.1103/PhysRevX.12.011059},
  issn    = {2160-3308},
  publisher = {American Physical Society (APS)}
}

@article{Porotti_2022,
  title   = {Deep Reinforcement Learning for Quantum State Preparation with Weak Nonlinear Measurements},
  author  = {Porotti, Riccardo and Essig, Antoine and Huard, Benjamin and Marquardt, Florian},
  journal = {Quantum},
  volume  = {6},
  pages   = {747},
  year    = {2022},
  month   = jun,
  doi     = {10.22331/q-2022-06-28-747},
  url     = {http://dx.doi.org/10.22331/q-2022-06-28-747},
  issn    = {2521-327X},
  publisher = {Verein zur F{\"o}rderung des Open Access Publizierens in den Quantenwissenschaften}
}

@article{Li_2025,
  title   = {Robust quantum control using reinforcement learning from demonstration},
  author  = {Li, Shengyong and Fan, Yidian and Li, Xiang and Ruan, Xinhui and Zhao, Qianchuan and Peng, Zhihui and Wu, Re-Bing and Zhang, Jing and Song, Pengtao},
  journal = {npj Quantum Information},
  volume  = {11},
  number  = {1},
  year    = {2025},
  month   = jul,
  doi     = {10.1038/s41534-025-01065-2},
  url     = {http://dx.doi.org/10.1038/s41534-025-01065-2},
  issn    = {2056-6387},
  publisher = {Springer Science and Business Media LLC}
}

@article{Daraeizadeh_2020,
  title   = {Machine-learning-based three-qubit gate design for the Toffoli gate and parity check in transmon systems},
  author  = {Daraeizadeh, S. and Premaratne, S. P. and Khammassi, N. and Song, X. and Perkowski, M. and Matsuura, A. Y.},
  journal = {Physical Review A},
  volume  = {102},
  number  = {1},
  pages   = {012601},
  year    = {2020},
  month   = jul,
  doi     = {10.1103/PhysRevA.102.012601},
  url     = {http://dx.doi.org/10.1103/PhysRevA.102.012601},
  issn    = {2469-9934},
  publisher = {American Physical Society (APS)}
}

@article{Baum_2021,
  title   = {Experimental Deep Reinforcement Learning for Error-Robust Gate-Set Design on a Superconducting Quantum Computer},
  author  = {Baum, Yuval and Amico, Mirko and Howell, Sean and Hush, Michael and Liuzzi, Maggie and Mundada, Pranav and Merkh, Thomas and Carvalho, Andre R. R. and Biercuk, Michael J.},
  journal = {PRX Quantum},
  volume  = {2},
  number  = {4},
  pages   = {040324},
  year    = {2021},
  month   = nov,
  doi     = {10.1103/PRXQuantum.2.040324},
  url     = {http://dx.doi.org/10.1103/PRXQuantum.2.040324},
  issn    = {2691-3399},
  publisher = {American Physical Society (APS)}
}

@inproceedings{Wright_2023,
  title     = {Fast Quantum Gate Design with Deep Reinforcement Learning Using Real-Time Feedback on Readout Signals},
  author    = {Wright, Emily and De Sousa, Rog{\'e}rio},
  booktitle = {2023 IEEE International Conference on Quantum Computing and Engineering (QCE)},
  publisher = {IEEE},
  pages     = {1295--1303},
  year      = {2023},
  month     = sep,
  doi       = {10.1109/QCE57702.2023.00146},
  url       = {http://dx.doi.org/10.1109/QCE57702.2023.00146}
}

@article{Nguyen_2024,
  title   = {Reinforcement learning pulses for transmon qubit entangling gates},
  author  = {Nguyen, Ho Nam and Motzoi, Felix and Metcalf, Mekena and Whaley, K. Birgitta and Bukov, Marin and Schmitt, Markus},
  journal = {Machine Learning: Science and Technology},
  volume  = {5},
  number  = {2},
  pages   = {025066},
  year    = {2024},
  month   = jun,
  doi     = {10.1088/2632-2153/ad4f4d},
  url     = {http://dx.doi.org/10.1088/2632-2153/ad4f4d},
  issn    = {2632-2153},
  publisher = {IOP Publishing}
}

@article{Sarma_2025,
  title   = {Designing fast quantum gates using optimal control with a reinforcement-learning ansatz},
  author  = {Sarma, Bijita and Hartmann, Michael J.},
  journal = {Physical Review Applied},
  volume  = {23},
  number  = {1},
  pages   = {014015},
  year    = {2025},
  month   = jan,
  doi     = {10.1103/PhysRevApplied.23.014015},
  url     = {http://dx.doi.org/10.1103/PhysRevApplied.23.014015},
  issn    = {2331-7019},
  publisher = {American Physical Society (APS)}
}

@article{Bhat_2025,
  title   = {Machine learning for arbitrary single-qubit rotations on an embedded device},
  author  = {Bhat, Madhav Narayan and Russo, Marco and Carloni, Luca P. and Di Guglielmo, Giuseppe and Fahim, Farah and Li, Andy C. Y. and Perdue, Gabriel N.},
  journal = {Quantum Machine Intelligence},
  volume  = {7},
  number  = {1},
  year    = {2025},
  month   = jan,
  doi     = {10.1007/s42484-024-00214-8},
  url     = {http://dx.doi.org/10.1007/s42484-024-00214-8},
  issn    = {2524-4914},
  publisher = {Springer Science and Business Media LLC}
}

@article{Liu_2025,
  title   = {Superconducting quantum computing optimization based on multi-objective deep reinforcement learning},
  author  = {Liu, Yangting},
  journal = {Scientific Reports},
  volume  = {15},
  number  = {1},
  year    = {2025},
  month   = jan,
  doi     = {10.1038/s41598-024-73456-y},
  url     = {http://dx.doi.org/10.1038/s41598-024-73456-y},
  issn    = {2045-2322},
  publisher = {Springer Science and Business Media LLC}
}

@article{Magesan_2015,
  title   = {Machine Learning for Discriminating Quantum Measurement Trajectories and Improving Readout},
  author  = {Magesan, Easwar and Gambetta, Jay M. and C{\'o}rcoles, A. D. and Chow, Jerry M.},
  journal = {Physical Review Letters},
  volume  = {114},
  number  = {20},
  pages   = {200501},
  year    = {2015},
  month   = may,
  doi     = {10.1103/PhysRevLett.114.200501},
  url     = {http://dx.doi.org/10.1103/PhysRevLett.114.200501},
  issn    = {1079-7114},
  publisher = {American Physical Society (APS)}
}

@article{Seif_2018,
  title   = {Machine learning assisted readout of trapped-ion qubits},
  author  = {Seif, Alireza and Landsman, Kevin A. and Linke, Norbert M. and Figgatt, Caroline and Monroe, C. and Hafezi, Mohammad},
  journal = {Journal of Physics B: Atomic, Molecular and Optical Physics},
  volume  = {51},
  number  = {17},
  pages   = {174006},
  year    = {2018},
  month   = aug,
  doi     = {10.1088/1361-6455/aad62b},
  url     = {http://dx.doi.org/10.1088/1361-6455/aad62b},
  issn    = {1361-6455},
  publisher = {IOP Publishing}
}

@article{Martinez_2020,
  title   = {Improving qubit readout with hidden Markov models},
  author  = {Martinez, Luis A. and Rosen, Yaniv J. and DuBois, Jonathan L.},
  journal = {Physical Review A},
  volume  = {102},
  number  = {6},
  pages   = {062426},
  year    = {2020},
  month   = dec,
  doi     = {10.1103/PhysRevA.102.062426},
  url     = {http://dx.doi.org/10.1103/PhysRevA.102.062426},
  issn    = {2469-9934},
  publisher = {American Physical Society (APS)}
}

@article{Flurin_2020,
  title   = {Using a Recurrent Neural Network to Reconstruct Quantum Dynamics of a Superconducting Qubit from Physical Observations},
  author  = {Flurin, E. and Martin, L. S. and Hacohen-Gourgy, S. and Siddiqi, I.},
  journal = {Physical Review X},
  volume  = {10},
  number  = {1},
  pages   = {011006},
  year    = {2020},
  month   = jan,
  doi     = {10.1103/PhysRevX.10.011006},
  url     = {http://dx.doi.org/10.1103/PhysRevX.10.011006},
  issn    = {2160-3308},
  publisher = {American Physical Society (APS)}
}

@article{Koolstra_2022,
  title   = {Monitoring Fast Superconducting Qubit Dynamics Using a Neural Network},
  author  = {Koolstra, G. and Stevenson, N. and Barzili, S. and Burns, L. and Siva, K. and Greenfield, S. and Livingston, W. and Hashim, A. and Naik, R. K. and Kreikebaum, J. M. and O'Brien, K. P. and Santiago, D. I. and Dressel, J. and Siddiqi, I.},
  journal = {Physical Review X},
  volume  = {12},
  number  = {3},
  pages   = {031017},
  year    = {2022},
  month   = jul,
  doi     = {10.1103/PhysRevX.12.031017},
  url     = {http://dx.doi.org/10.1103/PhysRevX.12.031017},
  issn    = {2160-3308},
  publisher = {American Physical Society (APS)}
}

@article{Lienhard_2022,
  title   = {Deep-Neural-Network Discrimination of Multiplexed Superconducting-Qubit States},
  author  = {Lienhard, Benjamin and Veps{\"a}l{\"a}inen, Antti and Govia, Luke C. G. and Hoffer, Cole R. and Qiu, Jack Y. and Rist{\`e}, Diego and Ware, Matthew and Kim, David and Winik, Roni and Melville, Alexander and Niedzielski, Bethany and Yoder, Jonilyn and Ribeill, Guilhem J. and Ohki, Thomas A. and Krovi, Hari K. and Orlando, Terry P. and Gustavsson, Simon and Oliver, William D.},
  journal = {Physical Review Applied},
  volume  = {17},
  number  = {1},
  pages   = {014024},
  year    = {2022},
  month   = jan,
  doi     = {10.1103/PhysRevApplied.17.014024},
  url     = {http://dx.doi.org/10.1103/PhysRevApplied.17.014024},
  issn    = {2331-7019},
  publisher = {American Physical Society (APS)}
}

@misc{Azad_2022,
  title         = {Machine Learning based Discrimination for Excited State Promoted Readout},
  author        = {Azad, Utkarsh and Zhang, Helena},
  year          = {2022},
  eprint        = {2210.08574},
  archivePrefix = {arXiv},
  primaryClass  = {quant-ph},
  url           = {https://arxiv.org/abs/2210.08574}
}

@inproceedings{Maurya_2023,
  title     = {Scaling Qubit Readout with Hardware Efficient Machine Learning Architectures},
  author    = {Maurya, Satvik and Mude, Chaithanya Naik and Oliver, William D. and Lienhard, Benjamin and Tannu, Swamit},
  booktitle = {Proceedings of the 50th Annual International Symposium on Computer Architecture (ISCA '23)},
  publisher = {ACM},
  pages     = {1--13},
  year      = {2023},
  month     = jun,
  doi       = {10.1145/3579371.3589042},
  url       = {http://dx.doi.org/10.1145/3579371.3589042}
}

@article{Luchi_2023,
  title   = {Enhancing Qubit Readout with Autoencoders},
  author  = {Luchi, Piero and Trevisanutto, Paolo E. and Roggero, Alessandro and DuBois, Jonathan L. and Rosen, Yaniv J. and Turro, Francesco and Amitrano, Valentina and Pederiva, Francesco},
  journal = {Physical Review Applied},
  volume  = {20},
  number  = {1},
  pages   = {014045},
  year    = {2023},
  month   = jul,
  doi     = {10.1103/PhysRevApplied.20.014045},
  url     = {http://dx.doi.org/10.1103/PhysRevApplied.20.014045},
  issn    = {2331-7019},
  publisher = {American Physical Society (APS)}
}

@article{Cosco_2023,
  title   = {Enhancing qubit readout with Bayesian learning},
  author  = {Cosco, F. and Lo Gullo, N.},
  journal = {Physical Review A},
  volume  = {108},
  number  = {6},
  pages   = {L060402},
  year    = {2023},
  month   = dec,
  doi     = {10.1103/PhysRevA.108.L060402},
  url     = {http://dx.doi.org/10.1103/PhysRevA.108.L060402},
  issn    = {2469-9934},
  publisher = {American Physical Society (APS)}
}

@misc{You_2023,
  title         = {Neural network based time-resolved state tomography of superconducting qubits},
  author        = {You, Ziyang and Duan, Jiheng and Huang, Wenhui and Zhang, Libo and Liu, Song and Zhong, Youpeng and Ian, Hou},
  year          = {2023},
  eprint        = {2312.07958},
  archivePrefix = {arXiv},
  primaryClass  = {quant-ph},
  url           = {https://arxiv.org/abs/2312.07958}
}

@misc{Cao_2024,
  title         = {Superconducting qubit readout enhanced by path signature},
  author        = {Cao, Shuxiang and Shao, Zhen and Zheng, Jian-Qing and Alghadeer, Mohammed and Fasciati, Simone D. and Piscitelli, Michele and Spring, Peter A. and Wang, Shiyu and Tamate, Shuhei and Vora, Neel and Xu, Yilun and Huang, Gang and Nowrouzi, Kasra and Nakamura, Yasunobu and Siddiqi, Irfan and Leek, Peter and Lyons, Terry and Bakr, Mustafa},
  year          = {2024},
  eprint        = {2402.09532},
  archivePrefix = {arXiv},
  primaryClass  = {quant-ph},
  url           = {https://arxiv.org/abs/2402.09532}
}

@misc{Vora_2024,
  title         = {ML-Powered FPGA-based Real-Time Quantum State Discrimination Enabling Mid-circuit Measurements},
  author        = {Vora, Neel R. and Xu, Yilun and Hashim, Akel and Fruitwala, Neelay and Nguyen, Ho Nam and Liao, Haoran and Balewski, Jan and Rajagopala, Abhi and Nowrouzi, Kasra and Ji, Qing and Whaley, K. Birgitta and Siddiqi, Irfan and Nguyen, Phuc and Huang, Gang},
  year          = {2024},
  eprint        = {2406.18807},
  archivePrefix = {arXiv},
  primaryClass  = {quant-ph},
  url           = {https://arxiv.org/abs/2406.18807}
}

@misc{Gautam_2024,
  title         = {Low-latency machine learning FPGA accelerator for multi-qubit-state discrimination},
  author        = {Gautam, Pradeep Kumar and Kalipatnapu, Shantharam and {Shankaranarayanan H} and Singhal, Ujjawal and Lienhard, Benjamin and Singh, Vibhor and Thakur, Chetan Singh},
  year          = {2024},
  eprint        = {2407.03852},
  archivePrefix = {arXiv},
  primaryClass  = {quant-ph},
  url           = {https://arxiv.org/abs/2407.03852}
}

@article{Chatterjee_2025,
  title   = {Enhanced qubit readout via reinforcement learning},
  author  = {Chatterjee, Aniket and Schwinger, Jonathan and Gao, Yvonne Y.},
  journal = {Physical Review Applied},
  volume  = {23},
  number  = {5},
  pages   = {054057},
  year    = {2025},
  month   = may,
  doi     = {10.1103/PhysRevApplied.23.054057},
  url     = {http://dx.doi.org/10.1103/PhysRevApplied.23.054057},
  issn    = {2331-7019},
  publisher = {American Physical Society (APS)}
}

@article{DiGuglielmo_2025,
  title   = {End-to-End Workflow for Machine-Learning-Based Qubit Readout With QICK and hls4ml},
  author  = {Di Guglielmo, Giuseppe and Du, Botao and Campos, Javier and Boltasseva, Alexandra and Dixit, Akash and Fahim, Farah and Kudyshev, Zhaxylyk and Lopez, Santiago and Ma, Ruichao and Perdue, Gabriel N. and Tran, Nhan and Yesilyurt, Omer and Bowring, Daniel},
  journal = {IEEE Transactions on Quantum Engineering},
  volume  = {6},
  pages   = {1--10},
  year    = {2025},
  doi     = {10.1109/TQE.2025.3604712},
  url     = {http://dx.doi.org/10.1109/TQE.2025.3604712},
  issn    = {2689-1808},
  publisher = {Institute of Electrical and Electronics Engineers (IEEE)}
}

@inproceedings{Guo_2025,
  title     = {{KLiNQ}: Knowledge Distillation-Assisted Lightweight Neural Network for Qubit Readout on FPGA},
  author    = {Guo, Xiaorang and Bunarjyan, Tigran and Liu, Dai and Lienhard, Benjamin and Schulz, Martin},
  booktitle = {2025 62nd ACM/IEEE Design Automation Conference (DAC)},
  publisher = {IEEE},
  pages     = {1--7},
  year      = {2025},
  month     = jun,
  doi       = {10.1109/DAC63849.2025.11132854},
  url       = {http://dx.doi.org/10.1109/DAC63849.2025.11132854}
}

@inproceedings{Mude_2025,
  title     = {Efficient and Scalable Architectures for Multi-level Superconducting Qubit Readout},
  author    = {Mude, Chaithanya Naik and Maurya, Satvik and Lienhard, Benjamin and Tannu, Swamit},
  booktitle = {2025 62nd ACM/IEEE Design Automation Conference (DAC)},
  publisher = {IEEE},
  pages     = {1--7},
  year      = {2025},
  month     = jun,
  doi       = {10.1109/DAC63849.2025.11133314},
  url       = {http://dx.doi.org/10.1109/DAC63849.2025.11133314}
}

@article{Cosco_2025,
  title   = {Bayesian mitigation of measurement errors in multiqubit experiments},
  author  = {Cosco, F. and Plastina, F. and Lo Gullo, N.},
  journal = {Physical Review A},
  volume  = {112},
  number  = {4},
  pages   = {042621},
  year    = {2025},
  month   = oct,
  doi     = {10.1103/d65d-x8lt},
  url     = {http://dx.doi.org/10.1103/d65d-x8lt},
  issn    = {2469-9934},
  publisher = {American Physical Society (APS)}
}

@article{Kent_2026,
  title   = {Superconducting-qubit readout using next-generation reservoir computing},
  author  = {Kent, Robert and Lienhard, Benjamin and Lafyatis, Gregory and Gauthier, Daniel J.},
  journal = {Physical Review Applied},
  volume  = {25},
  number  = {4},
  pages   = {044009},
  year    = {2026},
  month   = apr,
  doi     = {10.1103/bnwn-d2p4},
  url     = {http://dx.doi.org/10.1103/bnwn-d2p4},
  issn    = {2331-7019},
  publisher = {American Physical Society (APS)}
}

@article{Reuer_2023,
  title   = {Realizing a deep reinforcement learning agent for real-time quantum feedback},
  author  = {Reuer, Kevin and Landgraf, Jonas and F{\"o}sel, Thomas and O'Sullivan, James and Beltr{\'a}n, Liberto and Akin, Abdulkadir and Norris, Graham J. and Remm, Ants and Kerschbaum, Michael and Besse, Jean-Claude and Marquardt, Florian and Wallraff, Andreas and Eichler, Christopher},
  journal = {Nature Communications},
  volume  = {14},
  number  = {1},
  year    = {2023},
  month   = nov,
  doi     = {10.1038/s41467-023-42901-3},
  url     = {http://dx.doi.org/10.1038/s41467-023-42901-3},
  issn    = {2041-1723},
  publisher = {Springer Science and Business Media LLC}
}

@article{Bejanin_2021,
  title   = {Resonant Coupling Parameter Estimation with Superconducting Qubits},
  author  = {B{\'e}janin, J. H. and Earnest, C. T. and Sanders, Y. R. and Mariantoni, M.},
  journal = {PRX Quantum},
  volume  = {2},
  number  = {4},
  pages   = {040343},
  year    = {2021},
  month   = nov,
  doi     = {10.1103/PRXQuantum.2.040343},
  url     = {http://dx.doi.org/10.1103/PRXQuantum.2.040343},
  issn    = {2691-3399},
  publisher = {American Physical Society (APS)}
}

@article{Genois_2021,
  title   = {Quantum-Tailored Machine-Learning Characterization of a Superconducting Qubit},
  author  = {Genois, {\'E}lie and Gross, Jonathan A. and Di Paolo, Agustin and Stevenson, Noah J. and Koolstra, Gerwin and Hashim, Akel and Siddiqi, Irfan and Blais, Alexandre},
  journal = {PRX Quantum},
  volume  = {2},
  number  = {4},
  pages   = {040355},
  year    = {2021},
  month   = dec,
  doi     = {10.1103/PRXQuantum.2.040355},
  url     = {http://dx.doi.org/10.1103/PRXQuantum.2.040355},
  issn    = {2691-3399},
  publisher = {American Physical Society (APS)}
}

@misc{Qian_2022,
  title         = {Fast Quantum Calibration using Bayesian Optimization with State Parameter Estimator for Non-Markovian Environment},
  author        = {Qian, Peng and Qamar, Shahid and Xiao, Xiao and Gu, Yanwu and Chai, Xudan and Zhao, Zhen and Forcellini, Nicolo and Liu, Dong E.},
  year          = {2022},
  eprint        = {2205.12929},
  archivePrefix = {arXiv},
  primaryClass  = {quant-ph},
  url           = {https://arxiv.org/abs/2205.12929}
}

@article{Barrett_2023,
  title   = {Learning-Based Calibration of Flux Crosstalk in Transmon Qubit Arrays},
  author  = {Barrett, Cora N. and Karamlou, Amir H. and Muschinske, Sarah E. and Rosen, Ilan T. and Braum{\"u}ller, Jochen and Das, Rabindra and Kim, David K. and Niedzielski, Bethany M. and Schuldt, Meghan and Serniak, Kyle and Schwartz, Mollie E. and Yoder, Jonilyn L. and Orlando, Terry P. and Gustavsson, Simon and Grover, Jeffrey A. and Oliver, William D.},
  journal = {Physical Review Applied},
  volume  = {20},
  number  = {2},
  pages   = {024070},
  year    = {2023},
  month   = aug,
  doi     = {10.1103/PhysRevApplied.20.024070},
  url     = {http://dx.doi.org/10.1103/PhysRevApplied.20.024070},
  issn    = {2331-7019},
  publisher = {American Physical Society (APS)}
}

@article{Genois_2025,
  title   = {Quantum optimal control of superconducting qubits based on machine-learning characterization},
  author  = {Genois, {\'E}lie and Stevenson, Noah J. and Goss, Noah and Siddiqi, Irfan and Blais, Alexandre},
  journal = {Physical Review Applied},
  volume  = {24},
  number  = {3},
  year    = {2025},
  month   = sep,
  doi     = {10.1103/d9yg-d3qr},
  url     = {http://dx.doi.org/10.1103/d9yg-d3qr},
  issn    = {2331-7019},
  publisher = {American Physical Society (APS)}
}

@article{Berritta_2025,
  title   = {Efficient Qubit Calibration by Binary-Search Hamiltonian Tracking},
  author  = {Berritta, Fabrizio and Benestad, Jacob and Pahl, Lukas and Mathews, Melvin and Krzywda, Jan A. and Assouly, R{\'e}ouven and Sung, Youngkyu and Kim, David K. and Niedzielski, Bethany M. and Serniak, Kyle and Schwartz, Mollie E. and Yoder, Jonilyn L. and Chatterjee, Anasua and Grover, Jeffrey A. and Danon, Jeroen and Oliver, William D. and Kuemmeth, Ferdinand},
  journal = {PRX Quantum},
  volume  = {6},
  number  = {3},
  pages   = {030335},
  year    = {2025},
  month   = aug,
  doi     = {10.1103/77qg-p68k},
  url     = {http://dx.doi.org/10.1103/77qg-p68k},
  issn    = {2691-3399},
  publisher = {American Physical Society (APS)}
}

@misc{Kung_2025,
  title         = {Automatic Characterization of Fluxonium Superconducting Qubits Parameters with Deep Transfer Learning},
  author        = {Kung, Huan-Hsuan and Liu, Chen-Yu and Lee, Qian-Rui and Hu, Chiang-Yuan and Chang, Yu-Chi and Chen, Ching-Yeh and Wang, Daw-Wei and Lin, Yen-Hsiang},
  year          = {2025},
  eprint        = {2503.12099},
  archivePrefix = {arXiv},
  primaryClass  = {quant-ph},
  url           = {https://arxiv.org/abs/2503.12099}
}

@article{Berritta_2026,
  title   = {Real-Time Adaptive Tracking of Fluctuating Relaxation Rates in Superconducting Qubits},
  author  = {Berritta, Fabrizio and Benestad, Jacob and Krzywda, Jan A. and Krause, Oswin and Marciniak, Malthe A. and Kr{\o}jer, Svend and Warren, Christopher W. and Hogedal, Emil and Nylander, Andreas and Ahmad, Irshad and Osman, Amr and Bizn{\'a}rov{\'a}, Janka and Rommel, Marcus and Roudsari, Anita Fadavi and Bylander, Jonas and Tancredi, Giovanna and Danon, Jeroen and Hastrup, Jacob and Kuemmeth, Ferdinand and Kjaergaard, Morten},
  journal = {Physical Review X},
  volume  = {16},
  number  = {1},
  pages   = {011025},
  year    = {2026},
  month   = feb,
  doi     = {10.1103/gk1b-stl3},
  url     = {http://dx.doi.org/10.1103/gk1b-stl3},
  issn    = {2160-3308},
  publisher = {American Physical Society (APS)}
}

@article{Gentile_2021,
  title   = {Learning models of quantum systems from experiments},
  author  = {Gentile, Antonio A. and Flynn, Brian and Knauer, Sebastian and Wiebe, Nathan and Paesani, Stefano and Granade, Christopher E. and Rarity, John G. and Santagati, Raffaele and Laing, Anthony},
  journal = {Nature Physics},
  volume  = {17},
  number  = {7},
  pages   = {837--843},
  year    = {2021},
  month   = apr,
  doi     = {10.1038/s41567-021-01201-7},
  url     = {http://dx.doi.org/10.1038/s41567-021-01201-7},
  issn    = {1745-2481},
  publisher = {Springer Science and Business Media LLC}
}

@article{Flynn_2022,
  title   = {Quantum model learning agent: characterisation of quantum systems through machine learning},
  author  = {Flynn, Brian and Gentile, Antonio A. and Wiebe, Nathan and Santagati, Raffaele and Laing, Anthony},
  journal = {New Journal of Physics},
  volume  = {24},
  number  = {5},
  pages   = {053034},
  year    = {2022},
  month   = may,
  doi     = {10.1088/1367-2630/ac68ff},
  url     = {http://dx.doi.org/10.1088/1367-2630/ac68ff},
  issn    = {1367-2630},
  publisher = {IOP Publishing}
}

@article{Youssry_2024,
  title   = {Experimental graybox quantum system identification and control},
  author  = {Youssry, Akram and Yang, Yang and Chapman, Robert J. and Haylock, Ben and Lenzini, Francesco and Lobino, Mirko and Peruzzo, Alberto},
  journal = {npj Quantum Information},
  volume  = {10},
  number  = {1},
  year    = {2024},
  month   = jan,
  doi     = {10.1038/s41534-023-00795-5},
  url     = {http://dx.doi.org/10.1038/s41534-023-00795-5},
  issn    = {2056-6387},
  publisher = {Springer Science and Business Media LLC}
}

@article{Cao_2025,
  title   = {Automating quantum computing laboratory experiments with an agent-based AI framework},
  author  = {Cao, Shuxiang and Zhang, Zijian and Alghadeer, Mohammed and Fasciati, Simone D. and Piscitelli, Michele and Bakr, Mustafa and Leek, Peter and Aspuru-Guzik, Al{\'a}n},
  journal = {Patterns},
  volume  = {6},
  number  = {10},
  pages   = {101372},
  year    = {2025},
  month   = oct,
  doi     = {10.1016/j.patter.2025.101372},
  url     = {http://dx.doi.org/10.1016/j.patter.2025.101372},
  issn    = {2666-3899},
  publisher = {Elsevier BV}
}

@misc{Li_2026,
  title         = {Large Language Model-Assisted Superconducting Qubit Experiments},
  author        = {Li, Shiheng and Miller, Jacob M. and Lee, Phoebe J. and Andersson, Gustav and Conner, Christopher R. and Joshi, Yash J. and Karimi, Bayan and King, Amber M. and Malc, Howard L. and Mishra, Harsh and Qiao, Hong and Ryu, Minseok and Wu, Xuntao and Xing, Siyuan and Yan, Haoxiong and Shi, Jian and Cleland, Andrew N.},
  year          = {2026},
  eprint        = {2603.08801},
  archivePrefix = {arXiv},
  primaryClass  = {quant-ph},
  url           = {https://arxiv.org/abs/2603.08801}
}

@article{briegel1998quantum,
  title = {Quantum Repeaters: The Role of Imperfect Local Operations in Quantum Communication},
  author = {Briegel, H.-J. and D\"ur, W. and Cirac, J. I. and Zoller, P.},
  journal = {Phys. Rev. Lett.},
  volume = {81},
  issue = {26},
  pages = {5932--5935},
  numpages = {0},
  year = {1998},
  month = {Dec},
  publisher = {American Physical Society},
  doi = {10.1103/PhysRevLett.81.5932},
  url = {https://link.aps.org/doi/10.1103/PhysRevLett.81.5932}
}

@article{dur1999quantum,
  title = {Quantum repeaters based on entanglement purification},
  author = {D\"ur, W. and Briegel, H.-J. and Cirac, J. I. and Zoller, P.},
  journal = {Phys. Rev. A},
  volume = {59},
  issue = {1},
  pages = {169--181},
  numpages = {0},
  year = {1999},
  month = {Jan},
  publisher = {American Physical Society},
  doi = {10.1103/PhysRevA.59.169},
  url = {https://link.aps.org/doi/10.1103/PhysRevA.59.169}
}

@article{azuma2023quantum,
  title = {Quantum repeaters: From quantum networks to the quantum internet},
  author = {Azuma, Koji and Economou, Sophia E. and Elkouss, David and Hilaire, Paul and Jiang, Liang and Lo, Hoi-Kwong and Tzitrin, Ilan},
  journal = {Rev. Mod. Phys.},
  volume = {95},
  issue = {4},
  pages = {045006},
  numpages = {66},
  year = {2023},
  month = {Dec},
  publisher = {American Physical Society},
  doi = {10.1103/RevModPhys.95.045006},
  url = {https://link.aps.org/doi/10.1103/RevModPhys.95.045006}
}

@article{vardoyan2020exact,
title = {On the exact analysis of an idealized quantum switch},
journal = {Performance Evaluation},
volume = {144},
pages = {102141},
year = {2020},
issn = {0166-5316},
doi = {https://doi.org/10.1016/j.peva.2020.102141},
url = {https://www.sciencedirect.com/science/article/pii/S0166531620300614},
author = {Gayane Vardoyan and Saikat Guha and Philippe Nain and Don Towsley}
}

@article{vardoyan2023capacity,
author = {Vardoyan, Gayane and Nain, Philippe and Guha, Saikat and Towsley, Don},
title = {On the Capacity Region of Bipartite and Tripartite Entanglement Switching},
year = {2023},
issue_date = {June 2023},
publisher = {Association for Computing Machinery},
address = {New York, NY, USA},
volume = {8},
number = {1–2},
issn = {2376-3639},
url = {https://doi.org/10.1145/3571809},
doi = {10.1145/3571809},
journal = {ACM Trans. Model. Perform. Eval. Comput. Syst.},
month = mar,
articleno = {1},
numpages = {18}
}

@Article{Pant2019,
author={Pant, Mihir
and Krovi, Hari
and Towsley, Don
and Tassiulas, Leandros
and Jiang, Liang
and Basu, Prithwish
and Englund, Dirk
and Guha, Saikat},
title={Routing entanglement in the quantum internet},
journal={npj Quantum Information},
year={2019},
month={Mar},
day={13},
volume={5},
number={1},
pages={25},
issn={2056-6387},
doi={10.1038/s41534-019-0139-x},
url={https://doi.org/10.1038/s41534-019-0139-x}
}

@Article{Inesta2023,
author={I{\~{n}}esta, {\'A}lvaro G.
and Vardoyan, Gayane
and Scavuzzo, Lara
and Wehner, Stephanie},
title={Optimal entanglement distribution policies in homogeneous repeater chains with cutoffs},
journal={npj Quantum Information},
year={2023},
month={May},
day={06},
volume={9},
number={1},
pages={46},
issn={2056-6387},
doi={10.1038/s41534-023-00713-9},
url={https://doi.org/10.1038/s41534-023-00713-9}
}

@article{Haldar2023-dz,
  title   = {Fast and reliable entanglement distribution with quantum repeaters: principles for improving protocols using reinforcement learning},
  author  = {Haldar, Stav and Barge, Pratik J. and Khatri, Sumeet and Lee, Hwang},
  journal = {Physical Review Applied},
  volume  = {21},
  number  = {2},
  pages   = {024041},
  year    = {2024},
  doi     = {10.1103/PhysRevApplied.21.024041}
}

@article{loock2023,
  title = {Deep reinforcement learning for key distribution based on quantum repeaters},
  author = {Rei\ss{}, Simon D. and van Loock, Peter},
  journal = {Phys. Rev. A},
  volume = {108},
  issue = {1},
  pages = {012406},
  numpages = {25},
  year = {2023},
  month = {Jul},
  publisher = {American Physical Society},
  doi = {10.1103/PhysRevA.108.012406},
  url = {https://link.aps.org/doi/10.1103/PhysRevA.108.012406}
}

@article{Haldar2024-xm,
  title   = {Reducing classical communication costs in multiplexed quantum repeaters using hardware-aware quasi-local policies},
  author  = {Haldar, Stav and Barge, Pratik J. and Cheng, Xiang and Chang, Kai-Chi and Kirby, Brian T. and Khatri, Sumeet and Wong, Chee Wei and Lee, Hwang},
  journal = {Communications Physics},
  volume  = {8},
  pages   = {132},
  year    = {2025},
  doi     = {10.1038/s42005-025-02062-9}
}

@article{Li2024-bw,
  title   = {Optimising entanglement distribution policies under classical communication constraints assisted by reinforcement learning},
  author  = {Li, Jan and Coopmans, Tim and Emonts, Patrick and Goodenough, Kenneth and Tura, Jordi and van Nieuwenburg, Evert},
  journal = {Machine Learning: Science and Technology},
  volume  = {6},
  number  = {3},
  pages   = {035024},
  year    = {2025},
  doi     = {10.1088/2632-2153/adfa5e}
}

@inproceedings{casado2025,
author = {Casado, Andr\'{e}s Agust\'{\i} and Olivas, \'{A}lvaro Troyano and Faba, Javier and Robledo, Luis Miguel and Martin, Vicente and Ortiz, Laura},
title = {Reinforcement Learning for Entanglement Distribution in Quantum Networks},
year = {2025},
isbn = {9798400720970},
publisher = {Association for Computing Machinery},
address = {New York, NY, USA},
url = {https://doi.org/10.1145/3749096.3750028},
doi = {10.1145/3749096.3750028},
booktitle = {Proceedings of the 2nd Workshop on Quantum Networks and Distributed Quantum Computing},
pages = {66–68},
numpages = {3},
location = {Coimbra, Portugal},
series = {QuNet '25}
}

@misc{Yau2026-fx,
  title         = {Reinforcement learning for quantum network control with application-driven objectives},
  author        = {Yau, Guo Xian and Burushkina, Alexandra and Ferreira da Silva, Francisco and Maji, Subhransu and Thomas, Philip S. and Vardoyan, Gayane},
  year          = {2025},
  eprint        = {2509.10634},
  archivePrefix = {arXiv},
  primaryClass  = {quant-ph},
  note          = {arXiv:2509.10634}
}

@INPROCEEDINGS{islam2024,
  author={Islam, Tasdiqul and Arifuzzaman, Md and Arslan, Engin},
  booktitle={2024 International Conference on Quantum Communications, Networking, and Computing (QCNC)}, 
  title={Reinforcement Learning Based Proactive Entanglement Swapping for Quantum Networks}, 
  year={2024},
  volume={},
  number={},
  pages={135-142},
  doi={10.1109/QCNC62729.2024.00030}}

@ARTICLE{lenguyen2022,
  author={Le, Linh and Nguyen, Tu N.},
  journal={IEEE Transactions on Quantum Engineering}, 
  title={DQRA: Deep Quantum Routing Agent for Entanglement Routing in Quantum Networks}, 
  year={2022},
  volume={3},
  number={},
  pages={1-12},
  doi={10.1109/TQE.2022.3148667}}

@Article{Roik2024,
author={Roik, Jan
and Bartkiewicz, Karol
and {\v{C}}ernoch, Anton{\'i}n
and Lemr, Karel},
title={Routing in quantum communication networks using reinforcement machine learning},
journal={Quantum Information Processing},
year={2024},
month={Mar},
day={04},
volume={23},
number={3},
pages={89},
issn={1573-1332},
doi={10.1007/s11128-024-04287-z},
url={https://doi.org/10.1007/s11128-024-04287-z}
}

@misc{Taherpour2025,
  title         = {Robust belief-state policy learning for quantum network routing under decoherence and time-varying conditions},
  author        = {Taherpour, Amirhossein and Taherpour, Abbas and Khattab, Tamer},
  year          = {2025},
  eprint        = {2509.08654},
  archivePrefix = {arXiv},
  primaryClass  = {quant-ph},
  note          = {arXiv:2509.08654}
}

@misc{Meuser2026-bv,
  title         = {{SatQNet}: Satellite-assisted quantum network entanglement routing using directed line graph neural networks},
  author        = {Meuser, Tobias and Weil, Jannis and Lahiri, Aninda and Paraschiv, Marius},
  year          = {2026},
  eprint        = {2604.09306},
  archivePrefix = {arXiv},
  primaryClass  = {quant-ph},
  note          = {arXiv:2604.09306}
}

@misc{Meuser2025-kg,
  title         = {{RELiQ}: Scalable entanglement routing via reinforcement learning in quantum networks},
  author        = {Meuser, Tobias and Weil, Jannis and Lahiri, Aninda and Paraschiv, Marius},
  year          = {2025},
  eprint        = {2511.22321},
  archivePrefix = {arXiv},
  primaryClass  = {quant-ph},
  note          = {arXiv:2511.22321}
}

@INPROCEEDINGS{Mobayenjarihani,
  author={Mobayenjarihani, Mohammad and Vardoyan, Gayane and Towsley, Don},
  booktitle={2023 IEEE International Conference on Quantum Computing and Engineering (QCE)}, 
  title={Optimistic Entanglement Purification in Quantum Networks}, 
  year={2023},
  volume={01},
  number={},
  pages={1143-1153},
  doi={10.1109/QCE57702.2023.00129}}

@misc{Prielinger2024,
  title         = {Surrogate-guided optimization in quantum networks},
  author        = {Prielinger, Luise and I{\~n}esta, {\'A}lvaro G. and Vardoyan, Gayane},
  year          = {2024},
  eprint        = {2407.17195},
  archivePrefix = {arXiv},
  primaryClass  = {quant-ph},
  note          = {arXiv:2407.17195}
}

@INPROCEEDINGS{Vardoyan_qnum,
  author={Vardoyan, Gayane and Wehner, Stephanie},
  booktitle={2023 IEEE International Conference on Quantum Computing and Engineering (QCE)}, 
  title={Quantum Network Utility Maximization}, 
  year={2023},
  volume={01},
  number={},
  pages={1238-1248},
  doi={10.1109/QCE57702.2023.00140}}

@Article{Inesta2026,
author={I{\~{n}}esta, {\'A}lvaro G.
and Davies, Bethany
and Kar, Sounak
and Wehner, Stephanie},
title={Entanglement buffering with multiple quantum memories},
journal={npj Quantum Information},
year={2026},
month={Feb},
day={20},
volume={12},
number={1},
pages={64},
issn={2056-6387},
doi={10.1038/s41534-025-01161-3},
url={https://doi.org/10.1038/s41534-025-01161-3}
}

@misc{Mukherjee2024-mf,
  title         = {Quantum network tomography of Rydberg arrays by machine learning},
  author        = {Mukherjee, Kaustav and Schachenmayer, Johannes and Whitlock, Shannon and W{\"u}ster, Sebastian},
  year          = {2024},
  eprint        = {2412.05742},
  archivePrefix = {arXiv},
  primaryClass  = {quant-ph},
  note          = {arXiv:2412.05742}
}

@article{canonici2024,
author = {Canonici, Ettore and Martina, Stefano and Mengoni, Riccardo and Ottaviani, Daniele and Caruso, Filippo},
title = {Machine Learning based Noise Characterization and Correction on Neutral Atoms NISQ Devices},
journal = {Advanced Quantum Technologies},
volume = {7},
number = {1},
pages = {2300192},
doi = {https://doi.org/10.1002/qute.202300192},
url = {https://advanced.onlinelibrary.wiley.com/doi/abs/10.1002/qute.202300192},
year = {2024}
}

@misc{Wang2025,
  title         = {Quantum network tomography for general topology with {SPAM} errors},
  author        = {Wang, Xuchuang and De Andrade, Matheus Guedes and Avis, Guus and Chen, Yu-Zhen Janice and Hajiesmaili, Mohammad and Towsley, Don},
  year          = {2025},
  eprint        = {2511.01074},
  archivePrefix = {arXiv},
  primaryClass  = {cs.NI},
  note          = {arXiv:2511.01074}
}

@ARTICLE{GuedesDeAndrade2024,
  author={Guedes de Andrade, Matheus and Navas, Jake and Guha, Saikat and Montaño, Inès and Raymer, Michael and Smith, Brian and Towsley, Don},
  journal={IEEE Network}, 
  title={Quantum Network Tomography}, 
  year={2024},
  volume={38},
  number={5},
  pages={114-122},
  doi={10.1109/MNET.2024.3403805}}

@misc{DeRieux2024-io,
  title         = {{EQMARL}: Entangled quantum multi-agent reinforcement learning for distributed cooperation over quantum channels},
  author        = {DeRieux, Alexander and Saad, Walid},
  year          = {2024},
  eprint        = {2405.17486},
  archivePrefix = {arXiv},
  primaryClass  = {quant-ph},
  note          = {arXiv:2405.17486}
}

@article{huang2025continuous,
  title={Continuous-variable Quantum Diffusion Model for State Generation and Restoration},
  author={Huang, Haitao and Chen, Chuangtao and Zhao, Qinglin},
  journal={arXiv preprint arXiv:2506.19270},
  year={2025}
}

@article{de2024quantum,
  title={Quantum latent diffusion models},
  author={De Falco, Francesca and Ceschini, Andrea and Sebastianelli, Alessandro and Le Saux, Bertrand and Panella, Massimo},
  journal={Quantum Machine Intelligence},
  volume={6},
  number={2},
  pages={85},
  year={2024},
  publisher={Springer}
}

@article{parigi2025quantum,
  title={Quantum-Noise-Driven Generative Diffusion Models},
  author={Parigi, Marco and Martina, Stefano and Caruso, Filippo},
  journal={Advanced Quantum Technologies},
  volume={8},
  number={12},
  pages={2300401},
  year={2025},
  publisher={Wiley Online Library}
}

@article{zhang2026parameter,
  title={Parameter-efficient quantum denoising diffusion probabilistic models with temporal encoding},
  author={Zhang, Xuefen and Chen, Chuangtao},
  journal={Future Generation Computer Systems},
  volume={174},
  pages={107981},
  year={2026},
  publisher={Elsevier}
}

@article{liu2025measurement,
  title={Measurement-Based Quantum Diffusion Models},
  author={Liu, Xinyu and Zhuang, Jingze and Hou, Wanda and You, Yi-Zhuang},
  journal={arXiv preprint arXiv:2508.08799},
  year={2025}
}

@article{zhang2025scaling,
  title={Scaling Laws of Quantum Information Lifetime in Monitored Quantum Dynamics},
  author={Zhang, Bingzhi and Hu, Fangjun and Mo, Runzhe and Chen, Tianyang and T{\"u}reci, Hakan E and Zhuang, Quntao},
  journal={arXiv preprint arXiv:2506.22755},
  year={2025}
}

@article{kwun2025mixed,
  title={Mixed-state quantum denoising diffusion probabilistic model},
  author={Kwun, Gino and Zhang, Bingzhi and Zhuang, Quntao},
  journal={Physical Review A},
  volume={111},
  number={3},
  pages={032610},
  year={2025},
  publisher={APS}
}

@article{zhang2024generative,
  title={Generative quantum machine learning via denoising diffusion probabilistic models},
  author={Zhang, Bingzhi and Xu, Peng and Chen, Xiaohui and Zhuang, Quntao},
  journal={Physical Review Letters},
  volume={132},
  number={10},
  pages={100602},
  year={2024},
  publisher={APS}
}

@article{zhang2025energy,
  title={Energy-dependent barren plateau in bosonic variational quantum circuits},
  author={Zhang, Bingzhi and Zhuang, Quntao},
  journal={Quantum Science and Technology},
  volume={10},
  number={1},
  pages={015009},
  year={2025},
  publisher={IOP Publishing}
}

@article{volkoff2021efficient,
  title={Efficient trainability of linear optical modules in quantum optical neural networks},
  author={Volkoff, Tyler J},
  journal={Journal of Russian Laser Research},
  volume={42},
  number={3},
  pages={250--260},
  year={2021},
  publisher={Springer}
}

@article{liu2025stochastic,
  title={Stochastic noise can be helpful for variational quantum algorithms},
  author={Liu, Junyu and Wilde, Frederik and Mele, Antonio Anna and Jin, Xin and Jiang, Liang and Eisert, Jens},
  journal={Physical Review A},
  volume={111},
  number={5},
  pages={052441},
  year={2025},
  publisher={APS}
}

@incollection{kochen2011problem,
  title={The problem of hidden variables in quantum mechanics},
  author={Kochen, Simon and Specker, Ernst P},
  booktitle={Ernst Specker Selecta},
  pages={235--263},
  year={2011},
  publisher={Springer}
}

@article{spekkens2005contextuality,
  title={Contextuality for preparations, transformations, and unsharp measurements},
  author={Spekkens, Robert W},
  journal={Physical Review A—Atomic, Molecular, and Optical Physics},
  volume={71},
  number={5},
  pages={052108},
  year={2005},
  publisher={APS}
}

@book{vapnik1995nature,
  title={The Nature of Statistical Learning Theory},
  author={Vapnik, Vladimir N.},
  year={1995},
  publisher={Springer},
  address={New York}
}

@article{anshu2024survey,
  title   = {A Survey on the Complexity of Learning Quantum States},
  author  = {Anshu, Anurag and Arunachalam, Srinivasan},
  journal = {Nature Reviews Physics},
  volume  = {6},
  number  = {1},
  pages   = {59--69},
  year    = {2024},
  doi     = {10.1038/s42254-023-00662-4}
}

@article{mohseni2006direct,
  title={Direct characterization of quantum dynamics},
  author={Mohseni, Masoud and Lidar, DA},
  journal={Physical review letters},
  volume={97},
  number={17},
  pages={170501},
  year={2006},
  publisher={APS}
}

@article{altepeter2003ancilla,
  title={Ancilla-assisted quantum process tomography},
  author={Altepeter, Joseph B and Branning, David and Jeffrey, Evan and Wei, TC and Kwiat, Paul G and Thew, Robert T and O’Brien, Jeremy L and Nielsen, Michael A and White, Andrew G},
  journal={Physical Review Letters},
  volume={90},
  number={19},
  pages={193601},
  year={2003},
  publisher={APS}
}

@article{chuang1997prescription,
  title={Prescription for experimental determination of the dynamics of a quantum black box},
  author={Chuang, Isaac L and Nielsen, Michael A},
  journal={Journal of Modern Optics},
  volume={44},
  number={11-12},
  pages={2455--2467},
  year={1997},
  publisher={Taylor \& Francis}
}

@article{merkel2013selfconsistent,
  title   = {Self-Consistent Quantum Process Tomography},
  author  = {Merkel, Seth T. and Gambetta, Jay M. and Smolin, John A. and Poletto, S. and C{\'o}rcoles, A. D. and Johnson, B. R. and Ryan, Colm A. and Steffen, M.},
  journal = {Physical Review A},
  volume  = {87},
  pages   = {062119},
  year    = {2013},
  doi     = {10.1103/PhysRevA.87.062119}
}

@article{greenbaum2015introduction,
  title         = {Introduction to Quantum Gate Set Tomography},
  author        = {Greenbaum, Daniel},
  journal       = {arXiv preprint arXiv:1509.02921},
  year          = {2015},
  eprint        = {1509.02921},
  archivePrefix = {arXiv},
  primaryClass  = {quant-ph}
}

@article{magesan2011scalable,
  title   = {Scalable and Robust Randomized Benchmarking of Quantum Processes},
  author  = {Magesan, Easwar and Gambetta, Jay M. and Emerson, Joseph},
  journal = {Physical Review Letters},
  volume  = {106},
  pages   = {180504},
  year    = {2011},
  doi     = {10.1103/PhysRevLett.106.180504}
}

@article{haah2017sample,
  title   = {Sample-Optimal Tomography of Quantum States},
  author  = {Haah, Jeongwan and Harrow, Aram W. and Ji, Zhengfeng and Wu, Xiaodi and Yu, Nengkun},
  journal = {IEEE Transactions on Information Theory},
  volume  = {63},
  number  = {9},
  pages   = {5628--5641},
  year    = {2017},
  doi     = {10.1109/TIT.2017.2719044}
}

@article{gross2010quantum,
  title   = {Quantum State Tomography via Compressed Sensing},
  author  = {Gross, David and Liu, Yi-Kai and Flammia, Steven T. and Becker, Stephen and Eisert, Jens},
  journal = {Physical Review Letters},
  volume  = {105},
  number  = {15},
  pages   = {150401},
  year    = {2010},
  doi     = {10.1103/PhysRevLett.105.150401}
}

@article{flammia2012quantum,
  title   = {Quantum Tomography via Compressed Sensing: Error Bounds, Sample Complexity and Efficient Estimators},
  author  = {Flammia, Steven T. and Gross, David and Liu, Yi-Kai and Eisert, Jens},
  journal = {New Journal of Physics},
  volume  = {14},
  number  = {9},
  pages   = {095022},
  year    = {2012},
  doi     = {10.1088/1367-2630/14/9/095022}
}

@article{cramer2010efficient,
  title   = {Efficient Quantum State Tomography},
  author  = {Cramer, Marcus and Plenio, Martin B. and Flammia, Steven T. and Somma, Rolando and Gross, David and Bartlett, Stephen D. and Landon-Cardinal, Olivier and Poulin, David and Liu, Yi-Kai},
  journal = {Nature Communications},
  volume  = {1},
  pages   = {149},
  year    = {2010},
  doi     = {10.1038/ncomms1147}
}

@article{lanyon2017efficient,
  title   = {Efficient Tomography of a Quantum Many-Body System},
  author  = {Lanyon, B. P. and Maier, C. and Holz{\"a}pfel, M. and Baumgratz, T. and Hempel, C. and Jurcevic, P. and Dhand, I. and Buyskikh, A. S. and Daley, A. J. and Cramer, M. and Plenio, M. B. and Blatt, R. and Roos, C. F.},
  journal = {Nature Physics},
  volume  = {13},
  pages   = {1158--1162},
  year    = {2017},
  doi     = {10.1038/nphys4244}
}

@article{aaronson2020shadow,
  title = {Shadow Tomography of Quantum States},
  author = {Aaronson, Scott},
  year = {2020},
  journal = {SIAM Journal on Computing},
  volume = {49},
  number = {5},
  eprint = {https://doi.org/10.1137/18M120275X},
  pages = {STOC18-368-STOC18-394},
  doi = {10.1137/18M120275X}
}

@article{bairey2019learning,
  title   = {Learning a Local Hamiltonian from Local Measurements},
  author  = {Bairey, Eyal and Arad, Itai and Lindner, Netanel H.},
  journal = {Physical Review Letters},
  volume  = {122},
  number  = {2},
  pages   = {020504},
  year    = {2019},
  doi     = {10.1103/PhysRevLett.122.020504}
}

@article{wiebe2014hamiltonian,
  title   = {Hamiltonian Learning and Certification Using Quantum Resources},
  author  = {Wiebe, Nathan and Granade, Christopher and Ferrie, Christopher and Cory, D. G.},
  journal = {Physical Review Letters},
  volume  = {112},
  number  = {19},
  pages   = {190501},
  year    = {2014},
  doi     = {10.1103/PhysRevLett.112.190501}
}

@article{kokail2019self,
  title   = {Self-Verifying variational quantum simulation of lattice models},
  author  = {Kokail, Christian and Maier, Christoph and van Bijnen, Rick and Brydges, Tim and Joshi, Manoj K. and Jurcevic, Petar and Muschik, Christine A. and Silvi, Pietro and Blatt, Rainer and Roos, Christian F. and Zoller, Peter},
  journal = {Nature},
  volume  = {569},
  pages   = {355--360},
  year    = {2019},
  doi     = {10.1038/s41586-019-1177-4}
}

@article{stilckfranca2024efficient,
  title = {Efficient and Robust Estimation of Many-Qubit {{Hamiltonians}}},
  author = {Stilck Fran{\c c}a, Daniel and Markovich, Liubov A. and Dobrovitski, V. V. and Werner, Albert H. and Borregaard, Johannes},
  year = {2024},
  journal = {Nature Communications},
  volume = {15},
  number = {1},
  pages = {311},
  doi = {10.1038/s41467-023-44012-5}
}

@article{gu2024practical,
  title = {Practical {{Hamiltonian}} Learning with Unitary Dynamics and {{Gibbs}} States},
  author = {Gu, Andi and Cincio, Lukasz and Coles, Patrick J.},
  year = {2024},
  journal = {Nature Communications},
  volume = {15},
  number = {1},
  pages = {312},
  doi = {10.1038/s41467-023-44008-1}
}

@article{hangleiter2024robustly,
  title = {Robustly Learning the {{Hamiltonian}} Dynamics of a Superconducting Quantum Processor},
  author = {Hangleiter, Dominik and Roth, Ingo and Fuksa, Jon{\'a}{\v s} and Eisert, Jens and Roushan, Pedram},
  year = {2024},
  journal = {Nature Communications},
  volume = {15},
  number = {1},
  pages = {9595},
  doi = {10.1038/s41467-024-52629-3}
}

@article{torlai2018neural,
  author  = {Giacomo Torlai and Guglielmo Mazzola and Juan Carrasquilla and Matthias Troyer and Roger Melko and Giuseppe Carleo},
  title   = {Neural-network quantum state tomography},
  journal = {Nature Physics},
  year    = {2018},
  volume  = {14},
  pages   = {447--450},
  doi     = {10.1038/s41567-018-0048-5}
}

@article{wei2024neuralshadow,
  author  = {Victor Wei and Christopher Muir and Roger Melko and Pooya Ronagh},
  title   = {Neural-shadow quantum state tomography},
  journal = {Physical Review Research},
  year    = {2024},
  volume  = {6},
  pages   = {023250},
  doi     = {10.1103/PhysRevResearch.6.023250}
}

@article{levy2024shadowqpt,
  author  = {Ryan Levy and Di Luo and Bryan K. Clark},
  title   = {Classical shadows for quantum process tomography on near-term quantum computers},
  journal = {Physical Review Research},
  year    = {2024},
  volume  = {6},
  pages   = {013029},
  doi     = {10.1103/PhysRevResearch.6.013029}
}

\end{document}